\documentclass[sn-basic]{sn-jnl}% Basic Springer Nature Reference Style/Chemistry Reference Style

\usepackage{graphicx}%
\usepackage{multirow}%
\usepackage{amsmath,amssymb,amsfonts}%
\usepackage{amsthm}%
\usepackage{mathrsfs}%
\usepackage[title]{appendix}%
\usepackage{xcolor}%
\usepackage{textcomp}%
\usepackage{manyfoot}%
\usepackage{booktabs}%
\usepackage{algorithm}%
\usepackage{algorithmicx}%
\usepackage{algpseudocode}%
\usepackage{listings}%
\usepackage{lscape}
\usepackage[leftcaption]{sidecap}
\usepackage[normalem]{ulem}%
\makeatletter
\renewcommand{\orcid}[1]{%
  \,\href{https://orcid.org/#1}{\includegraphics[height=1.6ex]{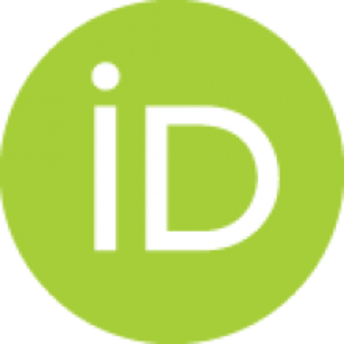}}%
}
\makeatother

\makeatletter
\input{aas_macros.sty}
\def\ref@jnl#1{{\jnl@style#1\ }}
\makeatother

\usepackage{color}
\definecolor{teal}{rgb}{0.0, 0.5, 0.5}
\begin{document}

\title[Sub-Millisecond Pulsars]{Sub-Millisecond Pulsars: Missing or Impossible?}

%%=============================================================%%
%% GivenName	-> \fnm{Joergen W.}
%% Particle	-> \spfx{van der} -> surname prefix
%% FamilyName	-> \sur{Ploeg}
%% Suffix	-> \sfx{IV}
%% \author*[1,2]{\fnm{Joergen W.} \spfx{van der} \sur{Ploeg} 
%%  \sfx{IV}}\email{iauthor@gmail.com}
%%=============================================================%%

\author*[1,2]{\fnm{Thomas M.} \sur{Tauris}\orcid{0000-0002-3865-7265}}\email{tauris@mp.aau.dk}
%Thomas, Paulo, Robert, Vivek, Scott, Alessandro, Christoph, Norbert L., Ed, Michael

\author[2]{\fnm{Vivek} \sur{Venkatraman Krishnan}\orcid{0000-0001-9518-9819}}\email{vkrishnan@mpifr-bonn.mpg.de}
%\equalcont{These authors contributed equally to this work.}

\author[2]{\fnm{Robert} \sur{Senzel}\orcid{0009-0002-0243-8199}}\email{rsenzel@mpifr-bonn.mpg.de}
%\equalcont{These authors contributed equally to this work.}

\author[2]{\fnm{Paulo C.~C.} \sur{Freire}\orcid{0000-0003-1307-9435}}\email{pfreire@mpifr-bonn.mpg.de}
%\equalcont{These authors contributed equally to this work.}

\author[3]{\fnm{Scott M.} \sur{Ransom}\orcid{0000-0001-5799-9714}}\email{sransom@nrao.edu}
%\equalcont{These authors contributed equally to this work.}

\author[4]{\fnm{Alessandro} \sur{Papitto}\orcid{0000-0001-6289-7413}}\email{alessandro.papitto@inaf.it}
%\equalcont{These authors contributed equally to this work.}

\author[5]{\fnm{Christophe A.~N.} \sur{Biscio}\orcid{0000-0002-2218-6825}}\email{christophe@math.aau.dk}
%\equalcont{These authors contributed equally to this work.}

\author[6,2]{\fnm{Norbert} \sur{Langer}\orcid{0000-0003-3026-0367}}\email{nlanger@uni-bonn.de}
%\equalcont{These authors contributed equally to this work.}

\author[7]{\fnm{Edward} \sur{van~den~Heuvel}\orcid{0000-0002-6417-2018}}\email{e.p.j.vandenheuvel@uva.nl}
%\equalcont{These authors contributed equally to this work.}

\author[2]{\fnm{Michael} \sur{Kramer}\orcid{0000-0002-4175-2271}}\email{mkramer@mpifr-bonn.mpg.de}
%\equalcont{These authors contributed equally to this work.}

%\author[8]{\fnm{Eleventh} \sur{Author}}\email{xauthor@gmail.com}
%\equalcont{These authors contributed equally to this work.}

\affil*[1]{\orgdiv{Department of Materials and Production}, \orgname{Aalborg University}, \orgaddress{\street{Fibigerstr{\ae}de 16}, \city{Aalborg}, \postcode{9220}, \country{Denmark}}}

\affil[2]{\orgdiv{Max-Planck-Institut f\"{u}r Radioastronomie}, \orgaddress{\street{Auf dem H\"{u}gel 69}, \city{Bonn}, \postcode{D-53121}, \country{Germany}}}

\affil[3]{\orgdiv{National Radio Astronomy Observatory}, \orgaddress{\street{520 Edgemont Road}, \city{Charlottesville}, \postcode{22903}, \state{VA}, \country{USA}}}

\affil[4]{\orgdiv{INAF Osservatorio Astronomico di Roma},  \orgaddress{\street{Via Frascati 33}, \city{Monte Porzio Catone, Roma}, \postcode{I-00078}, \country{Italy}}}

\affil[5]{\orgdiv{Department of Mathematical Sciences}, \orgname{Aalborg University}, \orgaddress{\street{Thomas Manns Vej 23}, \city{Aalborg}, \postcode{9220}, \country{Denmark}}}

\affil[6]{\orgdiv{Argelander Institut f\"{u}r Astronomie}, \orgname{University of Bonn}, \orgaddress{\street{Auf dem H\"{u}gel 71}, \city{Bonn}, \postcode{D-53121}, \country{Germany}}}

\affil[7]{\orgdiv{Anton Pannekoek Institute for Astronomy}, \orgname{University of Amsterdam}, \orgaddress{\street{PO Box 94249}, \city{Amsterdam}, \postcode{1090 GE}, \country{The Netherlands}}}

%{\affil[8]{\orgdiv{Department}, \orgname{Organization}, \orgaddress{\street{Street}, \city{City}, \postcode{610101}, \state{State}, \country{Country}}}
%}

%%==================================%%
%% Sample for unstructured abstract %%
%%==================================%%

\abstract{The minimum spin period attainable by a neutron star has been debated since the discovery of the first millisecond pulsar in 1982. A neutron star rotating faster than 1~ms would have far-reaching implications for the dense-matter equation of state, gravitational-wave emission, and the physics of accretion and spin-up. Yet despite the discovery of over 700 millisecond pulsars with spin periods between 1.4 and 10~ms, no sub-millisecond pulsar has been identified. Here we review the physical and observational constraints governing the formation, survival, and detectability of ultra-fast neutron stars. We discuss physical constraints imposed by the neutron star equation of state and their implications for the minimum attainable spin period, and we summarise gravitational-wave emission from both accreting and rotation-powered millisecond pulsars. We argue that sub-millisecond spins are not primarily excluded by equilibrium spin limits, magnetospheric physics, or selection effects, but instead by a combination of inefficient recycling, short mass-transfer lifetimes in the most favourable binaries, and rapid post-formation spin-down unless magnetic fields are exceptionally weak. Together, these effects make sub-millisecond pulsars intrinsically rare. If they exist at all, they are most likely to be detected transiently during accretion, rather than as long-lived radio pulsars. Finally, we show that stellar-mass black holes can attain sub-millisecond horizon spin periods, reflecting fundamentally different spin constraints.
}


%%\pacs[JEL Classification]{D8, H51}
%%\pacs[MSC Classification]{35A01, 65L10, 65L12, 65L20, 65L70}

\maketitle

\setcounter{tocdepth}{3} % TOC subsubsections
\tableofcontents

\newpage
%%%%%%%%%%%%%%%%%%%%%%%%%%%%%%%%%%%%%%%%%%%%%%%%%%%%%%%%%%%%%%%%%%%%%%%%%%%%%%%%%%%%%%%%%%%%%%%%%%%%
\section{Introduction} \label{sec:intro}
The discovery of the first millisecond pulsar (MSP), PSR~B1937+21 with a spin period of $P=1.6\;{\rm ms}$, nearly half a century ago \citep{bkh+82}, triggered intense debate over what ultimately limits how fast a neutron star (NS) can rotate.
Early discussions immediately followed related to the consequences for NS structure, stability and gravitational wave (GW) emission \citep[e.g.][]{rc83,har83}, the spin distribution of MSPs \citep{har84}, observational selection effects \citep{dss+84}, as well as formation scenarios \citep{acrs82,rs82}. A few years after the discovery, there was brief but intense excitement over a claimed detection of a 0.5~ms optical pulsation in SN~1987A \citep{kpm+89}. The signal was later shown to be noise, but even if it had been astrophysical, interpreting it as the NS’s rotation would be problematic --- most notably because such a period would violate all viable dense-matter equations of state (EoS) \citep[see, e.g.][]{wch+89,il89,stw89}. What, then, is the fastest spin a NS can achieve and maintain?
 
\begin{figure}[ht]
\centering
\vspace*{-1.0cm}\hspace*{-0.3cm}
\includegraphics[width=0.75\textwidth]{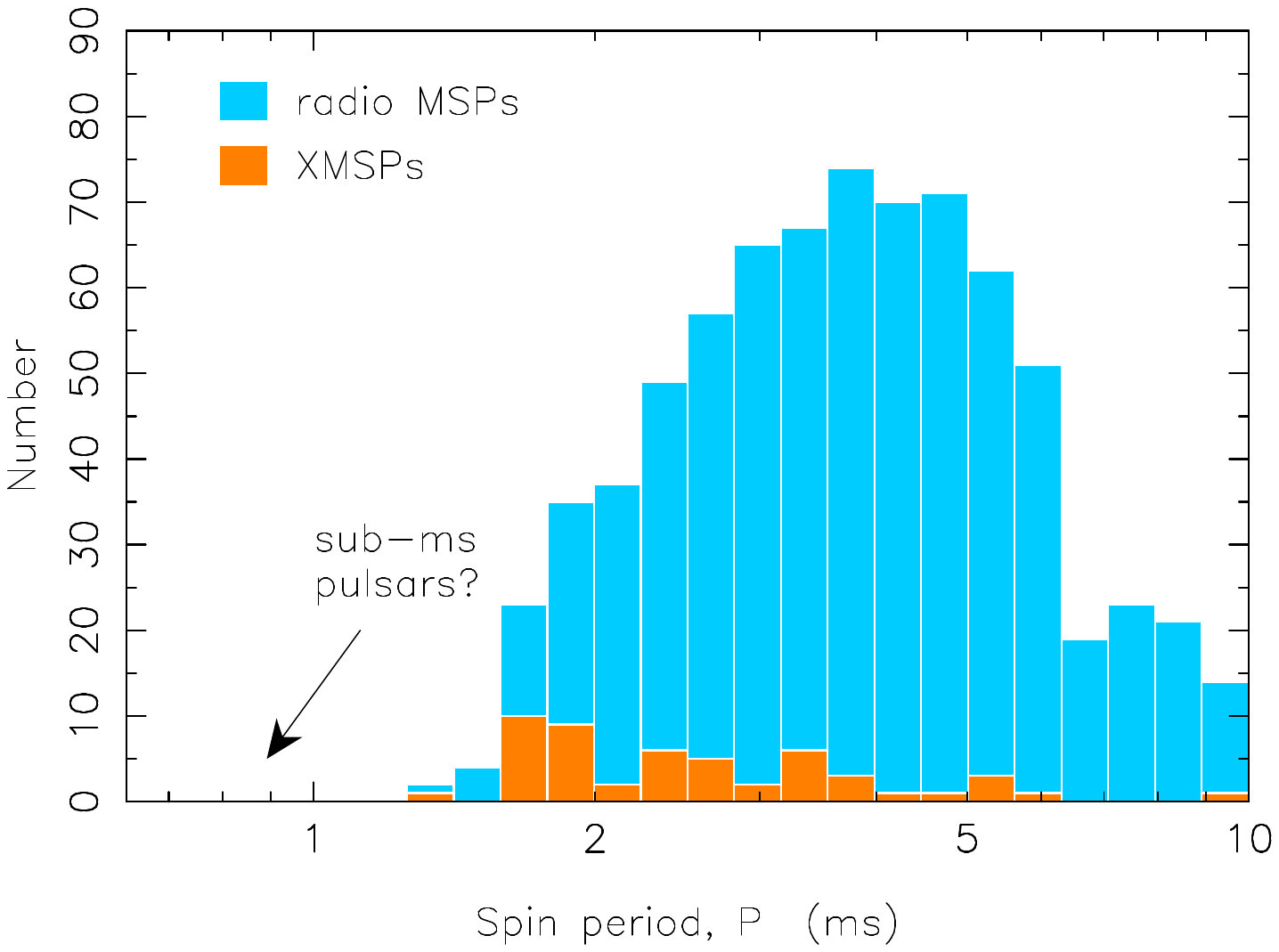} 
\caption{Distribution of measured spin periods of 694 radio MSPs (blue) and 51 XMSPs (orange). Here, XMSPs refers to both accretion-powered X-ray millisecond pulsars that show coherent pulsations during outbursts and nuclear-powered burst-oscillation sources. 
The X-ray data are taken from Table~7.5 in \cite{tv23}, \cite{fkg+24}, and from \cite{jbb+24} for 4U~1820$-$30 ($P=1.40\;{\rm ms}$, the fastest-spinning XMSP {\em candidate} to date; still to be confirmed). 
The radio data are from the \textit{ATNF Pulsar Catalogue}, version~2.7.0 (January~2026; \cite{mhth05}, \url{https://www.atnf.csiro.au/research/pulsar/psrcat}).} 
\label{fig:spin-histo-all}
\end{figure}

Despite the discovery of more than 700~MSPs with $P<10\;{\rm ms}$ since then (Fig.~\ref{fig:spin-histo-all}), intense efforts failed to detect sub-millisecond pulsars (sub-MSPs) in either radio, X-rays, optical or GWs \citep[e.g.][]{dam00,kjv+10,pat10,pwm18,lpi+26,aaa+22} and the fastest securely measured spin remains $P\simeq 1.4\;{\rm ms}$ \citep[PSRs~J1748--2446ad and J0952--0607,][]{hrs+06,bph+17}. 

\begin{figure}[ht]
\centering
%%\vspace*{+0.2cm}\hspace*{-0.0cm}
\includegraphics[width=0.90\textwidth]{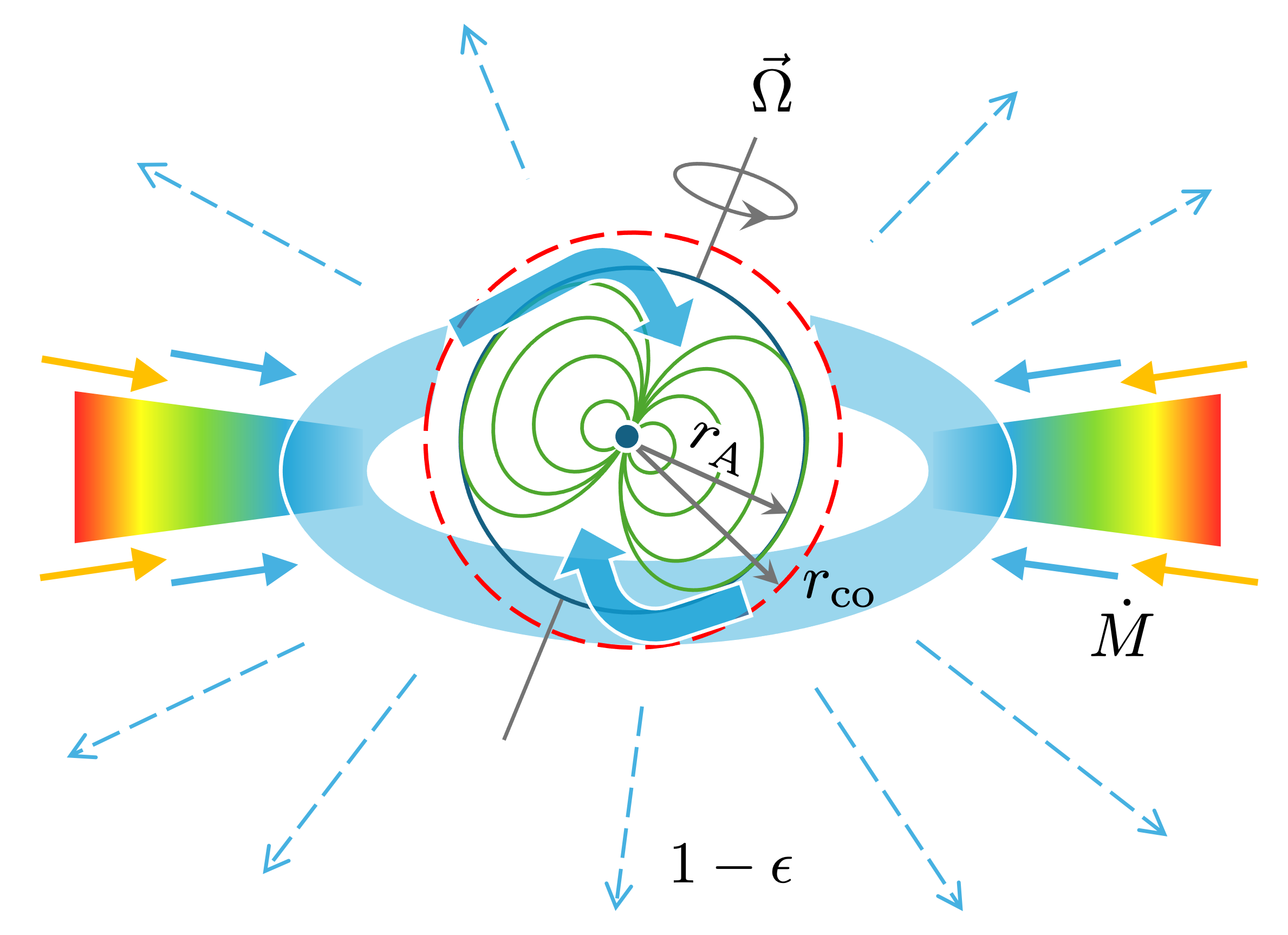} 
\caption{Illustration of a NS accretor. The inner edge of the accretion disc is truncated by the NS magnetosphere, roughly located at the Alfvén radius, $r_A$ (determined by its angular velocity, $\Omega$, B-field, and incoming ram pressure caused by accretion at rate, $\dot{M}$). For efficient accretion and spin-up of the pulsar, the magnetosphere must be within the so-called co-rotation radius, $r_{\rm co}$ for a substantial amount of time --- see Sect.~\ref{sec:theory}.}
\label{fig:illustration}
\end{figure}

The absence of sub-MSPs (or even $P\approx 1.0\;{\rm ms}$) is a long-standing puzzle in NS astrophysics. Physical limits --- set by the NS EoS and the mass-shedding (Keplerian break-up) limit, or by torques from gravitational-wave (GW) emission --- are obvious candidates. However, magnetospheric and accretion physics (Fig.~\ref{fig:illustration}) may also cap the equilibrium spin during recycling, and observational selection effects likewise warrant careful consideration.

Explanatory hypotheses for the non-detection of a sub-MSP include:
\begin{itemize}
  \item Physical limitations related to the NS EoS or GW torques.
  \item Nature prevents their formation because magnetospheric or accretion conditions during recycling cap the spin rate.
  \item Accreting fast-spinning NSs undergo a phase transition en route, after which they no longer emit radio pulses.
  \item Sub-MSPs do form but spin down on a short timescale.
  \item Intrinsic rarity --- there may simply be very few sub-MSPs in the Galaxy (small-number statistics).  
  \item Selection effects (propagation, scattering, sensitivity limits) prevent their detection.
\end{itemize}

Beyond their intrinsic importance for NS astrophysics, the existence (or absence) of sub-MSPs also has broader implications for relativistic astrophysics. Rapidly rotating NSs probe the dense-matter EoS under extreme conditions and approach the centrifugal break-up limit, thereby constraining fundamental aspects of strong-field gravity \citep{gle92,lp04,hzb09,ps17}. 
Moreover, the processes governing their formation and spin evolution are closely linked to those shaping compact object binaries, the primary sources of GWs detected by high-frequency ground-based interferometers. These are sensitive to (extragalactic) transient binary mergers \citep{LVK26} and potential continuous signals from rapidly rotating Galactic NSs \citep{bil98,hpp+15,aaa+22-known-MSPs,ril23}.
The distribution of NS spins, and in particular the maximum attainable spin rate, enters directly into population synthesis models --- through its impact on radio pulsar lifetimes and detectability --- that underpin predictions for merger rates and GW signal properties \citep{knst01,bdp03,tkf+17,pml19,mb22}. In addition, rapidly rotating NSs formed in core-collapse events \citep{mbm17} or binary mergers \citep{rbp20} may act as transient sources of gravitational radiation, for example through non-axisymmetric instabilities or during short-lived proto-NS phases. 
Tight Galactic MSP binaries may furthermore be detectable through their low-frequency GW emission arising from orbital motion, with space-based observatories such as LISA \citep{tau18,LISA2023} and TianQin \citep{llt+25} providing complementary probes of compact binary evolution.
Establishing whether sub-MSPs can form and persist therefore provides key input to models of compact object evolution and GW emission, with relevance for current and future observatories such as LISA, the Einstein Telescope, and Cosmic Explorer \citep{LISA2023,ET2026,CE2019}.
Ultimately, this places the study of sub-MSPs within the broader framework of general relativity, linking dense-matter physics, strong-field gravity, and compact object binaries as precision laboratories for relativistic effects and GW astrophysics \citep{hb23,fw24}.

In this review, we examine several of the above-mentioned hypotheses for the apparent absence of sub-MSPs. We begin by summarising key concepts and prior work (Sect.~\ref{sec:summary}). We then examine the current empirical spin distribution of pulsars, including a brief statistical assessment (Sect.~\ref{sec:NS-spins}), followed by an evaluation of possible observational selection effects against sub-MSPs (Sect.~\ref{sec:selection-effects}). In Sect.~\ref{sec:theory}, we develop the theoretical framework for pulsar recycling via mass transfer in X-ray binaries, including physical spin limits, empirical accretion efficiencies, and post-formation spin-down. Section~\ref{sec:discussions} provides a broader synthesis, including sub-ms birth spins of magnetars, as well as a separate comparison with observed black hole spins. Our conclusions are summarised in Sect.~\ref{sec:conclusions}.

%%%%%%%%%%%%%%%%%%%%%%%%%%%%%%%%%%%%%%%%%%%%%%%%%%%%%%%%%%%%%%%%%%%%%%%%%%%%%%%%%%%%%%%%%%%%%%%%%%%%%%%%%%%%%%%
\clearpage
\section{Summary of previous work and key concepts}\label{sec:summary}
In the late 1990s, as the number of observed radio MSPs increased and accreting X-ray millisecond pulsars \citep{wv98} and nuclear-powered burst-oscillation sources \citep{szs96} were discovered, the question of whether there exists a physical origin for the apparent cut-off in the observed NS spin distribution became relevant.
\cite{bil98} argued that the spin rate of XMSPs (including both accreting and nuclear-powered MSPs) is set by GW emission \citep[see also][]{cmm+03}. This was a commonly accepted explanation until \cite{phd12} argued that for accreting MSPs, the existence of spin equilibrium as set by the disc-magnetosphere interactions is sufficient to explain the observations. 
However, in a subsequent paper, \cite{pha17} re-investigated the spin-period distribution of XMSPs and radio MSPs and speculated that either the current distribution does mark the onset of GWs as an efficient mechanism to remove angular momentum or some of the NSs in the fast sub-population of XMSPs do not evolve into radio MSPs.

\cite{psb16} demonstrated that, for magnetic moments, spin frequencies, and accretion rates relevant to accreting MSPs, the spin-down torque from an enhanced pulsar wind can exceed that predicted by standard disc–magnetosphere models, and can in principle maintain spin equilibrium at periods longer than 1~ms.
\cite{ea21} investigated a possible correlation between mass-accretion rate and final surface B-field \citep[based on work by][]{kon17} and advocated that this explains the currently known minimum spin period of MSPs. 
\cite{ce23} argued that constraints from both disc-magnetosphere interactions, the change in the NS moment of inertia due to accretion of mass, and the torque from GW emission must be considered, and that all three processes are significant for explaining the absence of sub-MSPs.

Related to NS structure, \cite{hzf+18} ruled out centrifugal breakup as the mechanism preventing MSPs spinning faster than 1.4~ms, as the lowest breakup frequency allowed by their causal EoS is $f\approx 1200\;{\rm Hz}$, corresponding to a spin period of 0.83~ms.
\cite{gsi+25} investigated the use of massive and rapidly spinning MSPs (like the black widow PSR~J0952$-$0607)
as a useful probe of quark matter properties and the onset of a deconfinement phase transition, 
which has important implication for microphysical parameters and the dense matter EoS.
\cite{zl25} examined the spin-up of NSs and strangeon stars and found that B-fields hinder the spin-up of MSPs more strongly in NSs than in strangeon stars.

Finally, \cite{tv23} proposed that the cut-off in MSP spin periods can be explained by binary stellar evolution. Here, we explore this idea in greater depth.

\subsection{The equilibrium spin period}\label{subsec:equilibrium}
When an accreting NS in a binary system reaches its equilibrium spin period ($P_{\rm eq}$), the net torque vanishes: the magnetospheric radius is close to the co-rotation radius, where the stellar spin matches the Keplerian angular velocity of the disc at the magnetospheric boundary. For accreting MSPs \citep[see, e.g. the original work by][]{vdh77,bv91,tlk12}, $P_{\rm eq}$ can be expressed:
\begin{equation}\label{eq:spinup}
P_{\rm eq} \simeq  1.40\,{\rm ms}\quad B_8^{6/7}\left(\frac{\dot{M}}{0.1\,\dot{M}_{\rm Edd}}\right) ^{-3/7} M_{1.4}^{-5/7}\;R_{13}^{18/7} \;,
\end{equation}
where $B_8$ is the magnetic flux density at the NS surface in units of $10^8\,{\rm G}$, $\dot{M}$ is the mass-accretion rate (with $\dot{M}_{\rm Edd}$ the Eddington limit for a NS, a few $10^{-8}\;M_{\odot}\,{\rm yr}^{-1}$), $M_{1.4}$ is the NS mass in units of $1.4\,M_{\odot}$, and $R_{13}$ its radius in units of 13~km. 
This means that to reach sub-ms spin periods, the B-field of the NS must be very weak and the mass-transfer rate from the donor star must be very high --- two conditions that are mutually contradictory (Sect.~\ref{sec:theory}). 
Hence, according to this expression, $\sim$1.40~ms is, in principle, the shortest spin period a $1.4\;M_\odot$ NS can reach for $B_8=1$ and $\dot{M}=0.1\,\dot{M}_{\rm Edd}$ (with some dependence on $R$ via the EoS). 
As a consequence, roughly speaking \citep[for more detailed discussions, see][and references therein]{tv23}, if the NS rotates \emph{faster} than $P_{\rm eq}$, matter would be centrifugally \emph{expelled} from the magnetosphere and accretion would be prevented --- and vice versa for $P \ge P_{\rm eq}$, where matter can enter the magnetosphere and accretion is theoretically possible.

While the concept of spin equilibrium provides a useful first-order description of the recycling process, the long-term spin evolution of accreting NSs may be affected by a variety of disc--magnetosphere interactions, including propeller phases, trapped-disc states, and episodic accretion. Such effects can modify the torque balance and influence the approach to spin equilibrium \citep{wz97,ds12,dan17,bc17}.

In principle, if the long-term accretion rate could be sustained at the Eddington limit ($\dot{M}=1.0\,\dot{M}_{\rm Edd}$) for $B_8=1$, Eq.~(\ref{eq:spinup}) would naively predict the possibility of a $\sim 0.5\,{\rm ms}$ MSP (disregarding here limitations from the EoS, see Sect.~\ref{subsec:EoS}). However, such high accretion rates are difficult to sustain for more than a few Myr. To deliver them, either the donor must be relatively massive and/or the Roche-lobe overflow (RLO) becomes dynamically unstable. Hence, systems with high mass-transfer rates are short-lived (as we shall discuss in Sect.~\ref{sec:theory}), and the NS is unlikely to reach sub-ms spin periods within the lifetime of the mass-transferring X-ray phase.

A principal aim of this review is to combine empirical data (Sect.~\ref{sec:NS-spins}) with theoretical calculations (Sect.~\ref{sec:theory}) to investigate the recycling process of MSPs and to assess the possibility of producing a sub-MSP.
As we shall see, the accretion efficiency inferred from observations plays a key role.

\subsection{GW radiation}\label{subsec:GWs}
Several mechanisms have been proposed to explain the apparent empirical spin limit at $P\simeq1.40\,{\rm ms}$ (Fig.~\ref{fig:spin-histo-all}), corresponding to $\nu \simeq 716\,{\rm Hz}$, which is currently shared by the two fastest known radio pulsars (PSR~J1748$-$2446ad and PSR~J0952$-$0607; \citealt{hrs+06,bph+17}) and, interestingly, also by the candidate\footnote{If the spin frequency of 4U~1820$-$30 is confirmed, it would match that of the fastest known radio pulsars and become the fastest spin measured in an X-ray binary.} burst-oscillation spin reported for the accreting NS 4U~1820$-$30 \citep{jbb+24}.

As briefly mentioned, a leading class of explanations involves emission of continuous GWs (Appendix~\ref{app:GWs}) that remove spin angular momentum from the NS \citep{wag84,bil98,cmm+03}. One important mechanism is the r-mode instability, whose potential to limit the spin rates of rapidly rotating NSs was established in a series of seminal studies \citep{and98,fm98,lom98,olc+98,aks99}. 
Other candidate GW mechanisms include quadrupolar deformations caused by asymmetric accretion --- such as internal magnetic stresses, magnetically confined ``mountains'', or lateral variations in temperature, composition, or density \citep{ucb00,las15,gg18,gaj21}.
For example, considering continuous mass-quadrupole radiation from a rotating, non-axisymmetric NS, the GW luminosity scales steeply with spin period, $L_{\rm GW}\propto P^{-6}$, making GW radiation a very efficient process and an important ingredient when explaining the non-detection of sub-MSPs.

The sharp drop in observed spin periods for $P<1.40\,{\rm ms}$ ($\nu>716\,{\rm Hz}$; Fig.~\ref{fig:spin-histo-all}) may suggest some GW regulation (Appendix~\ref{app:GWs}). Direct evidence is still lacking (Sect.~\ref{subsubsec:scoX1}), but some population-synthesis studies \citep{ga19} support this hypothesis. In addition, \cite{hp17} proposed GWs to cause a spin-down of the transitional MSP J1023+0038 \citep[but see][who advocated that the enhanced spin-down in the disc state can be explained by magnetospheric/propeller torques rather than GWs]{ert18}. However, a reevaluation of the spin-down rate of J1023+0038 in the disc state indicates that the NS spin evolution during the sub-luminous X-ray phase is largely unchanged from that in the radio MSP state \citep{bzf+20,bcd+25}.

Conversely, the absence of continuous-wave detections in targeted GW searches of known MSPs, all-sky searches for isolated NSs, wide parameter-space searches for previously unknown NSs in binary systems \citep{cpp26}, narrowband searches of known pulsars, and searches of XMSPs, provides increasingly stringent constraints on strong, ubiquitous GW regulation at current sensitivities (see Appendix~\ref{app:GWs} for a summary of recent O3 and O4 continuous-wave searches).
Recent LIGO--Virgo--KAGRA (LVK) O3 and O4 searches have continued to improve upper limits on continuous GW emission from a variety of source classes. For nearby radio MSPs, targeted searches now constrain ellipticities to $\epsilon \lesssim 10^{-8}$ in the most favourable cases, while searches of XMSPs yield less restrictive limits \citep[see Appendix~\ref{app:GWs} and references therein]{aaa+22-known-MSPs,aaa+22-all-sky,aaa+25}.
Searches targeting XMSPs currently provide weaker constraints, with typical upper limits of order $\epsilon \sim 10^{-7}$ and $r$-mode amplitudes $\alpha \sim 10^{-5}$ \citep{aaa+22-AXMSPs}.

\subsection{Sub-MSP stability and the NS equation of state}\label{subsec:EoS}
Pioneering studies established that the centrifugal mass-shedding limit of rapidly rotating NSs is strongly dependent on the NS mass and the dense-matter EoS, with minimum spin periods of order $\sim 0.5-1\;{\rm ms}$ for realistic models, thereby providing the theoretical framework for assessing the possibility of sub-MSPs \citep{fip86,hz89,cst94a,cst94b}.
Self-bound quark-star models may permit stable rotation at periods as short as $\sim 0.1\;{\rm ms}$, substantially below the centrifugal break-up limit predicted for NSs \citep{dxqh09}.

Often in more recent discussions of constraints on the NS EoS, the minimum rotation period is discussed in a mass-independent way. This makes sense since, for the two previous spin-period record holders (PSR~B1937+21, \cite{bkh+82}, and PSR~J1748$-$2446ad, \cite{hrs+06}), no mass measurements could be obtained --- in the former case because it is an isolated pulsar, in the latter case because the pulsar is in a ``redback'' system (Sect.~\ref{subsubsec:spiders}), where measurements of orbital relativistic effects are not possible. Furthermore, the latter system has not been detected at optical wavelengths, where mass constraints could, in principle, be obtained. 

\begin{figure}[ht]
\centering
\includegraphics[width=0.99\textwidth]{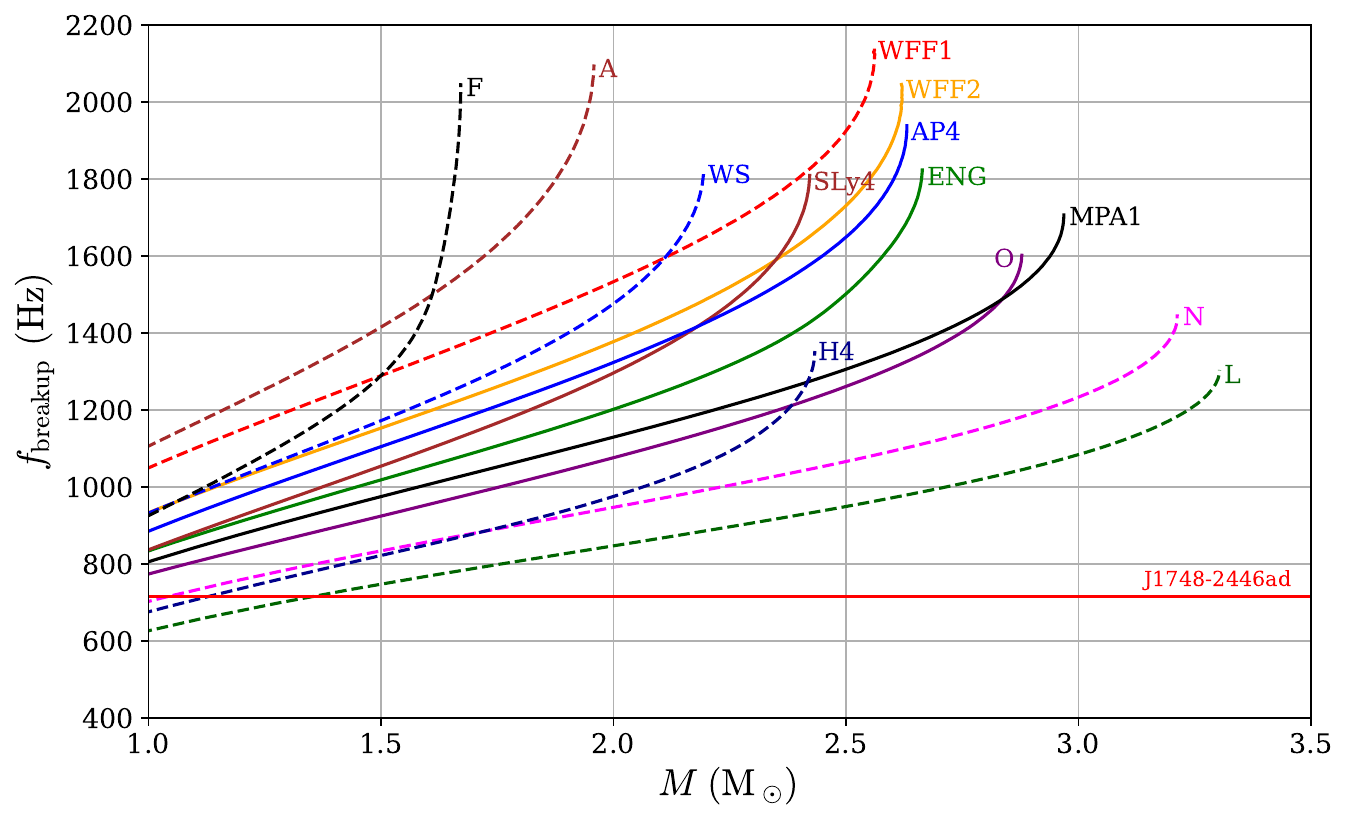} 
\caption{Maximum possible NS spin frequencies as a function of NS mass for a selected set of EoSs (coloured dashed and continuum lines), which does not include strong phase transitions, calculated using the \emph{RNS
code} (\url{https://github.com/cgca/rns}), see \cite{sf95} %\citep{Stergioulas_1996} 
for a detailed explanation.
The solid lines represent EoSs that are still consistent with all multi-messenger constraints \citep[radio pulsar mass measurements, X-ray measurements, LVK constraints from NS+NS mergers, see][]{Koehn2025}; the dashed lines represent other more extreme EoSs that were considered to be realistic until recently. The spin frequency of PSR~J1748$-$2446ad (the fastest rotating of all known pulsars) is indicated by the horizontal red line ($\nu = 716\;{\rm Hz}$). 
For some of the stiffer EoSs, which predict matter to be highly incompressible, this spin frequency already introduces interesting mass constraints. Figure and calculations by Norbert Wex.}
\label{fig:mass_maximum_spin_frequency}
\end{figure}

However, the minimum spin period for a particular NS strongly depends on its mass and also on the adopted EoS \citep[e.g.][]{lp04,hzb09,ps17}. 
In Fig.~\ref{fig:mass_maximum_spin_frequency}, the maximum break-up spin frequencies are calculated as a function of NS mass for a selected set of representative EoSs. The solid lines represent EoSs that are still consistent with all multi-messenger constraints (radio pulsar mass measurements, X-ray measurements, LVK constraints from GW detections of NS+NS mergers, see \citealt{Koehn2025}); the dashed lines represent other more extreme EoSs that were considered to be realistic until recently.

There are several consequences of this dependence. One of them is that, for all the EoSs that are still consistent with all current multi-messenger constraints, sub-MSPs are simply not stable below a mass threshold that lies well within the range of measured NS masses. Depending on the EoS being considered, the minimum mass for a sub-MSP varies from $1.2\;M_\odot$ for the WFF2 EoS to $1.75\; M_\odot$ for the ``O'' EoS, with most EoSs yielding a threshold above $1.4\;M_\odot$. As it turns out, 57\% of NS mass measurements from radio timing\footnote{For an updated list of NS mass measurements, see \url{https://www3.mpifr-bonn.mpg.de/staff/pfreire/NS_masses.html}.} are below $1.4\;M_\odot$. Consequently, for many mass–EoS pairs the mass-shedding limit lies below a spin frequency of 1000~Hz ($P=1\;{\rm ms}$) making such spins unattainable for a large fraction of the observed population.

More massive NSs can generally spin faster. However, even for the most massive NS measured via radio timing --- PSR~J0740+6620, with $M = 2.07 \pm 0.08\;M_\odot$ \citep{Fonseca_2021} --- 
the mass-shedding limit remains $\lesssim 1400\;{\rm Hz}$ for all EoSs considered here, and for some it is even below 1100~Hz. The absolute peak spin frequencies occur only at masses very near the EoS-dependent maximum NS mass, which are much higher than (or not represented by) any precisely measured pulsar masses to date. Potentially such configurations could occur after a NS+NS merger, but the subsequent loss of rotational energy would make such super-massive NSs unstable (Sects.~\ref{subsec:diff-rot} and \ref{subsec:born-MSP}).

%%%%%%%%%%%%%%%%%%%%%%%%%%%%%%%%%%%%%%%%%%%%%%%%%%%%%%%%%%%%%%%%%%%%%%%%%%%%%%%%%%%%%%%%%%%%%%%%%%%%%%%%%%%%%%%%%
\clearpage
\section{Empirical spin periods of NSs}\label{sec:NS-spins}
In this section, we analyse the distribution of radio MSP spin periods. Before doing so, we briefly examine the spin properties of their progenitors: NSs in X-ray binaries.

\subsection{NS spins in X-ray binaries}\label{subsec:data-X-rays}
In the following, we summarise the NS spins (and their closely related B-fields) in the two archetypal classes: high-mass X-ray binaries (HMXBs) and low-mass X-ray binaries (LMXBs). For reviews of NS spins in X-ray binaries, see e.g. \cite{ptrt14,pha17,bmz21,bpb22,tv23}.

\subsubsection{HMXBs}\label{subsubsec:HMXBs}
NSs in HMXBs \citep{fgsc23} typically have B-fields that are still fairly large \citep[$B\sim 10^{12}\;{\rm G}$, e.g.][]{rm15,lp18}, since these NSs are young and cannot have accumulated much material via wind accretion --- a commonly invoked mechanism for B-field decay is burial via mass accretion \citep[e.g.][and references therein]{tv86,gu94,kb97,pm04,ymm25}. 
Their spins are initially rapidly reduced by electromagnetic (``Gunn--Ostriker'') torques during the ejector phase, and subsequently by accretion (propeller) torques once the inflowing matter begins to interact with the NS magnetosphere, reaching $\sim10-1000\;{\rm s}$ within $\sim0.1\;{\rm Myr}$ \citep[e.g.][]{hwa+20}.
For example, for $B=10^{12}\;{\rm G}$ ($B_8 = 10^4$), $\dot{M}=0.001\;\dot{M}_{\rm Edd}$ ($\sim 10^{-11}\;M_\odot\,{\rm yr}^{-1}$), and assuming $M_{1.4}=1$ and $R_{13}=1$, the equilibrium spin period in Eq.~(\ref{eq:spinup}) becomes $P_{\rm eq}\simeq 30\;{\rm s}$ --- in agreement with many NS spin periods observed in HMXBs \citep{kfs+19}.
The ages of NSs in HMXBs, counted from their formation in a supernova (SN), are typically less than 20~Myr (a limit set by the nuclear evolution timescale of their massive companion star).

\subsubsection{LMXBs}\label{subsubsec:LMXBs}
NSs in LMXBs \citep{fkg+24}, by contrast, are typically fast-spinning (milliseconds to a few seconds). This reflects their much longer accretion history via RLO from a long-lived (Gyr), low-mass donor. LMXBs are either persistent or transient sources, mainly depending on their long-term mass-transfer rate, which is in turn related to the disc size \citep{las01}. Persistent systems are indeed often among bright atoll sources, Z~sources, and ultra-compact X-ray binaries \citep{cfd12}, as well as in systems with evolved donors or in globular clusters (GCs), also reflecting differences in binary evolution and donor type.
LMXBs characterized by a lower mass-transfer rate tend to be transient sources instead, detected in X-rays only during episodic outbursts.
Although most transient systems are undetectable in quiescence with current instruments, they likely remain faintly X-ray active. Outbursts are triggered by thermal-viscous instabilities in the accretion disc, temporarily boosting the accretion rate onto the NS \citep{par96,las01,dhl01,grg+20}.

Rapidly spinning NSs in LMXBs can be broadly divided into two observational classes:

\begin{enumerate}
    \item[(i)] \textit{Accretion-powered X-ray millisecond pulsars} --- a few dozen transient systems that, during outbursts, exhibit coherent millisecond X-ray pulsations tracing the NS spin period \citep{wv98}. These sources typically have relatively strong magnetic fields ($B \simeq 10^8-10^9\;{\rm G}$), and therefore retain a magnetosphere that channels the accretion flow onto the magnetic poles of the NS, producing surface hot spots and pulsed X-ray emission \citep[for reviews, see][]{pw21,dss22}.
    
    \item[(ii)] \textit{Burst oscillation sources}--- a comparable number of systems that reveal the NS spin period through nearly coherent oscillations during thermonuclear (Type~I) X-ray bursts, triggered by unstable nuclear burning on the NS surface \citep[][]{cmm+03,gk21}. These burst oscillations are thought to trace the stellar spin, although small frequency drifts are common \citep[see][for reviews]{w12,b22}. 
\end{enumerate}

LMXBs hosting NSs with weak or compressed magnetospheres (due to either very low magnetic fields $\lesssim 10^8\;{\rm G}$ or very high accretion rates, $\dot{M}_{\rm NS}\sim \dot{M}_{\rm Edd}$), or with near alignment between magnetic and spin axes, may fail to produce observable coherent pulsations. 
This includes many bright \textit{atoll} and \textit{Z~sources} \citep{hv89,vdk06}, which generally do not exhibit coherent X-ray pulsations. Whereas atoll sources are often burst oscillation sources, most Z~sources do not show bursts at all.
The absence of bursts and pulsations in Z~sources is likely due to their high accretion rates (near $\dot{M}_{\rm Edd} \sim 10^{-8}\;M_\odot\,{\rm yr}^{-1}$), which suppress thermonuclear ignition via stable burning and simultaneously compress the magnetospheric boundary close to or onto the NS surface (see Fig.~\ref{fig:mag-radius}). For these reasons, the NS spin periods in Z~sources remain unknown. Nevertheless, Z~sources are typically persistent X-ray emitters. Additional factors that may hinder the detection of X-ray pulsations include the accretion geometry, the viewing angle, and the degree of asymmetry in surface burning or radiation patterns.

\subsection{Radio pulsar spins}\label{subsec:data-radio}
\begin{figure}[ht]
\centering
\vspace*{-1.5cm}\hspace*{-0.5cm}
\includegraphics[width=1.10\textwidth]{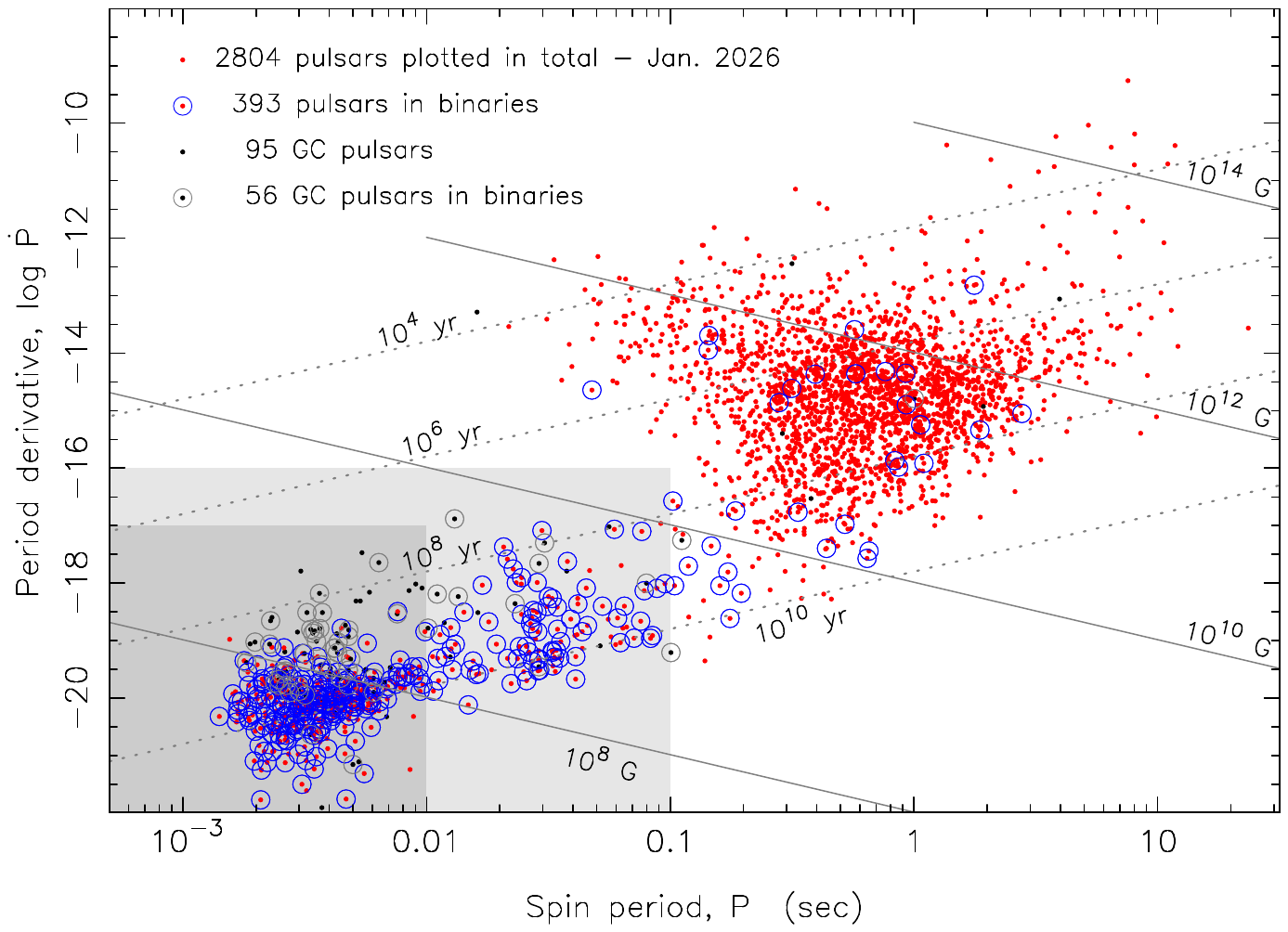} 
\vspace*{-1.0cm}
\caption{Distribution of 2804 radio pulsars in the $P\dot{P}$-diagram. 95 of these are found in GCs and have $\dot{P}_{\rm obs}>0$ (see text for a discussion of the interpretation of observed $\dot{P}$ values). A total of 393 binary pulsars are marked by a circle (of which 56 are located in GCs). Lines of constant characteristic age ($\tau\equiv P/(2\dot{P})$) or surface B-field flux density ($B\propto\sqrt{P\dot{P}}$) are shown, assuming $M=1.4\;M_\odot$ and magnetic inclination angle, $\alpha = 60^{\circ}$ \citep[cf. Eq.~5 in][]{tlk12}. The two grey-shaded regions include the mildly recycled pulsars ($10\;{\rm ms}<P<100\;{\rm ms}$) and the fully recycled MSPs ($P<10\;{\rm ms}$). Data taken from the \textit{ATNF Pulsar Catalogue} version 2.7.0 in January~2026 \citep[][\url{https://www.atnf.csiro.au/research/pulsar/psrcat}]{mhth05}.} 
\label{fig:PPdot}
\end{figure}

Figure~\ref{fig:PPdot} shows the current sample of 2804 radio pulsars with measured values of spin period ($P$) and its time derivative ($\dot{P}$), based on the ATNF Pulsar Catalogue \citep{mhth05}. The plot includes pulsars located in the Galactic field and in GCs. If $\dot{P}$ for a given pulsar is corrected for the Shklovskii (proper-motion) effect, we use that value. The inferred B-fields (surface magnetic flux densities, see Sect.~\ref{sec:theory}) span more than eight orders of magnitude ($10^7-10^{15}\;{\rm G}$), as shown in the plot. 

\subsubsection{Distribution of radio MSP spins}\label{subsubsec:Pdist}
The full distribution of fast radio MSP spin periods is shown in blue colour in Fig.~\ref{fig:spin-histo-all}. 
An interesting feature is the apparent deficit of detected MSPs around $P\sim 6-7\;{\rm ms}$ (e.g. relative to a log-normal distribution), which may indicate an overlap between populations associated with different formation histories.
In the following, however, we focus on comparing single versus binary MSPs, and Galactic-field versus GC MSPs.

\begin{figure*}
\centering
\hspace*{-0.7cm}
\includegraphics[width=0.55\textwidth]{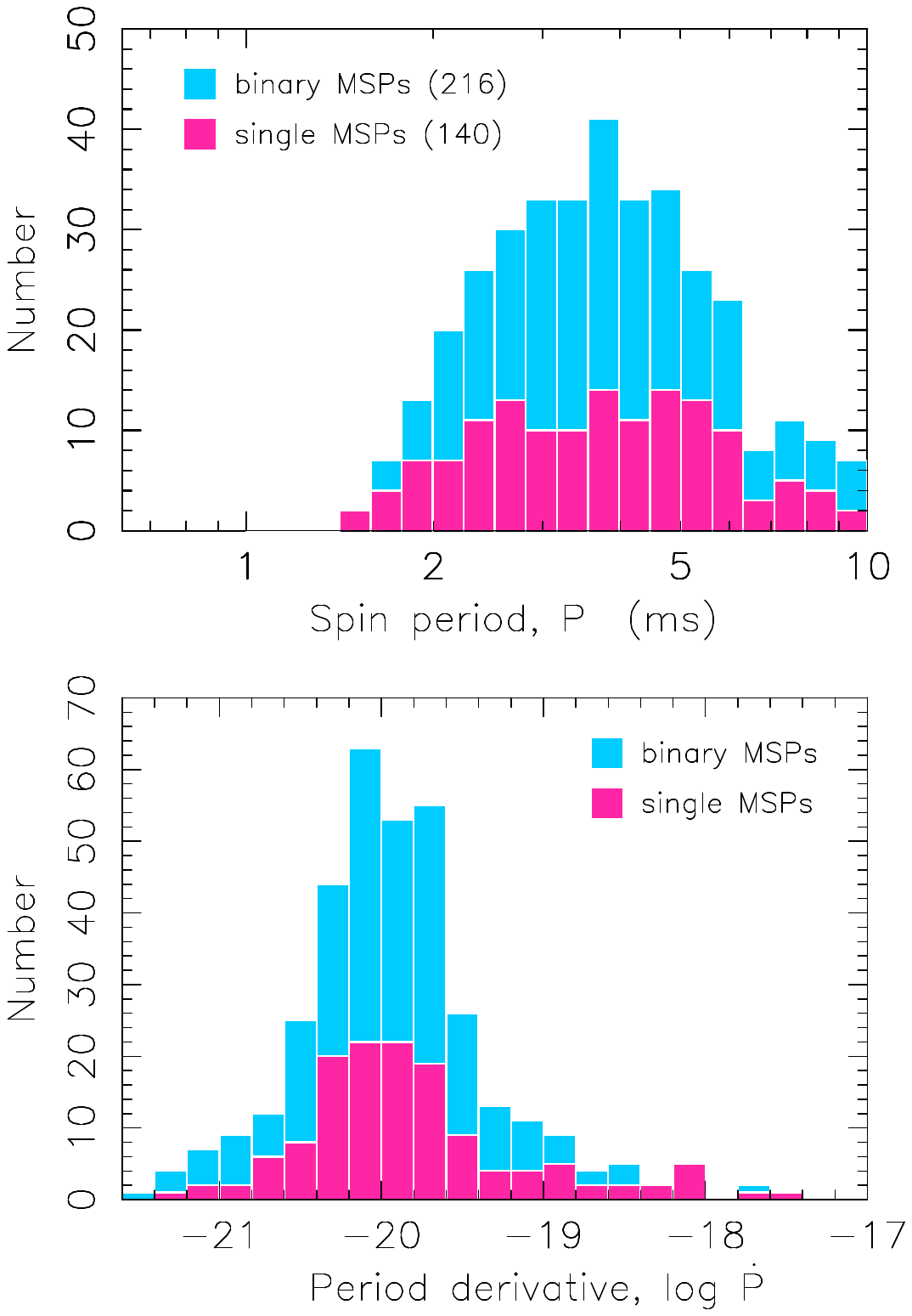}
\hspace*{-1.0cm}
\includegraphics[width=0.55\textwidth]{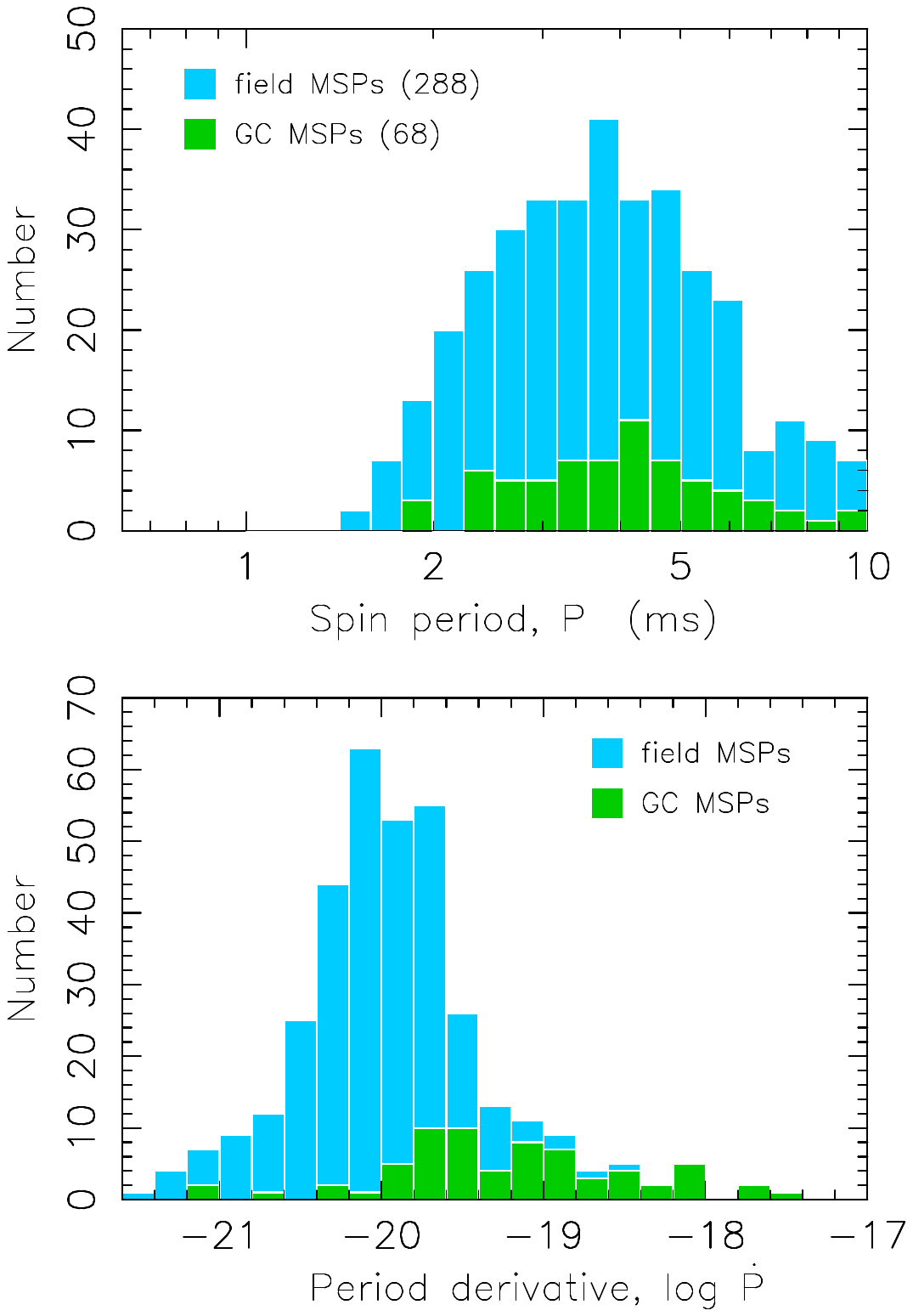} 
\caption{Distribution of spin periods (top panels) and spin period derivatives (bottom panels) of 356 radio MSPs with $\dot{P}>0$. Blue bins are binary MSPs (or triples), pink bins (left panels) are single, isolated MSPs. The distributions of $P$ and $\dot{P}$ for binary and isolated MSPs are roughly similar. Green bins (right panels) are GC MSPs. Whereas the distributions of $P$ are roughly similar, the $\dot{P}$ values are typically larger by an order of magnitude for the GC MSPs (a well-known effect; see text for interpretation of $\dot{P}$ values). The full $P$-distribution for all $\sim 700$ known MSPs is shown in Fig.~\ref{fig:spin-histo-all}. }
\label{fig:spin-histo}
\end{figure*}

In Fig.~\ref{fig:spin-histo}, we show the distributions of $P$ and $\dot{P}$ for all fully recycled MSPs ($P<10\,{\rm ms}$; i.e. the dark grey-shaded region in Fig.~\ref{fig:PPdot}). These figures compare the full sample with the subsets of single MSPs (Fig.~\ref{fig:spin-histo}, left) and GC MSPs (Fig.~\ref{fig:spin-histo}, right). Only the 356 MSPs with (positive) measured values of $\dot{P}$ are plotted here (see Appendix~\ref{app:PdotGC} for further discussion on the observed $\dot{P}_{\rm obs}$). If $\dot{P}$ for a given pulsar is corrected for the Shklovskii (proper-motion) effect, we use that value. An additional 58 MSPs with negative $\dot{P}_{\rm obs}$ values --- often in GCs --- are excluded from this analysis (but see Fig.~\ref{fig:spin-histo-all} for the full spin distribution, including sources without known $\dot{P}$ values). Negative $\dot{P}$ values are not intrinsic (non-accreting MSPs cannot spin up); they are most likely due to line-of-sight accelerations in the GC potential. Our sample includes single, binary, and triple MSPs. The fraction of single MSPs in the plotted sample is $140/356=0.393$ ($\sim 39\%$), and the fraction of GC MSPs (with $\dot{P}_{\rm obs}>0$) is $68/356=0.191$ ($\sim 19\%$).

The $P$ and $\dot{P}$ distributions are very similar for single and binary MSPs (Fig.~\ref{fig:spin-histo}, left). For MSPs located in GCs, the spin-period distribution is also similar to that of Galactic-field MSPs (Fig.~\ref{fig:spin-histo}, top-right), whereas Galactic-field MSPs typically have observed spin period derivatives of $\dot{P}_{\rm obs} \simeq 10^{-20\pm1}$, those of GC MSPs are often larger by about an order of magnitude, i.e. $\dot{P}_{\rm obs} \simeq 10^{-19\pm1}$ (see Appendix~\ref{app:PdotGC} for a discussion of this well-known difference, mainly caused by acceleration within the GC potential; see also \citealt{prf+17}).

\subsection{Statistical analysis of the MSP spin distribution}\label{subsec:statistics}
\label{subsec:stats}
In the following, we briefly describe a data-based approach to investigate the existence of a sub-MSP ($P<1\;{\rm ms}$). 
The empirical data consist of 694 radio MSPs (Fig.~\ref{fig:spin-histo-all}) with independent observations of spin on the range $[1.396,\,9.900]\;{\rm ms}$. 
We assume that the observations are realisations of random variables $X_1,\ldots, X_{694}$, independently distributed according to an unknown probability density function $f$. 
The problem of proving the existence of a pulsar below $1\;{\rm ms}$ is then equivalent to determining whether $f$ takes positive values on the range $[0,1]\;{\rm ms}$.

In statistics, determining the unknown distribution $f$ from independent observations without imposing a parametric model falls within the field of non-parametric density estimation \citep{tsy09}. A major theoretical limitation is that direct estimation techniques of $f$, such as kernel density estimation, are known to be imprecise outside the observed data range \citep{tsy09}. Therefore, they cannot be reliably applied to infer if $f$ takes positive values on $[0,1]\;{\rm ms}$, given that the minimum observed value is $1.396\;{\rm ms}$.

\subsubsection{Extreme value theory}
Another approach is to model only the tail of the distribution below $1\;{\rm ms}$, interpreting these values as extreme. The problem of estimating the behaviour of a distribution beyond a given lower threshold arises in many fields, and extreme value theory (EVT) has been developed for this purpose \citep{col01}. A central result of EVT, the Pickands--Balkema--De~Haan theorem, states that under broad conditions on the data, the conditional distribution of excesses or deficits around a threshold may be approximated by a generalized Pareto distribution (GPD).

Here, we follow \cite{kkr+25} and apply the Pickands--Balkema--De~Haan theorem.
Let $X$ be a random variable with cumulative distribution function (CDF) $F$ and consider the deficit: 
\begin{equation}
    Y = u - X \quad \text{given} \quad X < u\;,
\end{equation}
for some threshold $u$. The conditional CDF of $Y$ is: 
\begin{equation}
    F_u(y) = P(u - X \le y \mid X < u)\;.
\end{equation}
Then, for sufficiently low values of $u$ and for a wide class of distributions $F$, $F_u$ is well approximated by a GPD with CDF: 
\begin{equation}
    F_u(y;\sigma_u,\xi) = 1 - \left(1 + \frac{\xi y}{\sigma_u}\right)^{-1/\xi}\;,
\end{equation}
and probability density function: 
\begin{equation}
    f_u(y;\sigma_u,\xi) = \frac{1}{\sigma_u}
    \left(1 + \frac{\xi y}{\sigma_u}\right)^{-(1 + 1/\xi)}\;,
\end{equation}
defined on the domain: 
\begin{equation}
    \left\{ y > 0 \ \wedge \ 1 + \frac{\xi y}{\sigma_u} > 0 \right\}\;,
\end{equation}
where $\sigma_u > 0$ is a scale parameter and $\xi \in \mathbb{R}$ is a shape parameter.
A review of the theorem conditions and further details of EVT may be found in \cite{col01}.

\subsubsection{Application to the observed spin-period data}
We note that the EVT analysis applies strictly to the observed spin-period distribution
and does not correct for potential observational or evolutionary selection effects.

In practice, the main challenge in EVT is the choice of the threshold $u$ which defines what counts as an extreme value. Using the function \texttt{NC.diag} from the \texttt{mev} package in \texttt{R}, 
%we obtain a statistically justified threshold: 
we obtain an estimated threshold: 
\begin{equation}
    u = 1.75\;{\rm ms} \;.
\end{equation}
All observations with $X_i < u$ are therefore treated as extremes. A GPD is then fitted to estimate the parameters $\sigma_u$ and $\xi$, using both maximum-likelihood and Bayesian optimisation techniques implemented in \texttt{mev}\footnote{\url{https://cran.r-project.org/web/packages/mev/index.html}}.

Importantly, the threshold $u$ does not represent the fastest observed spin ($1.396\;{\rm ms}$), but is chosen slightly above it in order to ensure a sufficiently large and statistically stable sample for the EVT fit. For all fitting methods, we find that the conditional density of spin periods below $1\;{\rm ms}$ is effectively zero. To examine sensitivity to the threshold choice, we repeat the analysis for: 
\begin{equation}
    1.7\;{\rm ms} \le u \le 2.0\;{\rm ms} \;,
\end{equation}
and obtain consistent results.

Although statistical extrapolation cannot strictly exclude the existence of a pulsar with $P<1\;{\rm ms}$, the present data strongly imply that such an object would be an extremely rare outlier, if it exists at all. We stress, however, that the EVT analysis is based on the currently observed spin-period distribution and does not correct for observational or evolutionary selection effects. Consequently, while the analysis disfavors a sub-ms population, it cannot by itself exclude the existence of a small intrinsic sub-ms tail.

\subsubsection{Comparison to X-ray MSPs}\label{subsubsec:radio_vs_X}
Figure~\ref{fig:spin-histo-all} shows the spin distribution of radio MSPs compared to that of XMSPs. The latter are, on average, slightly faster-spinning than the radio MSPs \citep[see also discussions in][]{hes08,cha08}. This is not surprising, since after RLO decoupling radio MSPs spin down with time. Moreover, the RLO decoupling phase itself has been demonstrated \citep{tau12} to exert a modest braking torque as the NS magnetosphere expands when mass transfer ceases, thereby slowing the newly recycled radio MSP by a factor of about $\sqrt{2}$ (i.e. the MSP dissipates more than 50\% of its rotational energy during the RLO decoupling phase). The median spin periods of XMSPs and radio MSPs are 2.41~ms and 3.79~ms, respectively. More striking, however, is the difference in their overall distributions: XMSPs tend to cluster somewhat near their shortest spin periods.

As demonstrated by \cite{ptrt14}, accreting sources that exhibit oscillations exclusively during thermonuclear Type~I X-ray bursts are significantly faster than rotation-powered NSs (i.e. radio MSPs), whereas accreting systems with a magnetosphere that show coherent pulsations are not.
If this latter sample with magnetospheric pulsations and the eclipsing radio MSPs (spiders) represent a common class of transitional pulsars \citep{pddm22}, they have a spin distribution intermediate between the faster nuclear-powered MSPs and the slower non-eclipsing radio MSPs. 
\cite{ptrt14} also interpreted these findings as evidence for spin-down driven by a declining mass-accretion rate during the late stages of the accretion phase (RLO decoupling).
\cite{pha17} likewise analysed the spin distributions of the two LMXB sub-populations, concluding that the slower group resembles the radio MSPs, whereas the significantly faster group may potentially be more affected by GW spin-down.

A peculiarity of the distribution of all known NS spin periods in XMSPs (measured among the 339 LMXBs known in our Galaxy \citep{fkg+24}, and including both accretion-powered X-ray millisecond pulsars and nuclear-powered burst-oscillation sources) is the absence of NSs spinning between 10 and 467~ms; with one notable exception: IGR~J17480$-$2446 in Ter~5 (90~ms; \citealt{smps10,pdm+11}), likely a mildly recycled pulsar that has only recently begun spinning up \citep{pavv12}.
This may indicate that the accretion of
$\Delta M_{\rm eq}\simeq \mathcal{O}(0.01\,M_\odot)$ required to reach $P=10\;{\rm ms}$ (Sect.~\ref{sec:theory}) is typically quite rapid once RLO is initiated. At the Eddington accretion rate, this takes only $\sim 1$~Myr.

%%%%%%%%%%%%%%%%%%%%%%%%%%%%%%%%%%%%%%%%%%%%%%%%%%%%%%%%%%%%%%%%%%%%%%%%%%%%%%%%%%%%%%%%%%%%%%%%%%%%%%%%%%%%%%%%%
\clearpage
\section{Selection effects against sub-ms pulsars}\label{sec:selection-effects}
The detectability of sub-MSPs and their expected observational signatures have been investigated for many years \citep[e.g.][]{pcg+99}. Since the discovery of the first binary pulsar in the Arecibo Galactic-plane survey \citep{ht75} and of the first MSP \citep{bkh+82}, successive generations of large-scale surveys have progressively expanded the accessible discovery space for short-period and tight-binary pulsars, including the Parkes Multibeam Pulsar Survey \citep{mlc+01}, the PALFA survey \citep{cfl+06}, the HTRU surveys \citep{kjs+10}, the GBNCC survey \citep{slr+14}, the LOTAAS survey \citep{scb+19}, the FAST GPPS survey \citep{hww+21}, and most recently the MPIfR--MeerKAT Galactic Plane Survey \citep[MMGPS]{mmgps}. 
In parallel, motivated by the strong EoS constraints that a sub-MSP would provide \citep{bd97,hlz99}, several dedicated sub-ms experiments were conducted around the turn of this century: the Bologna Northern Cross survey \citep{dam00}, targeted searches of unidentified and highly polarized radio sources \citep{ckb00,hml+04}, a coherently dedispersed baseband search for sub-MSPs up to $\sim 0.5\;{\rm ms}$ spin periods in 19 GCs with $\sim 25\;\mu s$ sampling time \citep{evb01}, and blind plus targeted searches at Nan\c{c}ay \citep{dpf11}. Despite all these efforts, no sub-MSPs were found and the fastest known spin period has remained at about 1.4~ms \citep{hrs+06}. 
It is therefore natural to ask whether this apparent spin-period ceiling could, at least in part, result from survey selection effects.

Searches for sub-MSPs at radio frequencies are potentially subject to a number of significant radio-survey selection effects. Recently, a Bayesian analysis of MSP spin distributions, incorporating survey sensitivities, pulse broadening, and detection thresholds, supported the conclusion that the lack of sub-MSPs is not due to observational bias, but is an intrinsic property of the MSP population \citep{lyc+24}. In this section, we perform our own simulations to this effect, to understand these biases. Our simulations are complementary as they are performed using injections on real astrophysical data, and using the exact processing pipelines used in some of the ongoing large scale surveys. 

Firstly, any pulsar can only be potentially detected if and only if the effective observed pulse width $W_{\rm obs}$ is smaller than the pulse period $P$ \citep{lk12}. This effective width is related to the intrinsic width of the pulse $W_{\rm int}$ as: 
\begin{equation}
  W_{\rm obs} \approx \sqrt{W_{\rm int}^2 + \Delta t^2 + \tau_{\rm DM}^2  + \tau_{\rm sc}^2 + \tau_{B}^2} \;,
\end{equation}
where $\Delta t$, $\tau_{\rm DM}$, $\tau_{\rm sc}$, $\tau_{B}$
are pulse broadening due to finite sampling time of the data, residual intra- and inter-channel dispersion, interstellar scattering and uncorrected binary motion, respectively. As the pulse is broadened due to these effects, the power of the pulse gets redistributed to the background. In other words, this will eliminate power at higher harmonics of the pulsar, leading to reduced signal-to-noise ratio (S/N) of detection. While all pulsars are affected by these effects, this is exacerbated for shorter periods, especially at sub-ms. 

The second term, $\Delta t$, is due to the finite sampling of the time series. In  most modern surveys, this ranges between $64-128\;\mu{\rm s}$, which provides 8$-$16 unique samples across the phase of a 1~ms pulsar, enough to sufficiently resolve the pulse, unless sub-MSPs preferentially have either very narrow profiles (which are smeared by the finite sampling interval, $\Delta t$) or very wide profiles (which hinder robust estimation of the signal-to-noise ratio).

The third term, $\tau_{\rm DM}$ can in turn be split into intra-channel and inter-channel dispersion contributions. The intra-channel smear arises from the fact that most search data are not coherently dedispersed, and hence there is residual smear within a given channel that smears the pulse. This inter-channel dispersion on the other hand arises due to the fact that there is always a small difference between the ``trial'' DM searched and the actual DM of the pulsar. In large surveys, this smearing can become several $100\;\mu{\rm s}$ at high DMs, limiting the sensitivity to sub-MSPs. This generally sets a DM horizon for sub-MSP searches that is well below the conventional limit of the survey.

The fourth term, $\tau_{\rm sc}$, arises from multipath propagation of the signal through the inhomogeneous ionized interstellar medium, which convolves the intrinsic profile with a pulse broadening function that is often well approximated by a one-sided exponential of characteristic timescale $\tau_{\rm sc}$ \citep{wil72,lk12}. Crucially, and in contrast to $\Delta t$
and $\tau_{\rm DM}$, this broadening is a propagation effect and cannot be reduced by finer sampling, narrower channels, or coherent dedispersion: it sets an irreducible floor on $W_{\rm obs}$. 
Since $\tau_{\rm sc}$ scales steeply with both dispersion measure and observing frequency (with $\tau_{\rm sc} \propto \nu^{-4.4}$ expected for Kolmogorov turbulence and $\nu^{-3.9\pm0.2}$
measured empirically), the mean $\tau_{\rm sc}$--DM relations \citep{bcc+04,kmn+15,cordes2016radio,TPA-scattering} predict broadening comparable to a sub-ms spin period at $\mathrm{DM} \gtrsim 200\;\mathrm{pc\,cm^{-3}}$ at L~band, albeit with an order of magnitude or more of scatter at fixed DM. 
Scattering therefore imposes a hard DM (and hence distance) horizon for sub-MSPs that is substantially closer than both the survey's nominal DM limit and the $\tau_{\rm DM}$-set horizon discussed above. Observing at higher frequencies mitigates the scattering at the cost of the typically steep pulsar spectra \citep[mean spectral index $\approx -1.6$;][]{jhm+18}, while low-frequency surveys  are effectively blind to distant sub-MSPs. 

The smearing due to the final term, $\tau_{B}$ depends on the fraction of the orbit covered by the observation and on what kind of orbital corrections are applied. For integrations much smaller than the orbital period ($t_{\rm obs} \lesssim P_{\rm orb}/10$), the line-of-sight motion is well approximated by a constant acceleration, motivating acceleration searches implemented via time-domain resampling \citep{and90,jk91,clf+00} or matched filtering in the Fourier domain \citep{rem02}. 

In practice, Fourier-domain acceleration searches often adopt a constant search range in Fourier bins for all spin periods. This results in a progressively smaller coverage of the orbital parameter space for faster pulsars, further reducing the sensitivity to sub-MSPs (Appendix~\ref{app:radio-selection}; Fig.~\ref{fig:subms-UHF-presto}).
Alternatively, orbital modulation can be removed by coherent demodulation using banks of Keplerian orbital templates \citep{mpp09,hac09,bcb+22}. Although highly sensitive, these searches remain computationally prohibitive for blind large-scale pulsar surveys.

Short-period pulsars may also exhibit steeper radio spectra, with flux density scaling as $S_\nu \propto \nu^\alpha$ and $\alpha < 0$, which can cause them to be under-represented in surveys conducted at high observing frequencies (e.g.\ $\gtrsim 1.4$~GHz). For example, PSR~J1552+5437 ($P = 2.43$~ms) was discovered at 150~MHz and exhibits a spectral index of $\alpha = -2.8 \pm 0.4$, suggesting the existence of a population that may be missed by high-frequency searches.

Although there is evidence that the spectral indices of MSPs as a class are broadly similar to those of the wider pulsar population, there are indications that MSP spectra may become steeper at shorter spin periods. For instance, the Southern MSP census \citep{Spiewak+2022} reports a median spectral index of $\sim -1.9$ for MSPs with $P>2\;{\rm ms}$, compared to $\sim -2.8$ for those with $P<2\;{\rm ms}$. While these results may be influenced by small-number statistics and other selection effects, it is nevertheless noteworthy that, if this trend is real and extends to sub-ms periods, MSPs may be intrinsically brighter at lower radio frequencies. At the same time, low-frequency observations are increasingly affected by signal smearing due to interstellar dispersion and scattering. This provides strong motivation for pulsar searches that employ coherent dedispersion, which at least partly eliminates the adverse smearing effects compared to traditional incoherent search techniques. The computational complexity associated with coherent dedispersion will make this feasible for a small set of targets such as GCs. 

Beaming fractions of sub-MSPs are yet another unknown in this puzzle. In general, radio MSPs seem to exhibit much wider pulse profiles than their young counterparts. \cite{kxl+98} first proposed that the pulsar spin period is related to the radio-beam opening angle ($\rho$) as $ \rho \propto P_{spin}^{\alpha}$, where $\alpha \simeq -0.5$, which has since been supported by subsequent studies \citep{Posselt2021, VenkatramanKrishnan2019}. This would naturally lead to MSPs having wide profiles although whether this relation extends out to MSP (or sub-MSP periods) is still debated. 
Consistent with this picture, \cite{kxl+98} also found that MSPs exhibit systematically larger effective duty cycles, with a mean value of $\sim 20\%$, reflecting broad and frequently multi-component pulse profiles; this motivates the adoption of typical duty cycles of 10–-20\% in our detectability estimates (Appendix~\ref{subsec:scaling-eqn}). There have been recent hints that MSPs may also include emission from beyond the polar cap (cf.\ \citealt{Kramer2025}). If this is confirmed, this would be one explanation for the observed wider pulse profiles. 
Additionally, short duty cycles distribute power to higher harmonics of the spin frequency, allowing a sub-MSP to be detected through one of its harmonics at a nominal MSP period. Our failure to detect sub-MSPs in this way may suggest that they preferentially have very wide pulse profiles.

\subsection{Realistic estimates of selection effects in the radio spectrum}
Given the high dimensionality of the parameters affecting sub-MSP detection, as discussed above, introducing theoretical models with their own inherent biases would further complicate the picture. Hence, our approach to quantify the selection biases in this section is to only focus on concrete, observable parameters such as pulse dispersion, spin period and S/N. This way we remain agnostic to the assumptions of any luminosity functions, DM-distance relations and their variations for specific lines of sight. 
The reader can easily derive the corresponding conclusions for their model of choice by scaling the DM and S/N posteriors according to their preferred luminosity function, DM--distance relation, and other model assumptions, which we provide in full as supplementary material. 

We perform a simulation of a real-world pulsar survey, by iteratively injecting 40\,000 sub-MSPs and MSPs in total, with a variety of properties such as spin period, duty cycle, DM and orbital parameters into real data from the MeerKAT telescope (specific details on the injection methodologies will be reported in Senzel et~al., in~prep.). We use the data obtained as part of the MMGPS survey \citep{mmgps} with the L--band (856--1712\,MHz) and UHF receivers (544--1088\,MHz). 
Each observation consisted of 2048 frequency channels, with durations of 9.2~min (L-band) and 8.2~min (UHF-band), and corresponding time resolutions of 153~$\mu$s and 120~$\mu$s, respectively.

\begin{SCtable}[20][ht]
    \caption{Injected parameter ranges for ultra-fast-spinning pulsars. All parameters have a uniform prior distribution between these values.}
    \label{table:inject}
    \centering
    \begin{tabular}{lcc}
     \toprule
     Parameter & Min & Max \\
     \midrule
     Spin period (ms) & 0.5 & 2 \\
     Folded pulse S/N & 8 & 80 \\
     DM (pc\,cm$^{-3}$) & 3 & 350 \\
     Duty cycle & 0.01 & 0.8 \\
     Acceleration (m\,s$^{-2}$) & $-100$ & $+100$ \\
    \bottomrule
    \end{tabular}
\end{SCtable}

The pulsar properties were drawn from uniform distributions in the parameter ranges listed in Table~\ref{table:inject}. We note that these do not resemble the actual expected distribution of sub-MSP properties. However, these were chosen to be uniform so as to focus only on observable parameters, as opposed to inferred distributions based on any pulsar beaming or distance model. We also only provide outcomes in S/N, thereby taking the actual telescope used out of the equation. One can scale our results for a more sensitive telescope via the radiometer equation by simply scaling the S/N. The outcome of these simulations can be read as how much worse a given survey is at detecting sub-MSPs as opposed to conventional pulsars. Intra-channel DM smearing and scattering are applied to the pulse profile. For scattering, we use the Thousand-Pulsar-Array DM-scattering relation (see Eq.~4 from \citealt{TPA-scattering}).

We choose to employ standard data reduction pipelines that use two of the most widely used pulsar-search software packages: (1) \textsc{peasoup} \citep{Peasoup} and (2) \textsc{presto} \citep{Presto}. Further details on the implementations can be found in Appendix~\ref{app:radio-selection}.

\begin{figure}[ht]
\centering
% \vspace*{-0.7cm}\hspace*{-0.3cm}
\includegraphics[width=0.95\textwidth]{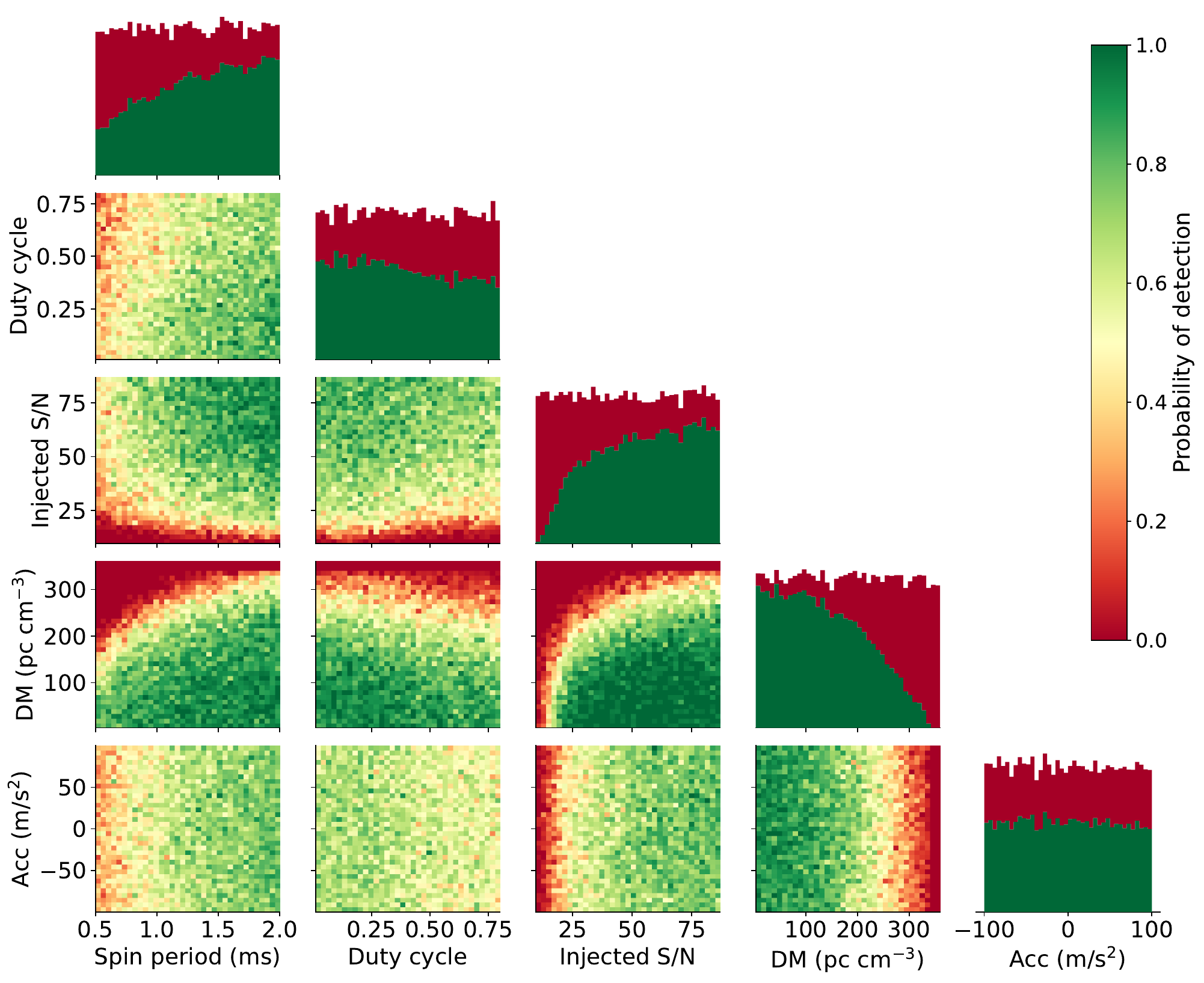}
\caption{Corner plot showing the results of the injection test performed at L-band. The diagonal histograms show the distribution of detected (green) and non-detected (red) pulsars. The off-diagonal 2D panels are colour-coded based on the probability of detecting the pulsar by the search pipeline.}
\label{fig:subms-Lband}
\end{figure}

Figure~\ref{fig:subms-Lband} illustrates the resulting regions of parameter space in which ultra-fast-spinning pulsars are successfully recovered by the search pipeline. The diagonal histograms show the full distribution of detected (green) and non-detected (red) pulsars, after running the injected pulsars through the time-domain acceleration search and the PICS AI classifier \citep{PICS}. Here, the AI serves as a proxy for the likelihood that a human observer would visually classify the (sub)-MSP candidate as astrophysical. PICS is agnostic to the pulsar candidate's physical parameters, such as spin period, dispersion measure, and acceleration. Instead, it relies exclusively on the structures seen in the candidate diagnostic plots, including the behaviour of the pulse profile as a function of time, frequency, and intensity.
It has been trained on a diverse population of pulsars spanning a wide range of pulse morphologies, and we therefore do not expect its performance to be intrinsically biased against sub-MSPs.
We note, however, that a classifier designed specifically to identify heavily smeared or scattered pulsars might perform marginally better. To date, no pulsar survey has employed such a classifier in its search pipeline.

The off-diagonal panels in Fig.~\ref{fig:subms-Lband} with 2D distributions are colour-coded based on the probability of detecting the pulsar by the search pipeline. This is computed by calculating the ratio of detected pulsars to the total number of pulsars in each bin. See Appendix~\ref{app:radio-selection} for a similar figure using the \textsc{presto} software.

\begin{figure}[ht]
\centering
\includegraphics[width=0.95\textwidth]{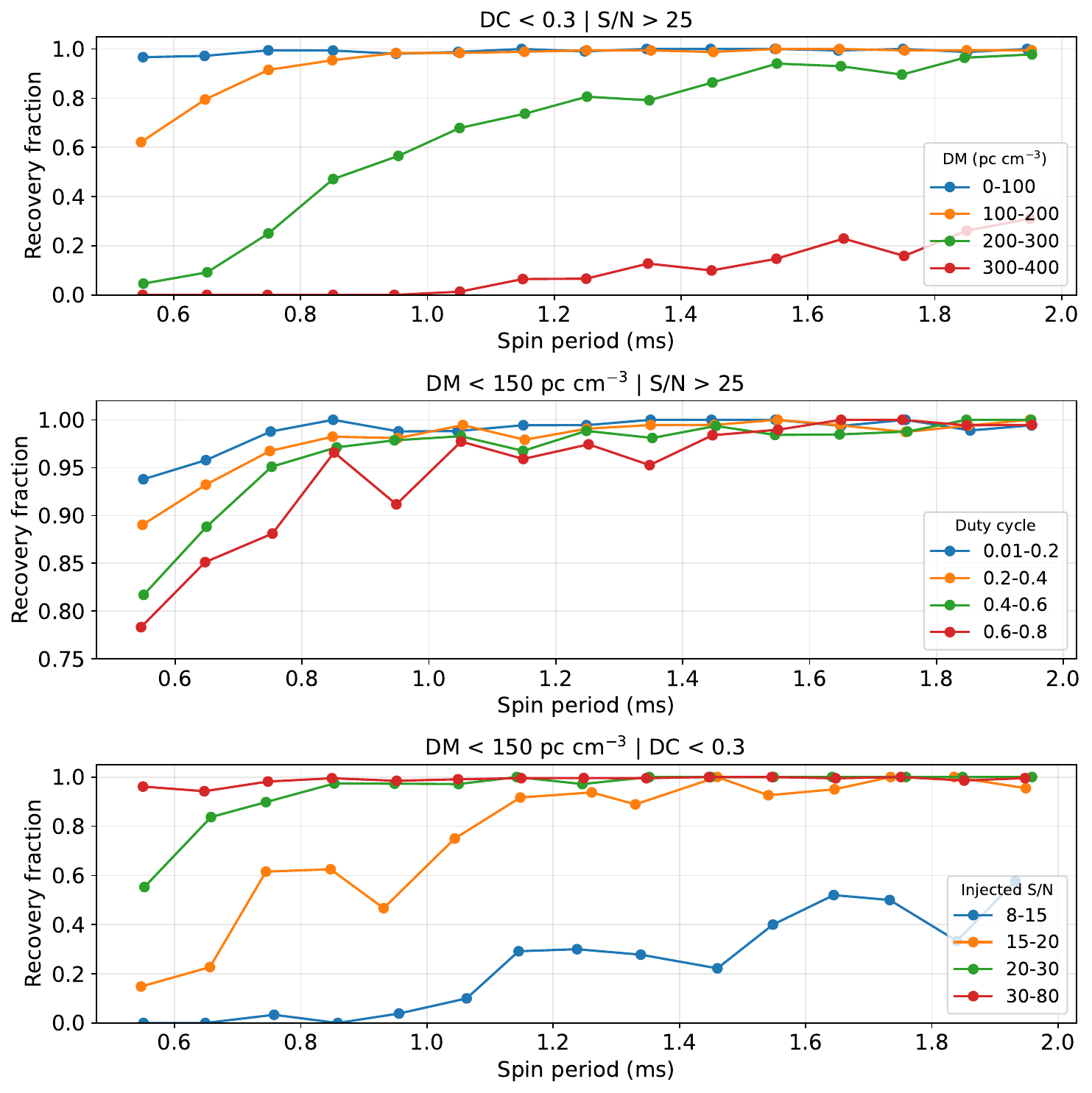}
\caption{Recovery fraction as a function of injected spin period for representative slices in dispersion measure (DM; top), duty cycle (middle), and injected S/N (bottom). In each panel, the remaining injection parameters are restricted to ranges where the recovery fraction is close to its maximum, thereby reducing the influence of those parameters and isolating the dependence on the quantity being varied. The curves show the fraction of injected pulsars recovered by the search pipeline within each spin-period bin.}
\label{fig:inj_summary}
\end{figure}

To provide a more concise overview of the injection-recovery analysis, Fig.~\ref{fig:inj_summary} shows the recovery fraction as a function of spin period for representative slices in dispersion measure, duty cycle, and injected S/N. In each panel, the remaining injection parameters are constrained to ranges over which they have little impact on the recovery fraction, allowing the dependence on the parameter of interest to be shown more clearly.

\begin{figure}[ht]
\centering
\includegraphics[width=0.95\textwidth]{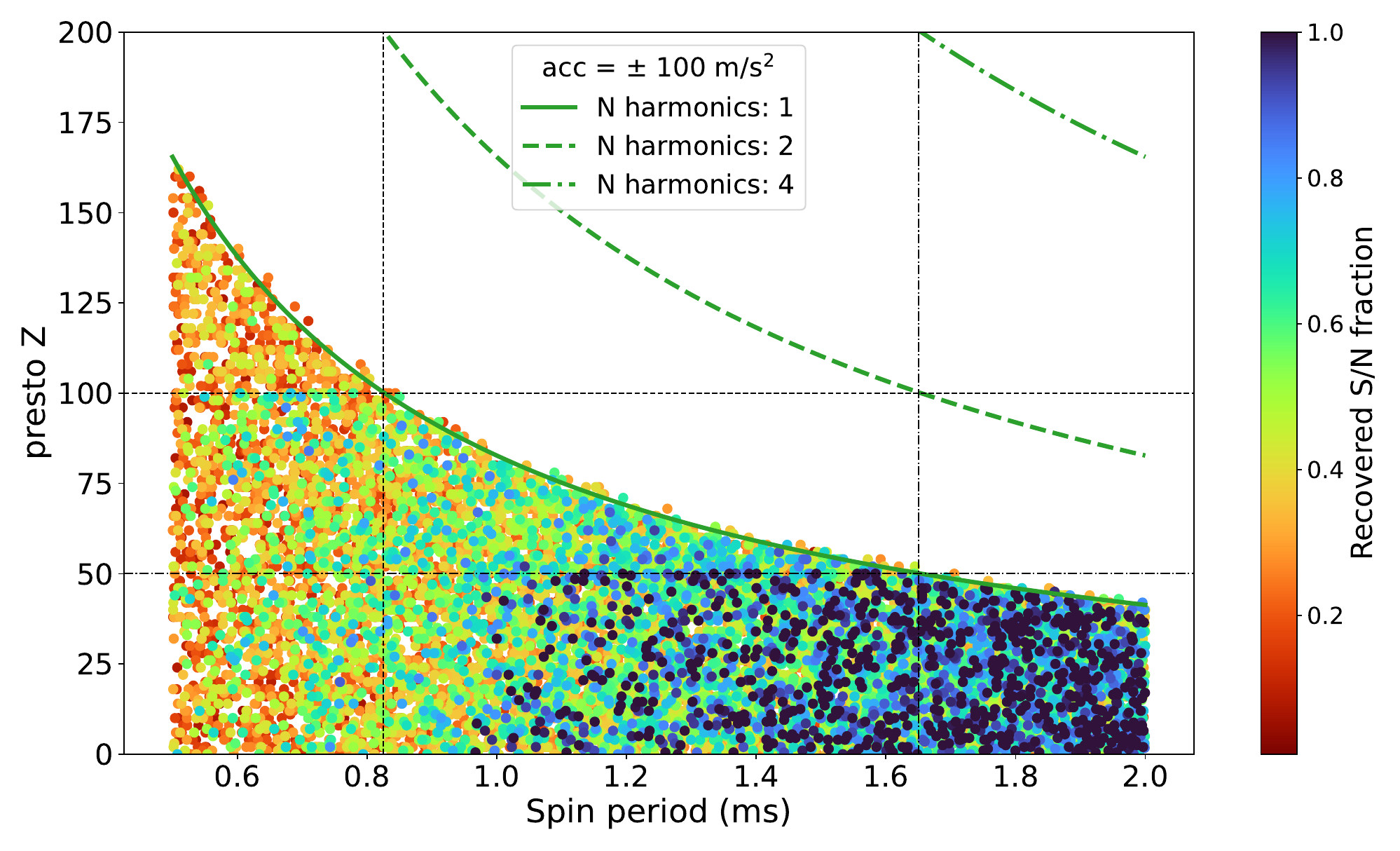}
\caption{\textsc{presto} $Z$-value (bins in Fourier drift; see Appendix~\ref{app:radio-selection}) as a function of spin period for each injected pulsar with the colour denoting the fractional recovery of S/N. UHF-band data was used for this analysis with a DM cut of $\leq 100\;{\rm cm^{-3}\,pc}$, thereby choosing injections where the reduction in sensitivity due to other effects is minimal. The colour scheme represents the recovered S/N by the \textsc{presto} search pipeline, using a maximum $Z$-value of 200. The green curves show the conversion between an acceleration of $100\;{\rm m\,s}^{-2}$ to \textsc{presto} $Z$-value as a function of spin period for different spin harmonics. For $Z=200$, the fundamental (solid green line) is fully covered. For the second harmonic (dashed line), $Z=200$ only covers up to a spin period of $\sim 0.8\;{\rm ms}$, while the fourth harmonic (dashed-dotted line) covers up to $\sim 1.6\;{\rm ms}$. This results in a discontinuous drop in recovered S/N at $Z=50$ and $Z=100$. The drop in recovered S/N going to faster spin periods is due to the finite time sampling of the data.}
\label{fig:subms-UHF-presto}
\end{figure}

\begin{figure}[ht]
\centering
\includegraphics[width=0.95\textwidth]{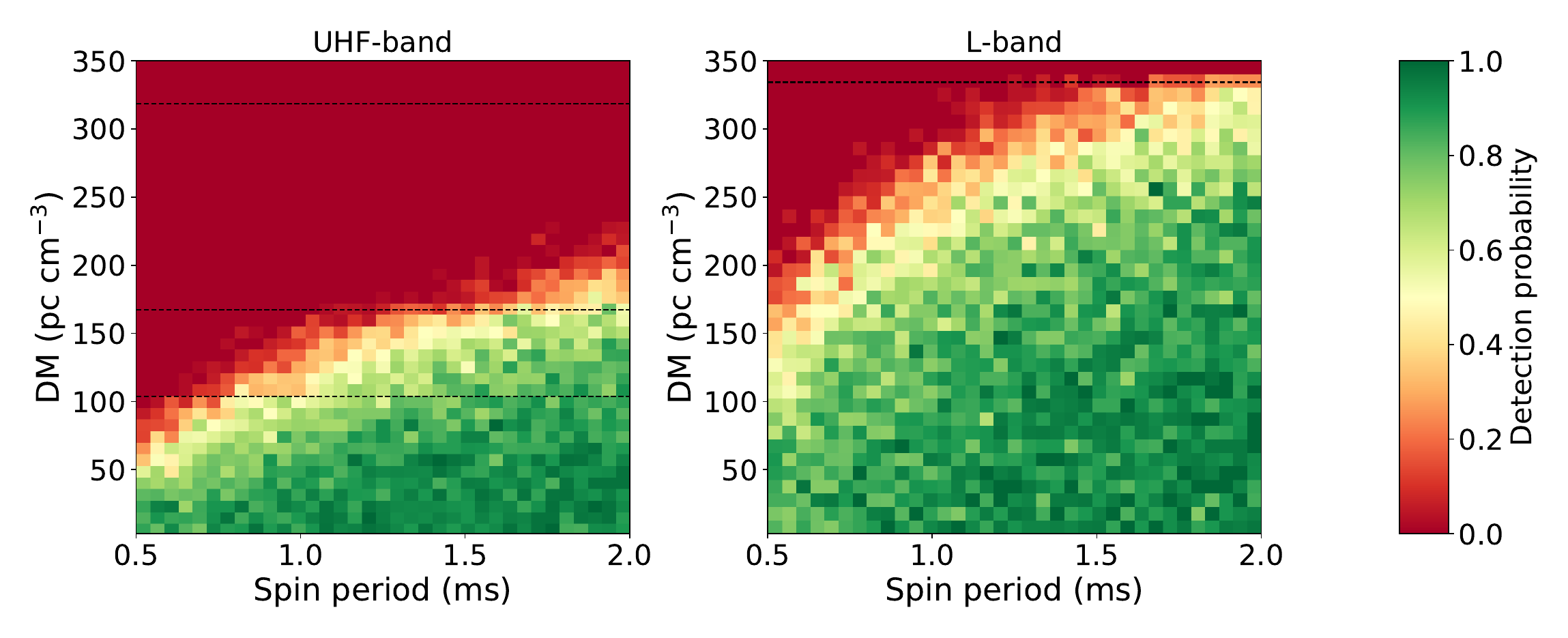}
\caption{Comparison between the detectability of pulsars as a function of spin period and dispersion measure. The left plot shows the results for UHF and the right shows the same plot for L--band. The dashed black lines show where the data is downsampled in time by a factor of two before the search.}
\label{fig:DM-band}
\end{figure}

Based on these simulations, the key qualitative conclusions are the following:
\begin{itemize}
    \item  There is a definite reduction in sensitivity towards sub-MSPs for a survey, compared to conventional MSPs. This is a strong function of period, S/N and DM, which mainly reflect the effects of intra-channel DM smearing and scattering.

    \item The reduction in sensitivity is also a function of duty-cycle (although weaker), with a bias against large duty-cycles. If sub-MSPs predominantly have very wide profiles (as most MSPs do), this may contribute to a systematic bias in the detection of sub-MSPs. This will also concentrate most of the Fourier power at the fundamental frequency, thereby prohibiting them from being detected at their harmonic periods where selection effects are less severe. Although the shortest injected duty-cycles are sampled by relatively few time bins, the effective pulse widths in our simulations are generally dominated by interstellar scattering rather than the intrinsic duty cycle. Consequently, the finite sampling introduces only a minimal reduction in sensitivity at low duty-cycle, as demonstrated by the weak dependence on low duty-cycle in Fig.~\ref{fig:subms-Lband}.
    
    \item As long as the orbit is wide enough (only $\lesssim 10\%$ of the orbit is recorded in a given observation, see Appendix~\ref{subsec:scaling-eqn} for details) to be approximated by a constant acceleration of the pulse profile and that this acceleration is covered in the range searched, there is no loss of sensitivity, other than the scalloping loss between acceleration trials. However, there may be degradation in S/N if the searched Fourier drifts in frequency-domain searches do not cover the full spread of higher harmonics of the pulsar, as is generally the case (see Fig.~\ref{fig:subms-UHF-presto}).
    If sub-MSPs predominantly occur in short period binary orbits where a significant fraction of the orbit ($> 10\%$) is covered in the observation, this might introduce further selection biases, unless higher order corrections like jerk or snap searches are performed. However, there are no \textit{a~priori} reasons to expect so. The two fastest-spinning MSPs, PSR~J1748$-$2446ad and PSR~J0952$-$0607, both reside in ``black widow'' systems (Sect.~\ref{subsubsec:spiders}) with low-mass, sub-stellar companions, although with very different orbital periods of 26~hr and 6.4~hr, respectively. At these orbital periods and companion masses, a bright 1~ms pulsar would be readily detectable under current search scenarios.

    \item  At L-band frequencies with MeerKAT, sub-MSPs can only be found with ${\rm DM}\lesssim 250\;{\rm cm^{-3}\,pc}$ and ${\rm S/N}\gtrsim 30$. Beyond this point, the predominant effects that affect the sensitivity are residual dispersion and interstellar scattering. If we compare this search to other wavelengths, such as UHF, we see an almost constant shift in the DM horizon (see Fig.~\ref{fig:DM-band}). Changing the scattering timescale or the scattering index also changes the DM horizon. An example is shown in Appendix~\ref{subsec:scattering_prescription}.

    \item For very bright pulsar signals, these selection effects are negligible. There is no reason to believe that any survey cannot detect bright sub-MSPs with ${\rm S/N}>30$. What this means in terms of luminosities and distance horizons depends on individual surveys and the luminosity distribution considered. For the L-band MMGPS survey, assuming pulsar luminosities similar to those of MSPs with known parallax-based distances, the limiting sensitivity for a spin period of 1~ms is about 0.23~mJy. 

    \item For GCs where the DM is known \textit{a~priori}, most modern observations exploit real-time coherent dedispersion at the pulsar's central DM, thereby largely mitigating intra-channel dispersion smearing. This, however, has no effect on interstellar scattering. At very high DMs, scattering therefore remains the dominant cause of sensitivity loss. An example of such a search is presented in Appendix~\ref{subsec:globular_cluster_search}.
\end{itemize}

\subsection{Selection effects against sub-ms X-ray MSPs}\label{subsec:selection-effects-X}
The time resolution of X-ray detectors and the number of photons detected in the given integration time are the main factors affecting the sensitivity of a search for a coherent signal from an X-ray pulsar. A number of past and currently active X-ray observatories feature a time resolution of a few tens of $\mu {\rm s}$ depending on the observing mode, with some (e.g., RXTE, NICER) attaining sub-$\mu {\rm s}$ resolution, so that no significant purely instrumental bias against the detection of a sub-ms signal exists.

If X-ray pulsars are rotation-powered, they are more energetic at fast spins. However, at very short spin periods the observable X-ray emission becomes increasingly affected by geometric averaging over extended emitting regions. As the light-cylinder radius ($r_{\rm lc}\equiv c/\Omega =cP/2\pi$) shrinks, magnetospheric X-ray emission originates from spatially compact regions that span a significant range of longitudes. Strong general-relativistic (GR) light bending --- which is largely independent of spin for realistic NS compactness --- further enhances the simultaneous visibility of large surface fractions throughout the spin cycle \citep[e.g.][]{pfc83,bel02,pb06,hpy07}. As a result, the observed flux becomes progressively less modulated in rotational phase and the pulsed fraction is reduced, which makes period detection, rather than photon detection, increasingly difficult. In the limit of extended emitting regions and strong geometric averaging, the emission becomes effectively phase-independent from the observer’s perspective, leading to a substantial suppression of coherent X-ray pulsations.

In addition, Doppler smearing makes sub-MSPs harder to detect in binaries. The Fourier power of a coherent signal increases linearly with the observed count rate and the integration time. If the integration time is a significant fraction of the binary period, the orbital motion spreads the power into neighbouring Fourier frequency bins, effectively reducing the sensitivity. Searching on a few hundred second-long intervals is the strategy that led to the detection of all the accreting MSPs discovered so far with amplitudes of 5--10\% \citep[see e.g.][]{pw21}. The amplitude of the frequency shift produced by the binary motion in a given time interval scales linearly with the frequency of the signal. As a result, the optimal integration time scales as $\sim \nu^{-1/2}$ \citep{jk91}, introducing a mild frequency-dependent sensitivity loss and slightly disfavouring the detection of a fast pulsar. However, the sensitivity decrease of a source spinning at 1~kHz compared to the fastest accreting MSP discovered so far ($\nu\simeq 599\;{\rm Hz}$) is just $\sim 20$\%. This is much smaller than the width of the distribution of distances and luminosities of XMSPs, and its impact on the search sensitivity can easily be balanced by observing slightly brighter sources.

To mitigate these orbital effects, acceleration searches increase the search sensitivity by lengthening the optimal integration time by a factor of $\sim$5--10 and weakening their dependence on pulsar frequency \citep[$\sim \nu^{-1/3}$,][]{jk91}. Semi-coherent strategies further allow for combining sparse observations performed over a wider time interval \citep[see e.g.][]{mp15,gwgm22}. However, employing these techniques to search for pulsations from a sample of 27 bright LMXBs did not succeed in detecting X-ray pulsations down to amplitude upper limits as low as 0.2\% \citep{vvdk94,pwm18}. Recent application of these search techniques in the optical band did not find pulsations from the brightest known LMXB (Sco~X-1) within an amplitude upper limit of $\sim 0.01$\% \citep{lpi+26}. Even though the sensitivity in these strategies still depend on the signal frequency\footnote{The larger frequency shifts produced by orbital motion of a fast pulsar increase the volume of the parameter space that has to be searched.}, the non-detection of any signal from bright LMXBs is most likely due to weak or absent magnetosphere around the NS in these systems \citep[see][and references therein]{pwm18}. Burial of the B-field below the NS surface \citep{czb01} and a higher optical depth of Compton scattering clouds around the NS \citep{tcw02} are possible explanations of why pulsations are not detected from systems with high mass-accretion rates (cf. Fig.~\ref{fig:mag-radius}). 
Furthermore, for a magnetised NS to channel the accretion flow onto its magnetic poles, the inner disc radius must exceed the NS radius, but not significantly exceed the co-rotation radius, whose size decreases with increasing spin frequency ($\sim \nu^{-2/3}$). The inner disc radius, which depends on the mass-accretion rate (e.g. $R_{\rm disc}\propto \dot{M}^{-2/7}$ when expressed in terms of the Alfv\'en radius; Sect.~\ref{sec:theory}), therefore has only a narrow allowed radial interval for a very rapidly spinning MSP ($\sim 12-19\;{\rm km}$ for a $2.0\;M_\odot$ NS spinning at 1~ms, see Fig.~\ref{fig:mag-radius}). This further selection effect reduces the likelihood of detecting accretion-powered coherent pulsations from a sub-ms MSP.

A similar argument applies to burst oscillations observed during Type~I X-ray bursts. These signals last a few tens of seconds at most, i.e. much shorter than the orbital period, and there is therefore no direct bias against the detection of a very fast signal. However, because burst oscillations are extremely short-lived phenomena with rapidly evolving properties, the number of trials employed in searches has a critical effect on the statistical significance of a detection, requiring particular care to distinguish between confirmed and candidate burst-oscillation sources \citep{bw19}. Furthermore, most of the brightest accreting NSs in LMXBs (the Z~sources) do not show bursts, likely because of stable burning of the accreted material. Similar to the case of accretion-powered X-ray pulsations, this reduces the likelihood of detecting the fastest pulsars, assuming that they are hosted in systems with the highest average mass-accretion rates (see Eq.~\ref{eq:spinup}).

\subsection{Selection effects against sub-ms MSPs in $\gamma$-rays}
Since the launch of the \textit{Fermi} satellite, an increasing number of rotation-powered pulsars are being found in $\gamma$-rays, owing to its Large Area Telescope (LAT, \citealt{2009_LAT}): in the latest $\gamma$-ray pulsar catalogue \citep{2023_Smith_pulsar_catalogue} about 340 pulsars are listed, and their number is approaching nearly 10\% of the population of all known rotation-powered pulsars \citep{mhth05}.

Searches for $\gamma$-ray pulsars \citep{2012_Pletsch_9gamma-ray_pulsars,2012_Pletsch_gamma-ray_binary} are especially relevant for the topic of this work because they select very strongly in favour of pulsars with a very large spin-down power:
\begin{equation}
  \dot{E} \equiv 4 \pi^2 I \frac{\dot{P}}{P^3}\;,
\end{equation}
which results in proportionally larger $\gamma$-ray luminosities. $\dot{E}$ is enhanced for very fast-spinning pulsars: despite having generally smaller values of $\dot{P}$, the $P^{-3}$ dependence implies that a large fraction of all MSPs are detected in $\gamma$-rays. In addition, searches for $\gamma$-ray pulsars are, in general, not affected by dispersive effects, multi-path scattering, and radio eclipses. For this reason, searches for $\gamma$-ray pulsars have found extreme MSPs that would have been very difficult to discover in radio surveys \citep{Nieder_2020}, some of them because they lack radio emission entirely \citep{Clark_2018}.

However, $\gamma$-ray searches also have their limitations. 
The first is that, for the faintest pulsars, the photon detection rates are of the order of one photon per month, or even lower. This implies that, for those sources, very long spans of data must be searched to achieve a statistically significant detection, resulting in a very large $P$-$\dot{P}$ search space. In addition, the relatively poor spatial resolution of $\gamma$-ray telescopes in general implies that there is always significant background contamination and poor {\em a~priori} source localisation. The latter in particular implies a large number of trials in sky position for every candidate $\gamma$-ray source; otherwise a poor correction of the Earth's orbital motion smears the timing signal.
Finally, the very limited photon statistics make it very difficult to detect binary pulsars. Searching for them requires sampling, in addition to the spin and position parameter spaces, a large parameter space describing the Keplerian orbit. In practice, this can almost exclusively be done for optically identified systems where good priors are available on parameters such as the orbital period and sky position \citep[e.g.][]{Nieder_2020_binary,2024_Thongmeearkom}.
This is particularly problematic if the fastest pulsars preferentially reside in binary systems, many of them without bright, strongly varying optical counterparts.

Finally, the computational cost of all these searches increases steeply for shorter spin periods, since the error tolerances in all parameters decrease as the spin period becomes shorter \citep{Clark_2017}. This means that the computational cost is the main barrier to the discovery of faster pulsars, especially in binary systems.

\subsection{Selection effects against sub-ms MSPs in optical and UV}
Very-fast optical and UV photometry has recently revealed coherent optical and UV pulsations at the spin period of a transitional MSP in a sub-luminous accretion-disc state \citep{ap17,bzf+20,j21,mza22} and from an accreting MSP in outburst \citep{amz21}. Evidence for an optical signal from a radio MSP in a tight orbit has also recently been reported, although the strength of the optical signal is orders of magnitude lower than that from NSs surrounded by an accretion disc \citep{pab25}. Searching for sub-MSPs in the optical band is therefore a very promising avenue. Even with a 4-m-class telescope, optical observations provide photon statistics that are much better than in the X-ray band, and hence a correspondingly higher sensitivity to pulsations \citep{lpp26}. At the same time, optical observations are unaffected by radio dispersion or scattering. On the other hand, optical pulsations may be diluted by the strong background emission from the accretion disc and/or companion star, may be affected by reprocessing and colour-dependent effects, and may only be present under certain physical or geometric conditions \citep{pas19,vnb19}.

%%%%%%%%%%%%%%%%%%%%%%%%%%%%%%%%%%%%%%%%%%%%%%%%%%%%%%%%%%%%%%%%%%%%%%%%%%%%%%%%%%%%%%%%%%%%%%%%%%%%%%%%%5
\clearpage
\section{Theoretical calculations}\label{sec:theory}
Basic stages of accretion (or more precisely: stages of \emph{mass transfer, ejection, accumulation}) can be identified depending on key properties that include: mass-transfer rate (wind vs. disc accretion and its geometry), the B-field, and the spin rate of the accretor. These together govern the relative locations of the so-called \emph{stopping radius}, the \emph{light cylinder}, the \emph{co-rotation radius} and the \emph{magnetospheric (Alfvén) boundary} --- for details, see e.g. \cite{lpp73,nag89,bv91,tv23,neg+26} --- which regulate the outcome of a given accretion stage. Here, we summarise the main results relevant for post-recycling spin periods (see Sect.~\ref{subsubsec:acc-uncertainties} for uncertainties related to the state-of-the-art in accretion theory). For an early investigation of the binary-evolution and accretion requirements for producing a sub-MSP, see \citet{bpc+99}.

\subsection{MSP spin-up bands in the $P\dot{P}$--diagram}\label{subsec:bands}
\begin{figure*}
\centering
\vspace*{-0.4cm}\hspace*{-0.0cm}
\includegraphics[width=0.98\textwidth]{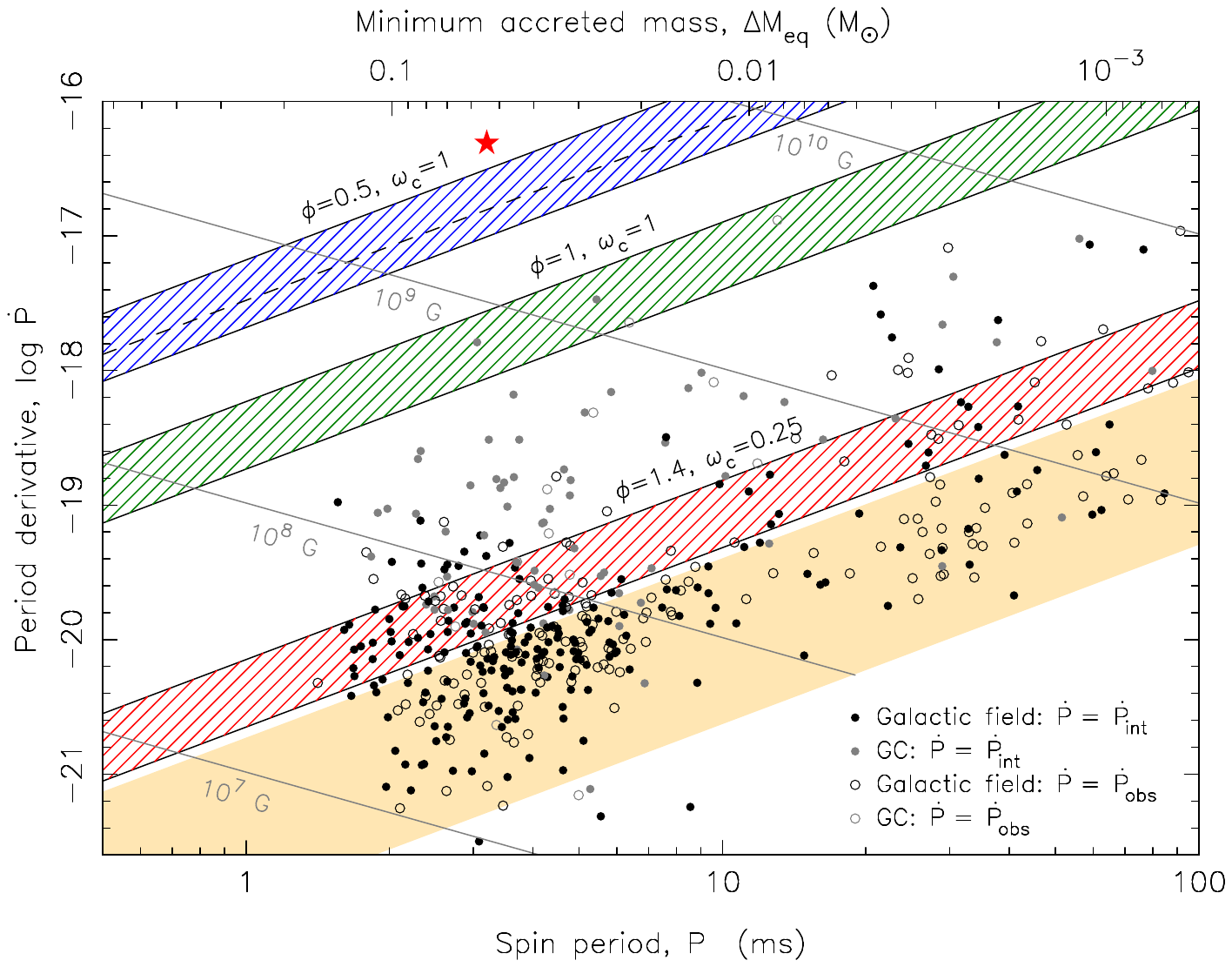}
\vspace{-0.7cm}
\caption{Theoretical spin-up lines, or ``spin-up bands'', calculated for three different cases of the parameters $\phi$ and $\omega_c$. In all cases shown (blue, green, red) the spin-up line is calculated assuming accretion at the Eddington limit, $\dot{M}=\dot{M}_{\rm Edd}$, and applying the Spitkovsky torque formalism. Thus these lines represent upper limits for the given set of parameters. The width of each ``line'' (band) results from using a spread in NS mass and magnetic inclination angle from ($2.0\,M_{\odot}$,~$\alpha=90^{\circ}$) to ($1.0\,M_{\odot}$,~$\alpha=0^{\circ}$), upper and lower boundary, respectively. The dashed line within the upper blue band is calculated from ($2.0\,M_{\odot}$,~$\alpha=0^{\circ}$), and hence the band width above this line reflects the dependence on $\alpha$, whereas the band width below this line shows the dependence on $M$. The yellow spin-up region below the red band is calculated for a $1.4\,M_{\odot}$ NS using the vacuum magnetic dipole model for the radio pulsar torque and assuming $\phi =1.4$, $\omega_{\rm c}=0.25$ and $\dot{M}=\dot{M}_{\rm Edd}$ for $\alpha=90^{\circ}$ and $\alpha=10^{\circ}$ (upper and lower boundary). The observed distribution of Galactic field and GC radio pulsars are plotted as black and grey filled circles, respectively. MSPs for which kinematic corrections of the observed value of $\dot{P}$ could not be provided (or were non-physical) are plotted as open circles. The upper horizontal axis shows the minimum amount of mass to be accreted (Eq.~\ref{eq:deltaM}) for reaching a given spin period. For PSR~J0435+3233 (red star), see Sect.~\ref{subsec:J0435+3233}. Updated after \cite{tlk12}.}
\label{fig:spinupline}
\end{figure*}

Figure~\ref{fig:spinupline} shows theoretical spin-up lines, or ``spin-up bands'', for given sets of magnetospheric conditions. The calculations can be briefly summarised as follows \citep[for a detailed derivation and discussion, see][]{tlk12}. An equilibrium configuration is obtained when the angular velocity of the NS equals the Keplerian angular velocity of matter at the magnetospheric boundary where the accreted matter enters the magnetosphere, i.e. $\Omega_\star=\Omega _{\rm eq} \equiv  \omega _c\,\Omega _{\rm K}(r_{\rm mag})$ or:
\begin{equation}\label{eq:equilibrium}
     P_{\rm eq} = 2\pi \sqrt{\frac{r_{\rm mag}^3}{GM}}\,\frac{1}{\omega _c} \;,
\end{equation}
where $\omega _c$ is the critical fastness parameter \citep[a threshold for torque reversal that depends on the dynamical importance of the NS spin and the magnetic pitch angle,][]{gl79b}, and $r_{\rm mag}=\phi\cdot r_A$ is the radius of the magnetospheric boundary, roughly equal to the Alfv\'en radius\footnote{Sometimes also defined as $r_{\rm A} \equiv \left( B^2 R^6/(4\dot{M}\sqrt{2GM}) \right) ^{2/7}$, see discussion and derivations in \cite{tv23}.}:
\begin{equation}\label{eq:alfven}
   r_{\rm A} \simeq \left( \frac{B^2 R^6}{\dot{M}\sqrt{2GM}} \right) ^{2/7} \;,
\end{equation}
defined as the location where the magnetic energy density begins to control the flow of matter (i.e. where the incoming material couples to the field lines and co-rotates with the NS magnetosphere). Here, $\phi = 0.5-1.4$ is a parameter based on the location of the inner edge of the accretion disc \citep{gl92,wan97,ds10}, although its precise value remains uncertain and is one of the main sources of uncertainty in torque prescriptions; see e.g. \citet{gs21}, who compared accreting MSPs spin-up episodes using several torque prescriptions and highlighted the uncertainty associated with the parameter $\phi$.
Spinning up the NS requires $r_{\rm mag}\lesssim r_{\rm co}$, where $r_{\rm co}=(GM/\Omega_\star^2)^{1/3}$ is the co-rotation radius.
Near spin equilibrium, trapped-disc states and time-dependent disc--magnetosphere interactions may lead to torque reversals \citep{ds12}. Numerical modelling \citep[e.g.][]{tau12} demonstrates that these torque reversals lead the NS spin period to undergo small oscillations around $P_{\rm eq}$ when $r_{\rm co}\simeq r_{\rm mag}$.

The spin-equilibrium picture is further complicated by the possibility that transient accretion may lead to substantially shorter equilibrium spin periods than persistent accretion with the same long-term average accretion rate \citep{bc17}, thereby increasing the likelihood of producing sub-MSPs, unless an additional spin-down mechanism intervenes (see Sect.~\ref{subsec:inefficient}).

By expanding the vacuum dipole model of the pulsar magnetosphere to include a plasma term in the spin-down torque, \cite{spi06} derived an expression for the B-field strength of a pulsar, $B$, depending on the magnetic inclination angle, $\alpha$ 
(where $B$ denotes the equatorial surface magnetic-field strength, such that the polar field strength is $B_p=2\,B$):
\begin{equation}\label{eq:Bspitkovsky} 
     B  = \sqrt{\frac{c^3IP\dot{P}}{4\pi ^2 R^6}\;\frac{1}{1+\sin^2\alpha}} \;,
\end{equation}
where $I$ is the NS moment of inertia. This expression is used for calculating the lines of constant $B$ in Figs.~\ref{fig:PPdot} and \ref{fig:spinupline}, assuming $M=1.4\;M_\odot$ and $\alpha = 60^\circ$.

Combining Eqs.~(\ref{eq:equilibrium})--(\ref{eq:Bspitkovsky}) and isolating $\dot{P}$, one finds \citep{tlk12}:
\begin{eqnarray}
     \dot{P} & = & \frac{2^{1/6}G^{5/3}}{\pi ^{1/3} c^3}\frac{\dot{M}M^{5/3}P_{\rm eq}^{4/3}}{I}
                \;\,(1+\sin^{2}\alpha)\;\,\phi^{-7/2}\,\omega_c^{7/3} \\ \nonumber 
             & \simeq & 1.85\times 10^{-19}\,\;P_{\rm ms}^{4/3}\,\left(\frac{M}{M_{\odot}}\right)^{2/3}\, \left(\frac{\dot{M}}{\dot{M}_{\rm Edd}}\right) \; g(\alpha,\phi,\omega_c) \;,
\label{eq:spinupline} 
\end{eqnarray}
where $P_{\rm ms}$ is the equilibrium spin period in units of milliseconds, and $g(\alpha,\phi,\omega_c)\equiv (1+\sin^{2}\alpha)\;\,\phi^{-7/2}\,\omega_c^{7/3}$. For the numerical expression, we assumed that $I_{45}\simeq M/M_{\odot}$, where $I_{45}$ is the moment of inertia in units of $10^{45}\;{\rm g\,cm}^2$ (Sect.~\ref{subsec:deltaM}).
For the Eddington accretion rate, we used $\dot{M}_{\rm Edd}=3.0\times 10^{-8}\;M_{\odot}\,{\rm yr}^{-1}$ since toward the end of the mass-transfer phase, the amount of hydrogen in the transferred matter is usually quite small ($X<0.20$); see discussions in \cite{tv23}.

From Fig.~\ref{fig:spinupline}, we notice that the location of the spin-up bands, depending on the assumed input physics, can explain all observed MSPs. The plotted blue and green bands agree with all observed MSPs --- which, after recycling, must lie to the right of these lines --- but the red band does not. \citep[The newly discovered PSR~J0435+3233,][marked with a red star, is discussed in Sect.~\ref{subsec:J0435+3233}]{wwy+26}.
One should keep in mind that whereas the three bands plotted here assume $\dot{M}=\dot{M}_{\rm Edd}$, and thus represent upper limits on $\dot{P}$ for a given value of $P$ (cf. Eq.~\ref{eq:spinupline}), each \emph{individual} pulsar has its own accretion history where, for example, $\langle \dot{M}\rangle = 0.001\;\dot{M}_{\rm Edd}$ and thus it appears at the bottom part of the diagram. Finally, and relevant for this work, all the plotted spin-up bands allow, in principle, for the existence of sub-MSPs. In other words, the maximum spin of a NS is, at first sight, not limited by the concept of equilibrium spin.

\subsubsection{Uncertainties in inferred quantities}\label{subsubsec:acc-uncertainties}
Several ingredients of disc–magnetosphere coupling remain uncertain and therefore warrant a caveat: (i) the inner-disc geometry and pressure regime (gas- versus radiation-pressure dominated); (ii) the angular-momentum exchange between the differentially rotating Keplerian disc and the NS, including how the NS B-field threads and torques the inner disc; (iii) the width and underlying physics of the transition from Keplerian flow to magnetospheric channelling; and (iv) the torque prescriptions --- that is, the relative contributions of material, magnetic, and viscous stresses \citep{pr72,vas79,gl92,wan97,fkr02,rfs04,ds10,ib12,spk+15,rbu+18,mt22}.
Furthermore, even the more simple definitions (such as the expression for the Alfvén radius) can be uncertain by factors of order unity. 
Consequently, the cumulative effect of these factors introduces non-negligible and difficult-to-quantify uncertainties in derived quantities --- for example, in the spin-up lines and in the accreted mass required for a NS to become a sub-MSP (see Sect.~\ref{subsec:deltaM} below). Nevertheless, as noted by \cite{gl92}, equilibrium spin periods computed under relatively simple assumptions remain close to those obtained from more detailed models.

\subsection{Minimum mass accreted, torque, spin and timescale}\label{subsec:deltaM}
Figure~\ref{fig:spinupline} also displays (upper horizontal axis) the \emph{minimum} mass, $\Delta M_{\rm eq}$, required to reach the equilibrium spin periods along the x-axis, following \cite{tlk12}:
\begin{eqnarray}\label{eq:deltaM}
     \Delta M_{\rm eq} & = & I\;\left( \frac{\Omega_{\rm eq}^4}{G^2M^2} \right) ^{1/3} \,f(\alpha,\xi,\phi,\omega_c) \\ \nonumber 
                       & \simeq & 0.22\,M_{\odot}\; \frac{(M/M_{\odot})^{1/3}}{P_{\rm ms}^{4/3}} \;,
\end{eqnarray}
assuming here the numerical factor $f(\alpha,\xi,\phi,\omega_c)=1$ \citep[see also][]{acrs82,bpc+99}. 
This expression follows from the addition of spin angular momentum to an accreting NS:
\begin{equation}
  \Delta J_\star = \int n(\omega,t)\,\dot{M}(t)\,\sqrt{GM(t)r_{\rm mag}(t)}\,\xi (t)\;dt \;,
  \label{eq:Jacc}
\end{equation}
where $n(\omega)$ is a dimensionless torque and $\xi \simeq 1$ is a numerical factor which depends on the flow pattern \citep{gl79b,gl92}. For details on the integration limits and the time evolution of $n(\omega)$, $\dot{M}(t)$, $r_{\rm mag}(t)$ and $\xi(t)$ during the spin-up phase, see \cite{tlk12,tv23}.

We repeat that the quantified prefactors in all expressions carry uncertainties that are non-trivial to assess. In addition to the aforementioned magnetospheric assumptions (encoded in $f(\alpha,\xi,\phi,\omega_c)$, which itself includes multiple factors of order unity), both the NS EoS and its moment of inertia remain uncertain, and internal structural changes during spin-up cannot be excluded. 
Recent GRMHD simulations further demonstrate that the torque depends on the magnetic-field geometry, the disc magnetic flux, and the accretion state, showing that simple torque prescriptions remain subject to order-unity uncertainties \citep{dpw22}.

For order-of-magnitude estimates, we adopt the canonical scaling $I_{45}=M/M_\odot$, which is accurate to within about 20\% for realistic NS masses, according to the nearly EoS-insensitive relation of \cite[][their Eq.~8 and Fig.~2]{ls05}.
As an example, when evaluating $\Delta M_{\rm eq}$ in Eq.~(\ref{eq:deltaM}), the combined uncertainties in $I$ and $\omega_c$ imply that the nominal prefactor of $0.22\;M_\odot$ should realistically lie in the range $0.18-0.30\;M_\odot$. This includes recent work refining the value of the fastness parameter that determines when the accretion torque saturates: whereas the analytic threaded-disc model of \cite{gl79b} predicted $\omega_c\simeq 0.35-0.40$, work by \cite{psb16}, based on axisymmetric relativistic force-free simulations, shows that the accretion disc opens additional B-field lines of the NS magnetosphere, thereby enhancing the electromagnetic spin-down (pulsar-wind) torque that arises from currents in a plasma-filled magnetosphere. This increase in open magnetic flux allows spin equilibrium to be maintained at higher fastness parameters, $\omega_c\simeq 0.7-1.0$, than predicted by standard disc–magnetosphere coupling models.

\subsubsection{Spin-up timescale}\label{subsubsec:timescale}
An additional requirement for producing a sub-MSP is that the accretion torque can impart the necessary spin-up on a timescale shorter than the mass-transfer timescale. To estimate the spin-relaxation timescale (the time needed to spin up a
slowly rotating NS to spin equilibrium) one can follow \cite{tlk12} and simply consider $t_{\rm torque}=J/N$, 
where $J=I\,\Omega_{\rm eq}$ and $N=\dot{M}\sqrt{GMr_{\rm mag}}\,\xi$, which yields:
\begin{eqnarray} 
\label{eq:timetorque}
    t_{\rm torque} & = & I\,\left(\frac{4G^2M^2}{B^8R^{24}\dot{M}^3}\right) ^{1/7} \frac{\omega _c}{\phi^2\,\xi} \\ \nonumber
           & \simeq & 50\;{\rm Myr}\quad B_8^{-8/7}\left(\frac{\dot{M}}{0.1\,\dot{M}_{\rm Edd}}\right) ^{-3/7} \left(\frac{M}{1.4\,M_{\odot}}\right) ^{17/7} \;,
\end{eqnarray} 
(equivalently $t_{\rm torque}=\Delta M/\dot{M}$) where the numerical prefactor is for $\phi=\xi=\omega_c=1$. 
If the duration of the mass-transfer phase $t_X$ is shorter than $t_{\rm torque}$, then the pulsar will not be fully recycled. 

When mass transfer ceases toward the end of RLO and the magnetosphere expands dramatically, the equilibrium spin period increases significantly ($P_{\rm eq} \propto r_{\rm mag}^{3/2}$), apparently preventing the formation of a fast MSP. The resolution to this paradox \citep{tau12} is that the pulsar decouples from strict equilibrium and that the detachment phase is relatively brief compared to the spin–relaxation timescale. Although the pulsar may typically lose $\sim 50\%$ of its rotational energy (i.e.\ $P$ increases by $\sqrt{2}$) during this phase, it nonetheless remains a MSP.

We note that the surface B-fields expected for potential sub-MSPs are small ($\sim 10^7\;{\rm G}$, unless $\dot{M}\gg \dot{M}_{\rm Edd}$; Fig.~\ref{fig:mag-radius}), so the RLO decoupling effect is less pronounced. Nevertheless, it cannot be ruled out that part of the non-detection of a radio sub-MSP is due to the decoupling phase --- supported by the comparison between XMSP spins and radio MSP spins (post-RLO decoupling), see also \cite{ptrt14}.

\subsection{Companion star and efficiency of recycling}\label{subsec:inefficient}
From Eq.~(\ref{eq:deltaM}) and Fig.~\ref{fig:spinupline} (top axis), it is immediately seen that, to produce a sub-MSP, a typical $1.4\;M_\odot$ NS must accrete at least $~\sim 0.25\;M_\odot$. However, this apparently straightforward condition might not be fulfilled in nature due to several effects that cause inefficient accretion, which we shall now discuss.

Because mass accretion (permanent accumulation of mass) is only possible for NS accretors in which the magnetosphere is located inside the co-rotation radius, $r_{\rm mag}\lesssim r_{\rm co}\equiv(GM/\Omega^2)^{1/3}$, efficient accretion for producing rapidly spinning NSs requires very weak B-fields or very high accretion rates. In other words, it is uneasy to keep accreting and to spin up a NS efficiently throughout a long-term RLO stage because the requirements on both $B$ and $\dot{M}$ are difficult to obtain due to stellar evolution constraints (see below). Furthermore, even if $r_{\rm mag}$ is pushed down onto the NS surface (to allow a small value of $P_{\rm eq}$), the resulting lever arm of the accretion torque would then be small, which challenges the timescale (Eq.~\ref{eq:timetorque}) available for recycling. 

Given Eddington-limited\footnote{Disregarding here the possibility for neutrino-cooled hypercritical accretion onto a NS in, for example, a common envelope environment --- a process originally proposed by \cite{che93}, but since then strongly challenged in a number of other papers \citep{che96,al00,rt12,mr15b}; see summary in \cite{tv23} --- and disregarding the possibility that ultra-luminous X-ray binaries are powered by highly super-Eddington accretion onto a NS rather than by geometric beaming \citep{kdw+01,kfm02}; however, see \citet{kl15}, who suggested that M82~X-2 may be a nascent MSP accreting at an ultra-high rate.} accretion onto a NS ($\dot{M}_{\rm Edd}\sim 10^{-8}\;M_\odot\,{\rm yr}^{-1}$), a steady source of RLO must be maintained for $\sim 0.1~{\rm Gyr}$, or more, depending on the accretion efficiency. Hence, NSs can only be fully recycled in LMXBs (and possibly some Case~A intermediate-mass X-ray binaries, IMXBs) where the nuclear timescale of the donor star is long enough to fulfil this minimum criterion. 
It is for this reason that binary MSPs with $P<5\;{\rm ms}$ and compact-object companions (WDs, NSs, or BHs\footnote{No MSP or other Galactic NS with a BH companion has yet been discovered.}) all have He or CO~WD companions, rather than ONeMg~WD, NS or BH companions.
Note, if the pre-RLO binaries are very tight and thus initiate mass transfer already on the main sequence, they often evolve into spiders (redbacks and black widows) or isolated MSPs \citep{ccth13,itl14,bdk+25}. Pulsars with massive ONeMg~WD or NS companions typically have spin periods $P\gtrsim 10\;{\rm ms}$ or $P\gtrsim 20\;{\rm ms}$, respectively, because of the much shorter RLO timescales in these systems --- typically from Case~BB RLO \citep{tlk12,tlp15}.

\begin{figure}[ht]
\centering
\vspace*{-0.7cm}\hspace*{-0.4cm}
\includegraphics[width=0.76\textwidth]{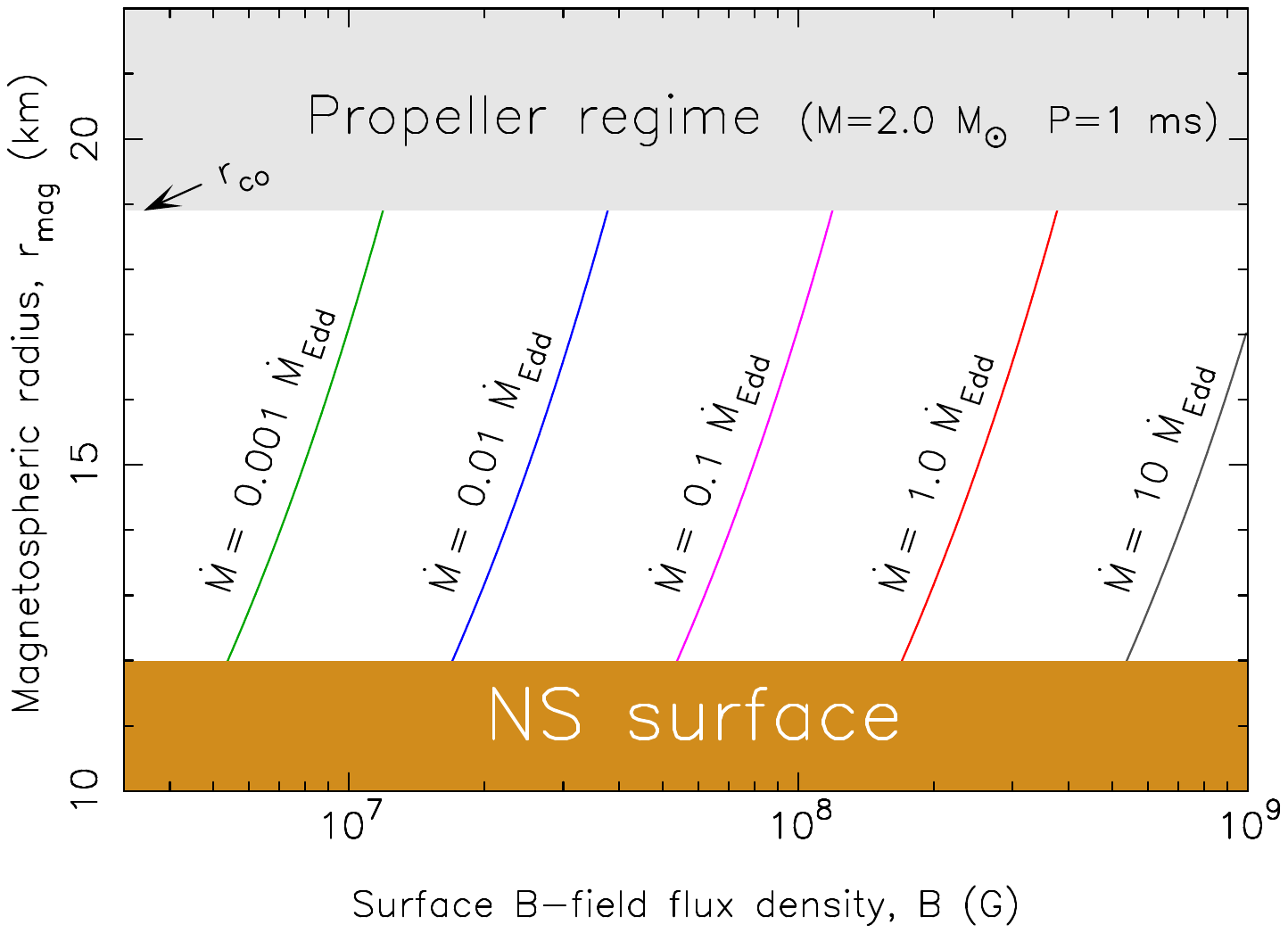} 
\caption{Accretion window for further spinning up a $2.0\;M_\odot$ NS rotating with $P = 1.0\;{\rm ms}$. The magnetospheric radius is shown as a function of surface B-field for accretion rates ranging from $10^{-3}$ to $10\;\dot{M}_{\rm Edd}$. Spin-up requires $r_{\rm mag} \lesssim r_{\rm co}$. When $r_{\rm mag} > r_{\rm co}$ (grey region), the NS enters the propeller regime and experiences a braking torque. The plotted accretion rates are illustrative and do not represent specific observed source classes. The formation of a sub-MSP is challenging: beyond the allowed pairs of ($B$,\,$\dot{M}$), it requires both a sufficiently massive donor to provide the necessary accreted mass and a long evolutionary timescale --- two conditions that are mutually contradictory (see text).}
\label{fig:mag-radius}
\end{figure}

\subsection{Empirical evidence for inefficient mass accretion}\label{subsec:efficiency}
The overall baryonic mass-accretion efficiency for an accreting NS, $\epsilon$, can take values between 0 and 1. Fully conservative mass transfer (no mass loss to the surroundings) yields $\epsilon=1$, whereas extreme mass transfer without any accretion onto the NS yields $\epsilon=0$. Possible mechanisms for mass ejection (and low $\epsilon$-values) discussed in the literature include propeller effects, accretion disc instabilities, and direct irradiation of the donor by an energetic pulsar \citep{is75,par96,dlhc99,dhl01,rbu+18}. An additional promising candidate is the so-called ``radio-ejection mechanism''.
As the mass-transfer rate continues to decrease toward the end of RLO and the radio MSP is activated, the presence of the remaining plasma wind may prevent further accretion from the now weak flow of material at low $\dot{M}$ \citep{krst88}. As discussed by \citet{bpd+01,bdb02,dbr+08}, this will be the case when the total spin-down pressure of the pulsar (from magnetodipole radiation and the plasma wind) exceeds the inward pressure of the material from the donor star, i.e. if: $\dot{E}_{\rm rot}/(4\pi r^2\,c) > P_{\rm disc}$.

The present mass of a MSP is related to its initial NS birth mass and the amount of mass transferred from the progenitor of its companion (typically a WD) via:
\begin{equation}\label{eq:Mns-final}
  M_{\rm NS} = M_{\rm NS,0} +(1-\eta)\,\epsilon \,\Delta M_2 \;,
\end{equation}
where $M_{\rm NS}$ and $M_{\rm NS,0}$ are the gravitational masses of the MSP and its original post-SN NS birth mass, respectively, $\Delta M_2 = M_2-M_{\rm WD}$ is the amount of mass transferred from the progenitor of the WD companion, $\epsilon$ is the (baryonic) accretion efficiency, and $\eta$ is the radiative efficiency, i.e. the fraction of the accreted baryonic rest-mass energy released as gravitational binding energy and escaping to infinity. 
In the following, we adopt $\eta=0.15$ as a representative gravitational binding-energy correction for the NS accretor. This choice is motivated by the approximate relation:
\begin{equation}
    1-\eta \simeq \sqrt{1-\frac{2GM}{Rc^2}}
    \simeq
    1-\frac{GM}{Rc^2}\;,
\end{equation}
where the second expression follows from a first-order Taylor expansion. Based on the compactness of a typical NS, this yields $\eta \simeq GM/(Rc^2) \approx 0.15$, whereas more compact NSs may have $\eta\simeq0.20$.

Table~\ref{table:NSmasses} lists fully recycled MSPs with well-determined NS masses, $M_{\rm NS}\lesssim 1.4\;M_\odot$. The list comprises six Galactic-field MSPs whose companion masses (within $1\sigma$) are consistent with the expected $(P_{\rm orb},\,M_{\rm WD})$--correlation following LMXB RLO \citep{rpj+95,ts99,imt+16}, and two GC MSPs that are clearly not the original recycled systems, given their very massive companions ($\sim 1.2\;M_\odot$) in highly eccentric orbits ($e=0.89$ and $e=0.75$, respectively). For each system, we can estimate the NS mass-accretion efficiency, $\epsilon$, from the prescription outlined below.

The minimum mass of the WD progenitor star is $M_2\simeq 1.0\,M_\odot$ because the available time for evolution is limited by the Hubble time (minus the WD cooling age, which can be several Gyr). The NS birth mass is at least $\sim 1.17\;M_\odot$ based on the smallest known NS mass \citep[J0453+1559,][]{msf+15}, close to the absolute lower limit in accordance with recent SN and NS formation theory \citep[e.g.][]{tj19,mhp25}. Combining the above values with Eq.~(\ref{eq:Mns-final}) we can calculate an upper limit for the accretion efficiency:
\begin{equation}\label{eq:epsilon-max}
  \epsilon_{\max}=\frac{(M_{\rm NS}/M_\odot)-1.17}{0.85\;\left(1.0-M_{\rm WD}/M_\odot\right)} \;.
\end{equation}
From Table~\ref{table:NSmasses} we see typical values $\epsilon_{\max}\simeq 0.20-0.30$, implying that \emph{at least} 70--80\% of all mass transferred was lost from their progenitor LMXB systems --- an interesting and somewhat surprisingly high fraction, indicating highly non-conservative mass transfer in LMXBs that produce MSPs. 
We emphasize that the $\epsilon_{\max}$ values inferred from the Galactic-field MSPs in Table~\ref{table:NSmasses} should be regarded as empirical upper limits based on this selected sample of low-mass recycled NSs with He~WD companions. Whether they are representative of the population-wide maximum accretion efficiency remains to be established.

\begin{landscape}
\begin{table}
    \caption{Galactic field MSPs and GC MSPs with well-determined NS masses, $M_{\rm NS} \lesssim 1.4\;M_\odot$. The columns show MSP name, spin period, orbital period, pulsar mass, companion mass, max and min accretion efficiency (the latter is based on the current spin period), minimum amount of mass accreted, and reference.}
    \label{table:NSmasses}
    \centering
%    \footnotesize
    \begin{tabular}{ccccccccc}
        \toprule
        Pulsar name & $P$ (ms) & $P_{\rm orb}$ (d) & $M_{\rm NS}$ ($M_\odot$) & $M_{\rm WD}$ ($M_\odot$) & $\epsilon_{\max}$ & $\epsilon_{\rm min}$ & $\Delta M_{\rm eq}$  ($M_\odot$) & Ref.\\
        \midrule
         J1543$-$5149  & 2.06 & 8.06  & 1.349(69)  & 0.2233(55) & 0.27 & 0.08 & 0.093 & \cite{cae+25}\\ 
         J1713+0747    & 4.57 & 67.8  & 1.35(7)    & 0.292(11)  & 0.30 & 0.03 & 0.032 & \cite{abb+18}\\
%         J1738+0333    & 5.85 & 0.35  & 1.47(7)    & 0.181(6)   & 0.43 & 0.02 & 0.024 & \cite{avk+12}\\ exclude or include 0437 and triple MSP
         B1855+09      & 5.36 & 12.3  & 1.37(12)   & 0.244(13)  & 0.31 & 0.02 & 0.026 & \cite{abb+18}\\
         J1918$-$0642  & 7.65 & 10.9  & 1.29(10)   & 0.231(10)  & 0.18 & 0.01 & 0.016 & \cite{abb+18}\\
         J2234+0611    & 3.58 & 32.0  & 1.353(17)  & 0.298(14)  & 0.31 & 0.04 & 0.044 & \cite{sfa+19}\\
         \midrule
         J0514$-$4002A & 4.99 & 18.8  & 1.39(3)    & 1.08(3)$^{\star}$    & N/A & N/A & 0.028 & \cite{dfg+25}\\
         J1807$-$2500B & 4.19 & 10.0  & 1.3655(21) & 1.2064(20)$^{\star}$ & N/A & N/A & 0.036 & \cite{lfrj12}\\
         \bottomrule
    \end{tabular}
        \smallskip
       {\small $^{\star}$ Note, the companions of the two GC MSPs (bottom) may be NSs rather than massive WDs.}
\end{table}
\end{landscape}

We can also place a strict lower limit on the accretion efficiency by assuming that (i) the NS accreted just sufficient mass ($\Delta M_{\rm eq}$, Eq.~\ref{eq:deltaM}) to reach equilibrium at its current spin period, $P=P_{\rm eq}$, and (ii) the WD progenitor star was relatively massive for a LMXB with stable RLO, say $M_2=1.6\;M_\odot$:
\begin{equation}\label{eq:epsilon-min}
  \epsilon_{\rm min}=\frac{\Delta M_{\rm eq}}{0.85\;\left(1.6-M_{\rm WD}/M_\odot\right)} \;.
\end{equation}
The inferred lower limits, $\epsilon_{\rm min}=0.01-0.08$ in Table~\ref{table:NSmasses}, are not very constraining. However, if, for example, PSR~J1918$-$0642 had been spinning at $1.40\,{\rm ms}$ (like PSR~J1748$-$2446ad), yielding $\Delta M_{\rm eq}=0.16\;M_\odot$, we would obtain $\epsilon _{\rm min}=0.14$ and thus constrain a realistic accretion efficiency $\epsilon = 0.14-0.18$.

A caveat should be mentioned here. We assume all MSPs in nature to be NSs rather than more exotic self-bound compact objects --- bound by the strong interaction rather than gravity --- such as fully deconfined strange stars \citep{wit84} or partially deconfined strangeon stars \citep{xu03}, or hybrid stars containing quark cores. Such hypothetical quark stars might form with masses as low as $10^{-3}-10^{-2}\;M_\odot$ \citep{wit84,fj84} and could therefore, in principle, permit post-accretion sub-MSPs with masses below $1.0\;M_\odot$ \citep{zl25}. A phase transition occurring during recycling is also a possibility.
However, so far there is no compelling evidence in nature in favour of exotic quark stars.

\subsection{Expected MSP spin periods and donor star masses}\label{subsec:expected-spins}
\subsubsection{Requirements for a sub-MSP}\label{subsubsec:requirements}
\begin{figure}[ht]
\centering
\vspace*{-0.6cm}\hspace*{-0.4cm}
\includegraphics[width=0.9\textwidth]{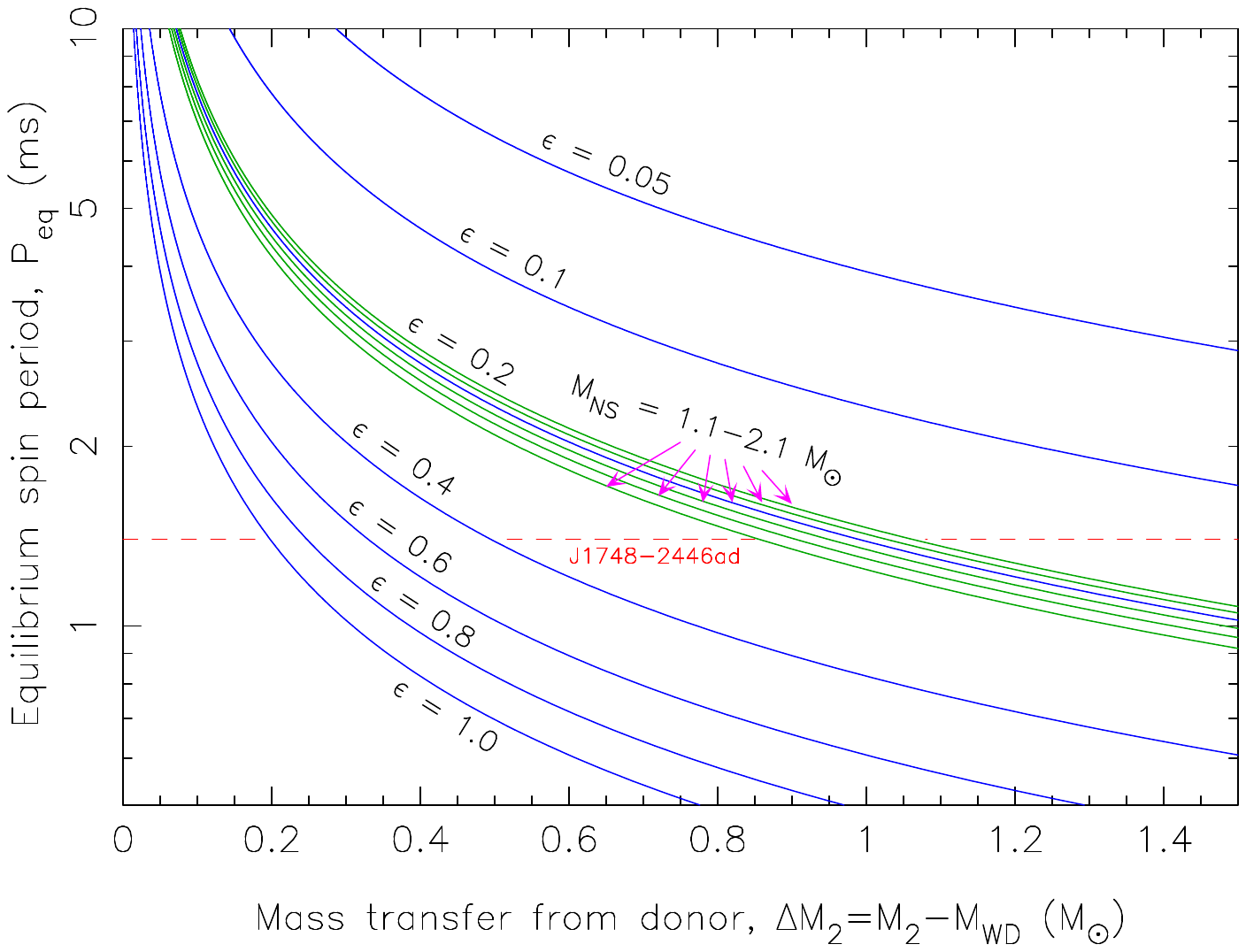} 
\caption{Equilibrium spin period as a function of the amount of mass transferred from the donor star in a LMXB system. The various lines are for different baryonic mass-accretion efficiencies, $\epsilon$, between 0.05 (top) and 1.0 (bottom), assuming $\eta=0.15$. The dependence on final NS mass is shown for $\epsilon = 0.2$, cf. green lines representing NS mass in steps of $0.2\;M_\odot$ between $1.1$ and $2.1\;M_\odot$ (for the blue lines $M_{\rm NS}=1.70\;M_\odot$ is assumed). The horizontal dashed red line marks the $1.4~{\rm ms}$ spin period of the fastest known MSP: J1748$-$2446ad.}
\label{fig:eta}
\end{figure}

Figure~\ref{fig:eta} shows (minimum) equilibrium spin period versus the amount of mass \emph{transferred} from a donor star under the assumption of different accretion efficiencies, $\epsilon=0.05-1.0$. The dependence on final NS mass is shown in an example case for $\epsilon = 0.2$, cf. green lines indicating NS mass in steps of $0.2\;M_\odot$ between 1.1 and $2.1\;M_\odot$ (the default value for the blue lines is $M_{\rm NS}=1.70\;M_\odot$). From the strong empirical constraints among the known population of MSPs (Table~\ref{table:NSmasses}) we found $\epsilon_{\max}\simeq 0.20-0.30$. These values are in good agreement with the fastest-spinning MSP known to date \citep[J1748$-$2446ad with $P=1.40\;{\rm ms}$,][]{hrs+06} for a typical LMXB transferring $\Delta M_2\simeq 0.8-1.1\;M_\odot$ (cf. Fig.~\ref{fig:eta} for $\epsilon\simeq 0.20-0.30$ and $P_{\rm eq}=1.40\;{\rm ms}$).

To reach $P_{\rm eq}= 1.0\;{\rm ms}$ a NS must {\em accrete} at least $\Delta M_{\rm eq}\simeq 0.25\;M_\odot$ according to Eq.~(\ref{eq:deltaM}), yielding $M_{\rm NS}\ge1.42\;M_\odot$ for a minimum NS birth mass of $M_{\rm NS,0}=1.17\;M_\odot$. For this reason, it is not surprising that none of the MSPs in Table~\ref{table:NSmasses} are sub-MSPs, since they all have $M_{\rm NS}\lesssim 1.4\;M_\odot$. (They were selected solely to constrain $\epsilon_{\max}$ based on their small post-recycling NS masses.)

If we consider a hypothetical MSP with $M_{\rm NS}=1.42\;M_\odot$ and assume, conservatively, a realistic accretion efficiency of $\epsilon\simeq 0.18$ (as inferred for PSR~J1918$-$0642), then Eq.~(\ref{eq:Mns-final}) yields $\Delta M_2=1.63\;M_\odot$, which means an original donor star mass of $M_2=\Delta M_2+M_{\rm WD}\simeq 1.8-2.0\;M_\odot$, given that the remnant He~WD has a typical mass of $\sim 0.20-0.30\;M_\odot$ \citep[an absolute minimum for a detached NS+WD system is $M_{\rm WD}\simeq 0.16\;M_\odot$,][]{tau18}, unless we are talking about black widow systems (Sect.~\ref{subsubsec:spiders}) which often have companion masses of just a few $0.01\;M_\odot$.

Such high donor-star masses of $M_2\gtrsim 2\;M_{\odot}$ approach the regime where the stability of the RLO becomes increasingly challenging, depending on the initial orbital period. Furthermore, the mass-transfer rate often becomes super-Eddington (Fig.~\ref{fig:IMXB-mdot}; see also \citealt{tvs00,prp02}), causing the accretion efficiency, $\epsilon$, to decrease further, leading to slower post-recycling spins (Fig.~\ref{fig:donor-mass-MESA} and Appendix~\ref{app:RLO-MESA}). As discussed in more detail below, producing a sub-MSP therefore requires an even larger donor star mass. 
More importantly, for such high donor star masses, the bound on $\epsilon_{\max}$ inferred from Eq.~(\ref{eq:epsilon-max}) is too optimistic (i.e. too large), because that expression assumes a conservative $M_2=1.0\;M_\odot$ to maximise $\epsilon$ --- for larger values of $M_2$ the upper limit tightens (i.e. $\epsilon_{\max}$ decreases). Similarly, the values of $\epsilon_{\rm min}$ from Eq.~(\ref{eq:epsilon-min}) can be even smaller if the donor star mass exceeds $M_2=1.6\;M_\odot$ adopted in that expression, thereby increasing the required minimum donor star mass.
The Eqs.~(\ref{eq:deltaM}) and (\ref{eq:Mns-final}) must instead be solved simultaneously:
\begin{equation}\label{eq:Mdonor}
  M_2=\frac{0.22\,M_{\odot}}{(1-\eta)\,\epsilon}\; \frac{(M_{\rm NS}/M_{\odot})^{1/3}}{P_{\rm ms}^{4/3}} + M_{\rm WD} \;.
\end{equation}

\begin{figure}[ht]
\centering
\vspace*{-0.4cm}\hspace*{-0.4cm}
\includegraphics[width=0.9\textwidth]{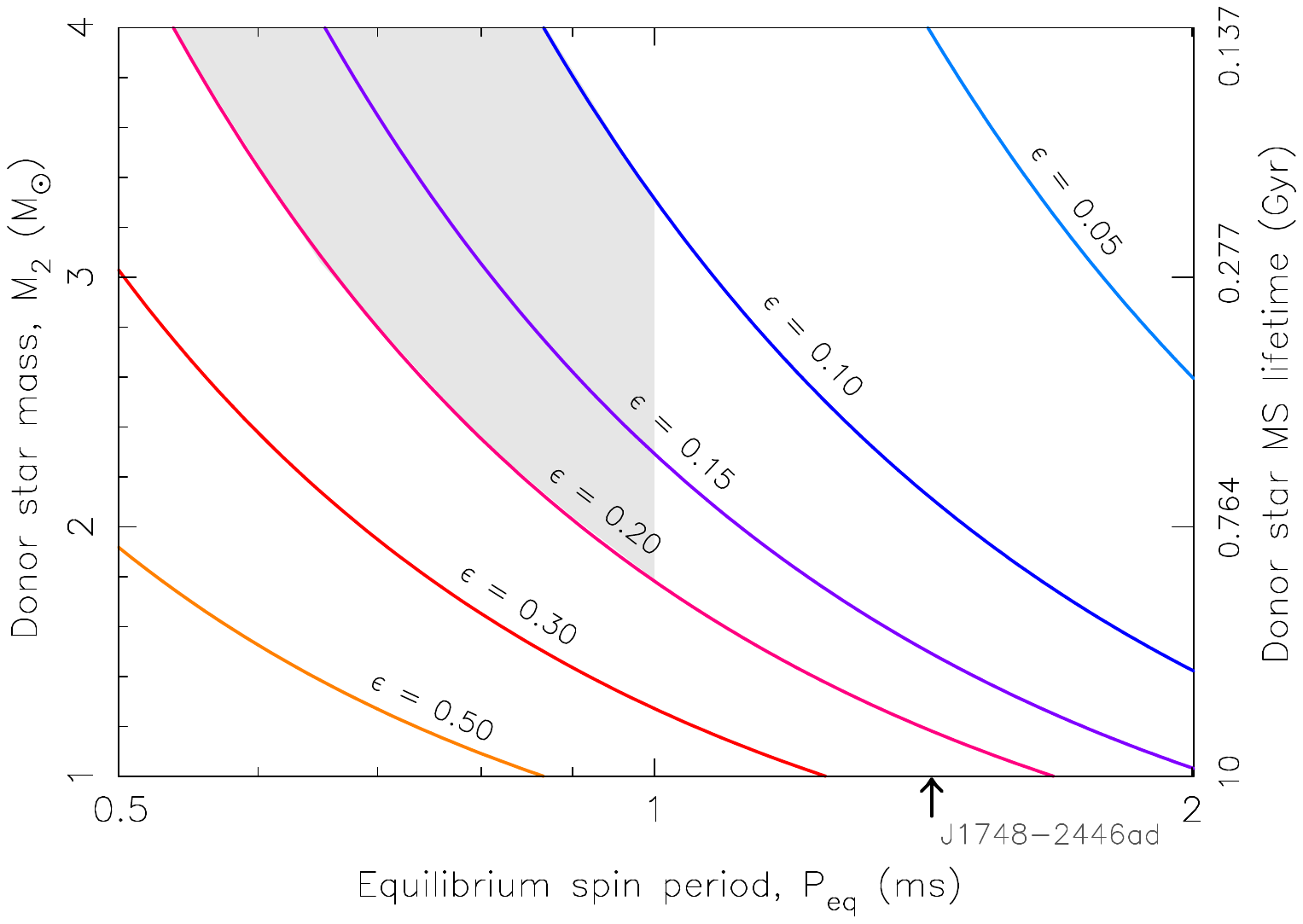} 
\caption{Donor star mass as a function of equilibrium spin period for given mass-accretion efficiencies, $\epsilon$, assuming $M_{\rm NS,0}=1.40\;M_\odot$, $\eta=0.15$ and a WD remnant mass of $M_{\rm WD}=0.25\;M_\odot$. The grey-shaded zone indicates necessary donor star masses ($\sim 2-4\;M_\odot$) to produce a sub-MSP for ``realistic'' mass-accretion efficiencies of $\epsilon=0.10-0.20$. The black arrow at the bottom marks the spin period of the fastest known MSP: J1748$-$2446ad.}
\label{fig:donor-mass}
\end{figure}

Figure~\ref{fig:donor-mass} displays the required initial donor-star mass, $M_2$, as a function of $P_{\rm eq}$ for different accretion efficiencies, $\epsilon$. The production of sub-MSPs for pragmatic values of $\epsilon \simeq 0.1-0.2$ requires initial donor stars of $M_2\simeq 2-4\;M_\odot$, i.e. progenitor systems that are IMXBs.
However, such massive donor stars evolve on relatively short timescales compared with $1.0-1.6\;M_\odot$ donor stars. For example, MESA calculations (Appendix~\ref{app:RLO-numerical}) yield main-sequence lifetimes of 764~Myr, 277~Myr and 137~Myr for $2.0\;M_\odot$, $3.0\;M_\odot$ and $4.0\;M_\odot$ stars (with a metallicity of $Z=0.0142$), respectively, approximately following:
\begin{equation}
\tau_{\rm nuc}\simeq 4.3\;{\rm Gyr}\;\left(\frac{M_2}{M_\odot}\right)^{-2.5}\;.
\end{equation}
These lifetimes are shorter than those of $1.0-1.6\;M_\odot$ donor stars by roughly one to two orders of magnitude.

\begin{figure}[ht]
\centering
\vspace*{-0.4cm}\hspace*{-0.4cm}
\includegraphics[width=0.9\textwidth]{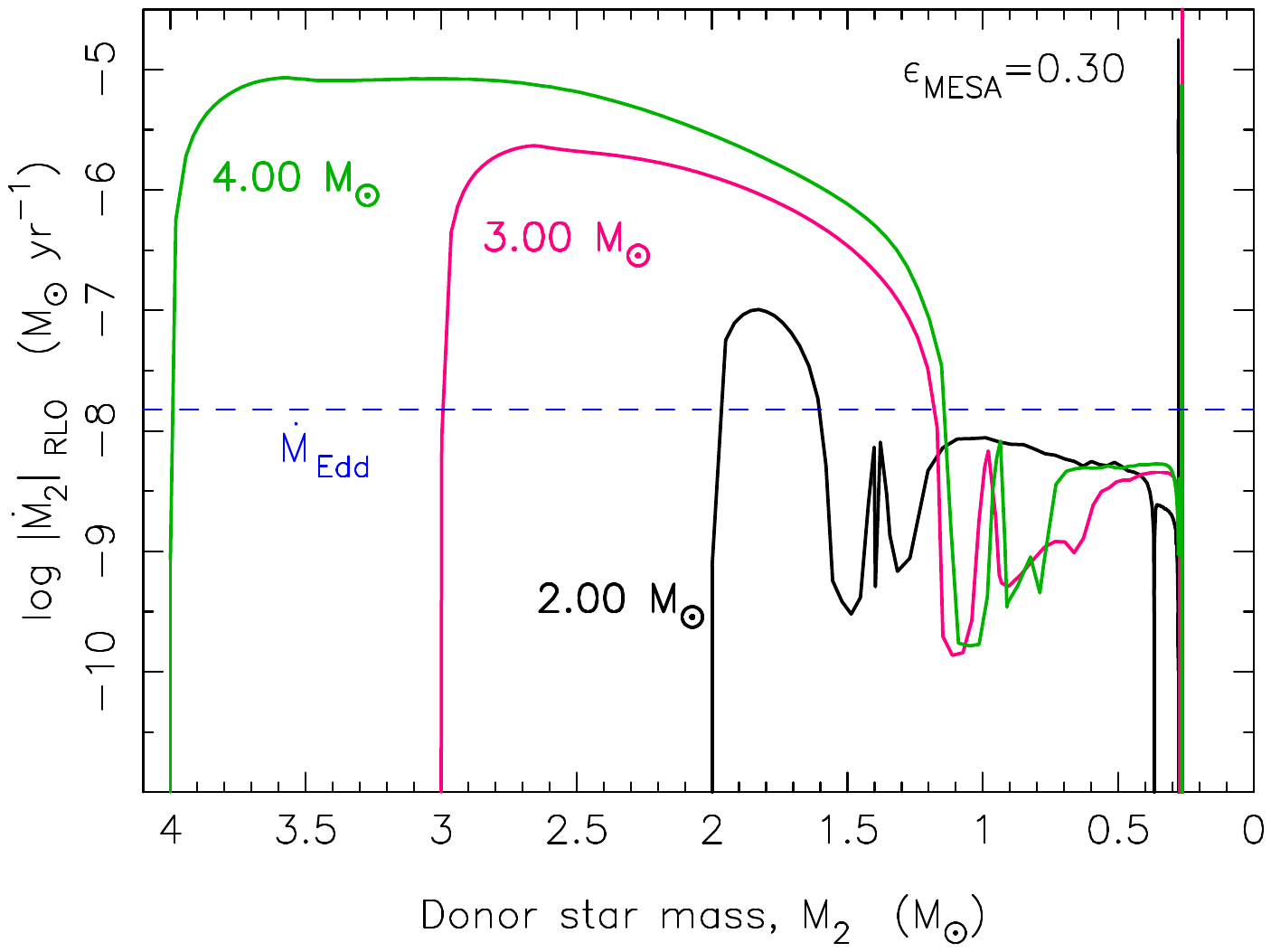} 
\caption{Mass-transfer rate as a function of decreasing donor-star mass for our sample MESA models with $2.0-4.0\;M_\odot$ main-sequence donor stars (Case~A RLO) transferring mass to a $1.4\;M_\odot$ NS ($1.7\;M_\odot$ for the $M_2=4.0\;M_\odot$ model). The initial orbital period is $P_{\rm orb}=1.0\;{\rm d}$ and the adopted MESA mass-accretion efficiency is $\epsilon_{\rm MESA}=0.30$. For donor-star masses $M_2\gtrsim2.0\;M_\odot$, the RLO is highly super-Eddington, causing the {\em effective} mass-accretion efficiency ($\epsilon_{\rm eff}$) to become significantly smaller than the adopted MESA-inlist value ($\epsilon_{\rm MESA}$). See Table~\ref{table:MESA-IMXB} and Appendix~\ref{app:RLO-MESA} for further details and discussion.}
\label{fig:IMXB-mdot}
\end{figure}

Furthermore, and more important here, depending on the initial orbital period of the X-ray progenitor system, the Kelvin-Helmholtz (thermal) timescale of a subgiant or Hertzsprung-gap (HG) donor star undergoing early Case~B RLO is only $\tau_{\rm KH}= GM^2/(2RL)\simeq 0.1-0.3\;{\rm Myr}$ (Appendix~\ref{app:RLO-timescale} and Table~\ref{table:KH}). 
For such stars with radiative envelopes, $\tau _{\rm KH}$ represents a lower limit for the duration of the mass-transfer phase ($\tau _X$) responsible for recycling of the NS. 

Detailed numerical calculations of the mass-transfer process in IMXBs with subgiant/HG donors \citep[e.g.][]{tvs00,prp02,mft+20} typically yield $\tau _X\simeq 1-10\;{\rm Myr}$ \citep[this deviation from $\tau _{\rm KH}$ is mainly explained by the luminosity of the donor star dropping 1--2 orders of magnitude after onset of RLO,][]{tvs00,tlk11}. 
If the NS accretion rate is Eddington limited, the amount of mass accreted by the NS is at most $\Delta M_{\rm NS}\simeq \dot{M}_{\rm Edd}\;\tau _X\,(1-\eta)\sim 0.20\;M_\odot$, which is insufficient to spin it up to a sub-MSP.
For LMXB and IMXB orbits that are wider at the onset of RLO --- eventually producing MSPs with final $P_{\rm orb}\gg10\;{\rm d}$ --- one can even think of $\tau_{\rm KH}$ as an upper limit for a (sub)thermal-timescale RLO episode.

A note of concern about using the Kelvin-Helmholtz timescale as a proxy for RLO duration is that if RLO starts on the main sequence (Case~A RLO), it could have a significantly longer duration than the thermal timescale. However, for mass ratios $q\equiv M_2/M_{\rm NS}>1.28$ the orbit always \emph{shrinks} upon mass transfer \citep{tv23}, and such systems may be dynamically unstable and coalesce in the process. Even if they survive, however, the RLO is initiated in an intense phase of mass transfer with $|\dot{M}_2|\gg \dot{M}_{\rm Edd}$, leading to very limited NS accretion anyway.

This situation is illustrated in Fig.~\ref{fig:IMXB-mdot}, which shows that even for Case~A RLO from $2-4\;M_\odot$ donor stars, a large fraction of the transferred mass is lost at highly super-Eddington transfer rates. In the specific MESA models shown here, the baryonic mass-accretion efficiency was set to $\epsilon_{\rm MESA}=0.30$. 
However, because the NS accretion rate was capped at the Eddington limit, the {\em effective} accretion efficiency, $\epsilon_{\rm eff}$, was only $0.08-0.28$ (see Appendix~\ref{app:RLO-MESA} and Table~\ref{table:MESA-IMXB} for further details).
Furthermore, the smaller the value of $\epsilon$, the more the IMXB orbit shrinks during RLO \citep{tv23}, thereby increasing the likelihood of common-envelope evolution, which drastically reduces the amount of mass that can be accreted by the NS. Hence, with such low effective accretion efficiencies, the formation of sub-MSPs is expected to be exceedingly difficult, consistent with their non-detection to date.

For $1.0-1.3\;M_\odot$ subgiants, on the other hand, one would expect $\tau _X \simeq 0.2-3\;{\rm Gyr}$ and (often or partly) sub-Eddington mass-transfer rates. However, as concluded above, such donor stars cannot supply sufficient mass to spin~up a NS to a sub-MSP, given the inefficient accretion inferred earlier (Sect.~\ref{subsec:efficiency}). This conclusion may change if the low accretion efficiencies inferred from the systems listed in Table~\ref{table:NSmasses} are not representative of the general population of NS binaries.

Although the uncertainties discussed in Sects.~\ref{subsec:bands} and \ref{subsec:deltaM} --- for example, those associated with the parameters entering $f(\alpha,\xi,\phi,\omega_c)$ --- modify the quantitative values of $P_{\rm eq}$, $\Delta M_{\rm eq}$, and the inferred donor-star masses by factors of order unity, they do not alter the qualitative conclusion that producing sub-MSPs requires a combination of unusually efficient mass accretion and sufficiently long-lived mass-transfer episodes.

\subsubsection{Further comparison with observational data}\label{subsubsec:mass-spin-obs}
\begin{figure}[ht]
\centering
\vspace*{-1.2cm}\hspace*{-0.4cm}
\includegraphics[width=0.92\textwidth]{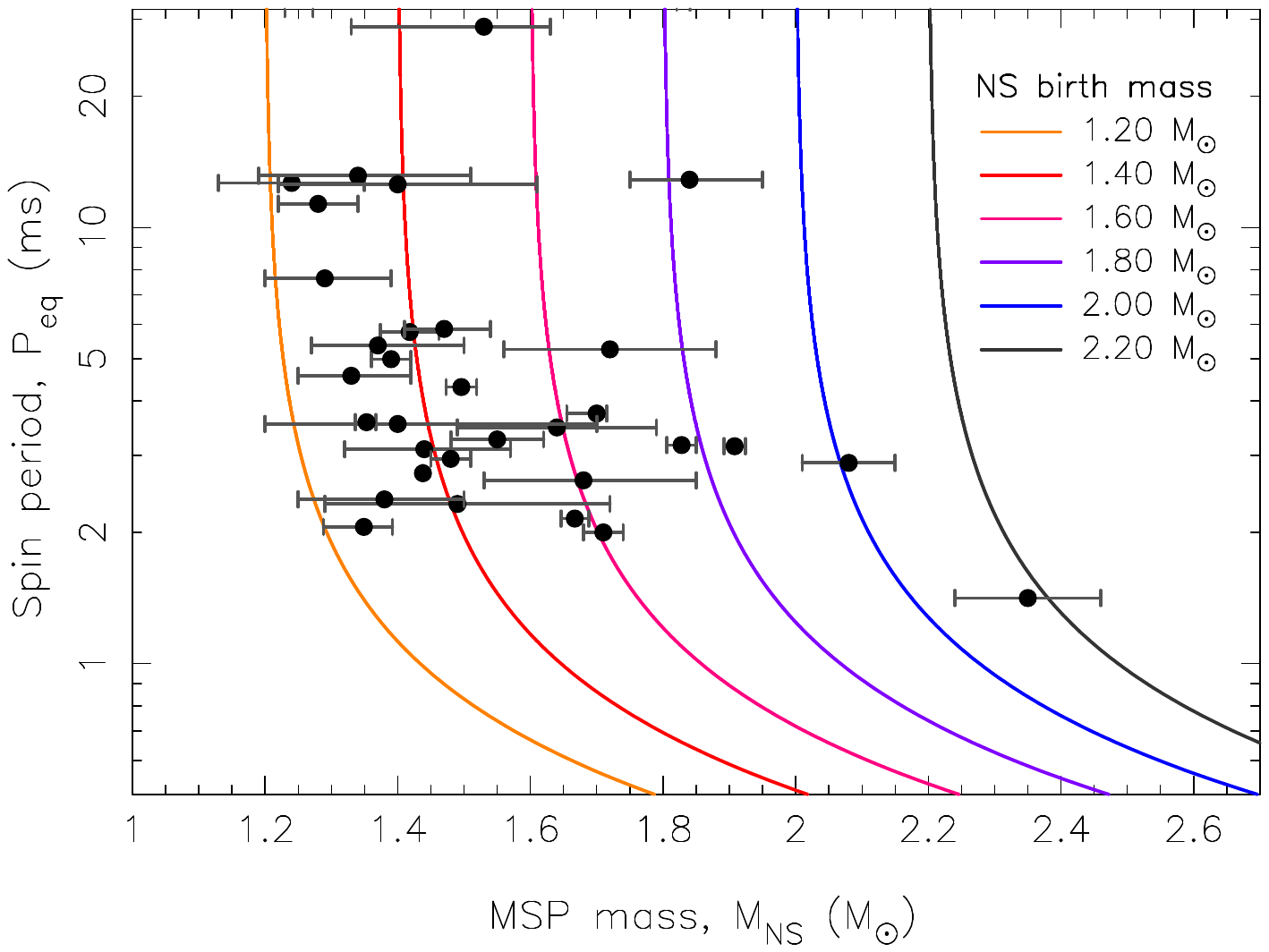} 
\caption{Recycled spin period as a function of final MSP mass. The coloured lines are for different NS birth masses between $1.20-2.20\;M_\odot$ and where the recycled (final) MSP masses are calculated from Eq.~(\ref{eq:deltaM}). Note, these lines are \emph{independent} of the accretion efficiency and show the (minimum) recycled MSP mass for a given spin period and NS birth mass. Observational NS mass data are provided by Paulo Freire's database: \url{https://www3.mpifr-bonn.mpg.de/staff/pfreire/NS_masses.html}}
\label{fig:mass-spin-obs}
\end{figure}

Figure~\ref{fig:mass-spin-obs} compares updated observational data with theoretical expectations from the recycling theory outlined in this section. 
These NS masses are almost exclusively from observations of MSPs with He~WD companions and based on constraints derived from GR and radio timing data. The uncertainties shown on the plot are $1\sigma$ error bars.
For NSs born with masses, $M_{\rm NS,0}/M_\odot\in [1.20,\,2.20]$ we expect all observed MSPs to fall between these plotted lines. The final (recycled) MSPs masses were calculated from Eq.~(\ref{eq:deltaM}). 
Note, these lines are independent of the accretion efficiency, $\epsilon$. Based on these calculations, we advocate that the record-high mass MSP~J0952$-$0607 \citep[$M_{\rm NS}=2.35\pm 0.11\;M_\odot$,][]{rbf+25} was born with a mass $2.00 \lesssim M_{\rm NS,0}/M_\odot \lesssim 2.20$.  
We do not expect the existence of sub-MSPs with masses below $1.40\;M_\odot$ given that NS are most likely all born with a mass $M_{\rm NS,0} \gtrsim 1.17\;M_\odot$. If such a low-mass sub-MSP is discovered it would provide evidence for a strangeon star (Sect.~\ref{subsec:efficiency}). At the other end, NSs born massive ($M_{\rm NS,0}> 2.0\;M_\odot$) may collapse to form a black hole (BH) before reaching sub-ms spin periods \citep{mly25}.

\subsubsection{Sub-MSPs from spiders?}\label{subsubsec:spiders}
\begin{figure}[ht]
\centering
\vspace*{-1.2cm}\hspace*{-0.4cm}
\includegraphics[width=0.75\textwidth]{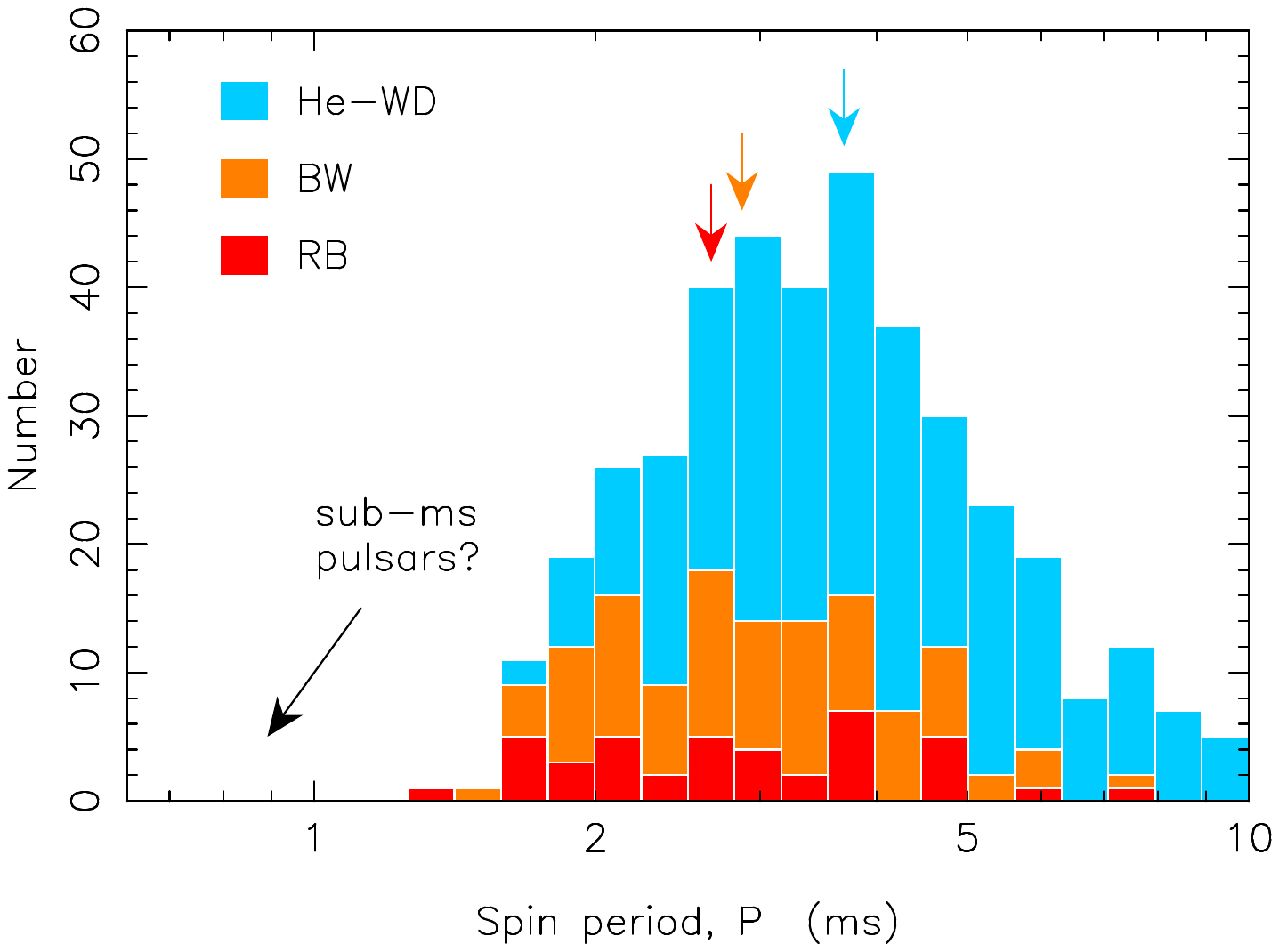}
\caption{Spin distributions of MSPs with He~WD companions (blue color) compared to black widows (BWs, orange color) and redbacks (RBs, red color). Median values (arrows) are shown for systems with $P<10\;{\rm ms}$. 
Data taken from the ATNF Pulsar Catalogue, version~2.7.0 accessed January~2026 \citep[][{\url{https://www.atnf.csiro.au/research/pulsar/psrcat}}]{mhth05}, augmented with newly discovered MSPs and updated BinComp classifications (see Appendix~\ref{app:companion-nature} and Table~\ref{table:BinComp}).}
\label{fig:spiders}
\end{figure}

MSPs in tight binaries whose radio signals are eclipsed for a fraction of the orbit spin faster on average than normal non-eclipsing MSPs \citep{ptrt14}. These eclipsing MSPs (``spiders'') are subdivided into \emph{black widows} and \emph{redbacks}, according to companion mass \citep{rob13,kl25}.
The spin periods of spiders are shown in Fig.~\ref{fig:spiders} for comparison with MSPs having He~WD companions. The median periods are shorter for redbacks (2.66~ms) and black widows (2.87~ms) than for MSPs with He~WDs (3.69~ms).  
Black widows are characterised by very small companion masses (typically a few $0.01\;M_\odot$), so their rapid spins could reflect that the NS accreted more mass than MSPs with a $\sim\!0.25\,M_\odot$ He~WD remaining as their companion. Hence, their overall accretion efficiencies may well be consistent with the range derived above, $\epsilon\simeq 0.10-0.30$.
Redbacks, however, have larger companion masses ($\sim\!0.2-0.4\,M_\odot$). Although the formation of both black widows and redbacks can be reproduced by standard LMXB evolution in tight orbits with varying irradiation and evaporation efficiencies and geometries \citep{ccth13}, in some cases \citep[e.g. PSR~J1417$-$4402;][]{dhb19} the progenitors have been suggested to be more massive than the typical $1.0-1.3\;M_\odot$. The accretion efficiency of redbacks therefore remains uncertain.
For further theoretical and multiwavelength observational studies of MSPs in spider systems, see e.g. \citet{bvr+13,bdh14,bkb+16,lsc18,ssc+19,gq20,gq21,ckk+23}. A recently proposed additional subclass, the {\em huntsman} spiders, is discussed by \citet{sru+25}.

Finally, in spider systems the ``radio-ejection mechanism'' \citep{krst88,bpd+01,bdb02,dbr+08} has been proposed as a possible means of suppressing accretion, particularly in connection with the {\em transitional} MSPs \citep[e.g.][]{a09,pfb+13,pddm22}.

\subsection{MSP spin-down since formation}\label{subsec:spindown}
The observed radio MSPs lose rotational energy and spin down with time. Hence, their current spin period, $P>P_{\rm eq}$, may have increased substantially. Could this explain the non-detection of sub-MSPs? To address this question, we calculate the spin evolution of putative sub-MSPs born with various $B$-fields and assuming different braking indices. 

The spin-down of pulsars is characterised by the braking index, $n\equiv \Omega\ddot{\Omega}/\dot{\Omega}^2$ (where $\Omega = 2\pi/P$), which relates the braking torque ($N=\dot{J}=I\,\dot{\Omega}$) to the angular velocity via $\dot{\Omega}\propto -\Omega ^{n}$. By integrating this spin-down equation, one obtains evolutionary tracks and isochrones in the $P\dot{P}$--diagram \citep{tk01,tlk12,ltk+14}. For a pulsar evolving with a constant value of $n$, the kinematic solution at time $t$ (positive in the future, negative in the past) is:
\begin{equation}
      P = P_0 \left[ 1 + (n-1) \frac{{\dot P}_0}{P_0}\,t \right]^{1/(n-1)} \;.
  \label{eq:Pevol}
\end{equation}

\begin{figure}[ht]
\centering
\vspace*{-1.2cm}\hspace*{-0.4cm}
\includegraphics[width=0.82\textwidth]{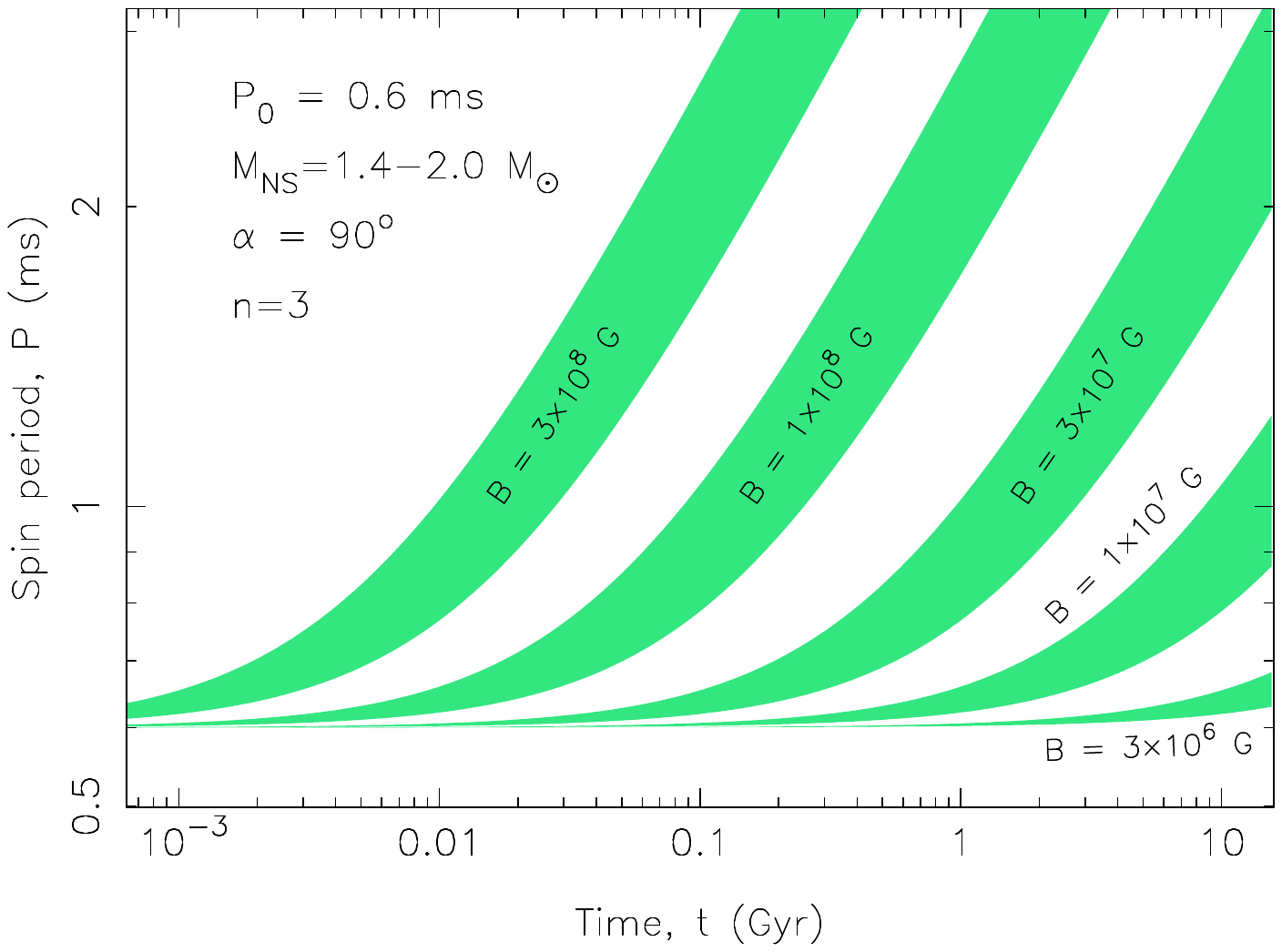} 
\caption{Spin evolution of a 0.6~ms pulsar with a constant B-field (assuming $M_{\rm NS}=1.4-2.0\;M_\odot$ (left and right boundary of each track) and magnetic inclination angle $\alpha = 90^{\circ}$), corresponding to a constant braking index $n=3$, cf. Eq.~(\ref{eq:Pevol}). Only pulsars with $B\lesssim 10^7\;{\rm G}$ will remain sub-MSPs on a Gyr timescale. Normal matter sub-MSPs spun-up to 0.6~ms are expected to have $M_{\rm NS}\gtrsim 1.7\;M_\odot$.}
\label{fig:Pevol-subMSP}
\end{figure}

Figure~\ref{fig:Pevol-subMSP} shows the spin evolution of a $0.6$~ms pulsar assuming a constant surface B-field between $3\times 10^6$ and $3\times 10^8\;{\rm G}$, a fixed magnetic inclination angle $\alpha = 90^\circ$, and a pulsar mass in the interval of $M_{\rm NS}=1.4-2.0\;M_\odot$. Note, sub-MSPs with $M_{\rm NS}=1.4\;M_\odot$ and $P=0.6\;{\rm ms}$ may not be stable according to most NS EoS (Sect.~\ref{subsec:EoS}). Furthermore, $P=0.6\;{\rm ms}$ requires accumulation of $>0.50\;M_\odot$ (Eq.~\ref{eq:deltaM}), yielding $M_{\rm NS}\gtrsim 1.70\;M_\odot$. Values of $M_{\rm NS}=1.4\;M_\odot$ are nevertheless included here for clarity of pulsar mass dependence. It is evident that \emph{only} if the sub-MSP is recycled with a very weak field, $B\lesssim 1\times 10^{7}\;{\rm G}$, will it remain a sub-MSP on a Gyr timescale. 

\begin{figure}[ht]
\centering
\vspace*{-1.2cm}\hspace*{-0.4cm}
\includegraphics[width=0.82\textwidth]{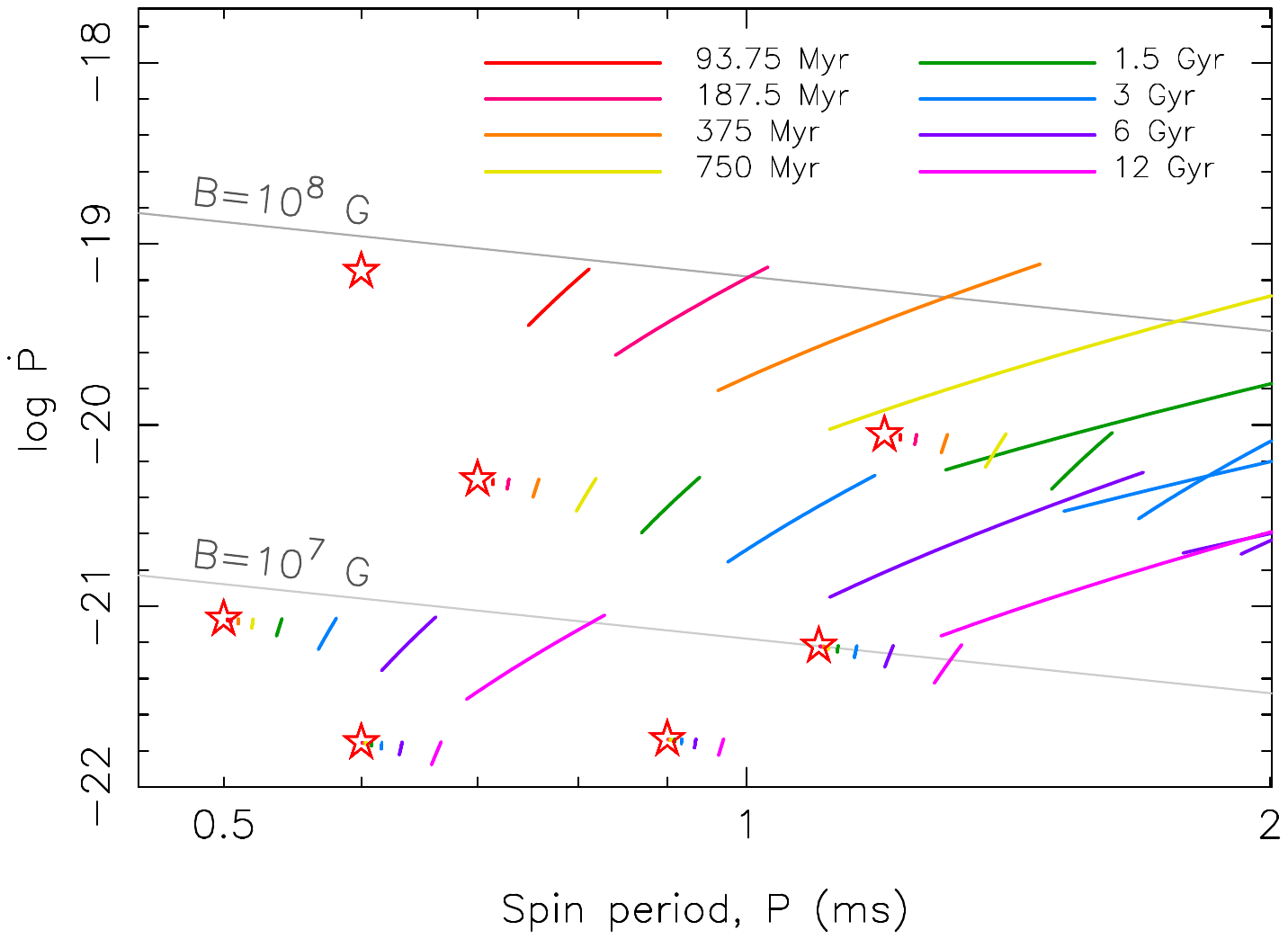} 
\caption{Isochrones of seven hypothetical recycled MSPs born at the locations of the red stars. The isochrones were calculated for a $1.7\;M_\odot$ NS with braking indices $2 \le n \le 5$, following the recipe in \cite{tlk12}. Thin grey lines indicate $B=10^7\;{\rm G}$ and $B=10^8\;{\rm G}$.}
\label{fig:isochrones}
\end{figure}

In case the spin evolution is affected by other torques, in addition to magnetodipole radiation, such as GWs ($n=5$ for quadrupole radiation) or related to differential rotation (see Sect.~\ref{subsec:diff-rot}) we made a broader test of the spin evolution.
Figure~\ref{fig:isochrones} displays evolutionary tracks in the $P\dot{P}$--diagram for a range of sub-MSPs with various B-fields, following the methodology outlined in Section~8 of \cite{tlk12}. The coloured isochrones cover evolution with constant braking index $2\le n\le 5$ (top and bottom of each isochrone is for $n=2$ and $n=5$, respectively).

We conclude that the spin-down timescale of sub-MSPs strongly depends on their B-field. In general, in case the spin-down is dipole dominated, sub-MSPs spin down to become ``normal'' MSPs on a timescale of a few 100~Myr ($B\sim 10^8\;{\rm G}$) to several Gyr ($B\lesssim 10^7\;{\rm G}$). A key question is whether the residual B-fields after recycling can reach sufficiently low values, $B=10^6 - 10^7\;{\rm G}$, to allow sub-MSP spin periods to persist over a Hubble timescale. 

We emphasise that determining small B fields of pulsars accurately is challenging in practice because the observed $\dot{P}_{\rm obs}$ values (from which $B$ is inferred) are severely affected by dynamical acceleration effects (Appendix~\ref{app:PdotGC}) such as the Shklovskii effect \citep{shk70}. For example, consider a pulsar with a spin period of $P=1\;{\rm ms}$ at a distance of $d=1000\;{\rm pc}$ and having a transverse velocity of $v_\perp=100\;{\rm km\,s}^{-1}$. 
This pulsar will experience a contribution to its observed $\dot{P}_{\rm obs}$ of $\dot{P}_{\rm shk}=P\,v_{\perp}^2/(c\,d)\simeq 1.1\times 10^{-21}$ --- quite a significant contribution (see e.g. Fig.~\ref{fig:spinupline}), which makes the empirical determination of B-fields below $\sim 10^7\;{\rm G}$ difficult, let alone the additional uncertainties from kinematic corrections caused by Galactic vertical acceleration and differential rotation \citep{lk12}.

\subsubsection{Critical ellipticity}
The critical ellipticity for which GW quadrupole radiation (Appendix~\ref{app:GWs}) dominates over electromagnetic (EM) radiation is found by equating the corresponding spin-down torques, $N_{\rm GW}=N_{\rm EM}$. Throughout this subsection we adopt the force-free magnetospheric torque of \citet{spi06}, for which the angular dependence is proportional to $1+\sin^2\alpha$. This implies:
\begin{equation}
  \frac{32}{5c^5}\,GI^2\varepsilon^2\Omega^5
  =
  \frac{\mu^2}{c^3}\,\Omega^3\left(1+\sin^2\alpha\right) \;,
\end{equation}
where $\mu=BR^3$ is the magnetic moment. Solving for the critical ellipticity yields: 
\begin{equation}
    \varepsilon_{\rm crit}\simeq 1.3\times 10^{-10}\;\sqrt{1+\sin^2 \alpha}\;\left( \frac{R_{12}^3}{I_{45}}\right)\; B_7\,P_{\rm ms} \;,
    \label{eq:epsilon_crit}
\end{equation}
where $R_{\rm NS}$ is the NS radius in units of 12~km and $B_7$ is the magnetic flux density at the NS surface in units of $10^7\,{\rm G}$. 
Whereas the current spin-down of radio MSPs can be explained within the EM dipole model, there has been some suggestive indications of a minimum MSP ellipticity of $\varepsilon\sim 10^{-9}$ in the current MSP population \citep{wph+18} which, according to Eq.~(\ref{eq:epsilon_crit}), means that the spin-down should be dominated by GW emission rather than EM spin~down. This is still an open question --- and probably will be until continuous GWs are detected directly in fast-spinning NSs. 
Such a quadrupolar deformation (``mountain'') would still be too weak to be detected in current searches (see \citealt{hb23} for a review).

%%%%%%%%%%%%%%%%%%%%%%%%%%%%%%%%%%%%%%%%%%%%%%%%%%%%%%%%%%%%%%%%%%%%%%%%%%%%%%%%%%%%%%%%%%%%%%%%%%%%%%%%%%
%%\clearpage
\section{Discussion}\label{sec:discussions}
So, do sub-MSPs exist? In this work, we have shown why sub-MSPs are difficult to form in nature. The key limitation is the strong empirical evidence for inefficient accretion, inferred from mass measurements of post-recycled MSPs. This implies that the donor stars responsible for delivering mass to the accreting NS must initially be relatively massive, of order $2-4\;M_\odot$. Such donors evolve and transfer mass on short timescales, compared with typical $1.0-1.3\;M_\odot$ donors in LMXBs, leading to super-Eddington mass-transfer rates. This further suppresses accretion efficiency, increases the likelihood of dynamical instability, and leaves insufficient time for accretion torques to spin the NS up to sub-ms periods.

Even if sub-MSPs were able to form in nature, we have shown that they would spin down to become ordinary MSPs ($P>1\;{\rm ms}$) on a timescale of a few 100~Myr, unless their magnetic fields are very weak ($B\lesssim 3\times 10^7\;{\rm G}$). It remains uncertain whether such low residual magnetic fields are attainable, and whether such NSs would remain active as radio pulsars.

We have shown that sub-MSP masses would need to exceed roughly $1.42\;M_\odot$ and $1.70\;M_\odot$ for spin periods of 1.0~ms and 0.6~ms, respectively. However, these masses may still be too small to avoid tension with modern NS EoS constraints (Sect.~\ref{subsec:EoS}).

\subsection{Radio luminosity vs. spin}
\begin{figure}[ht]
\centering
\vspace*{-0.8cm}\hspace*{-0.4cm}
\includegraphics[width=0.75\textwidth]{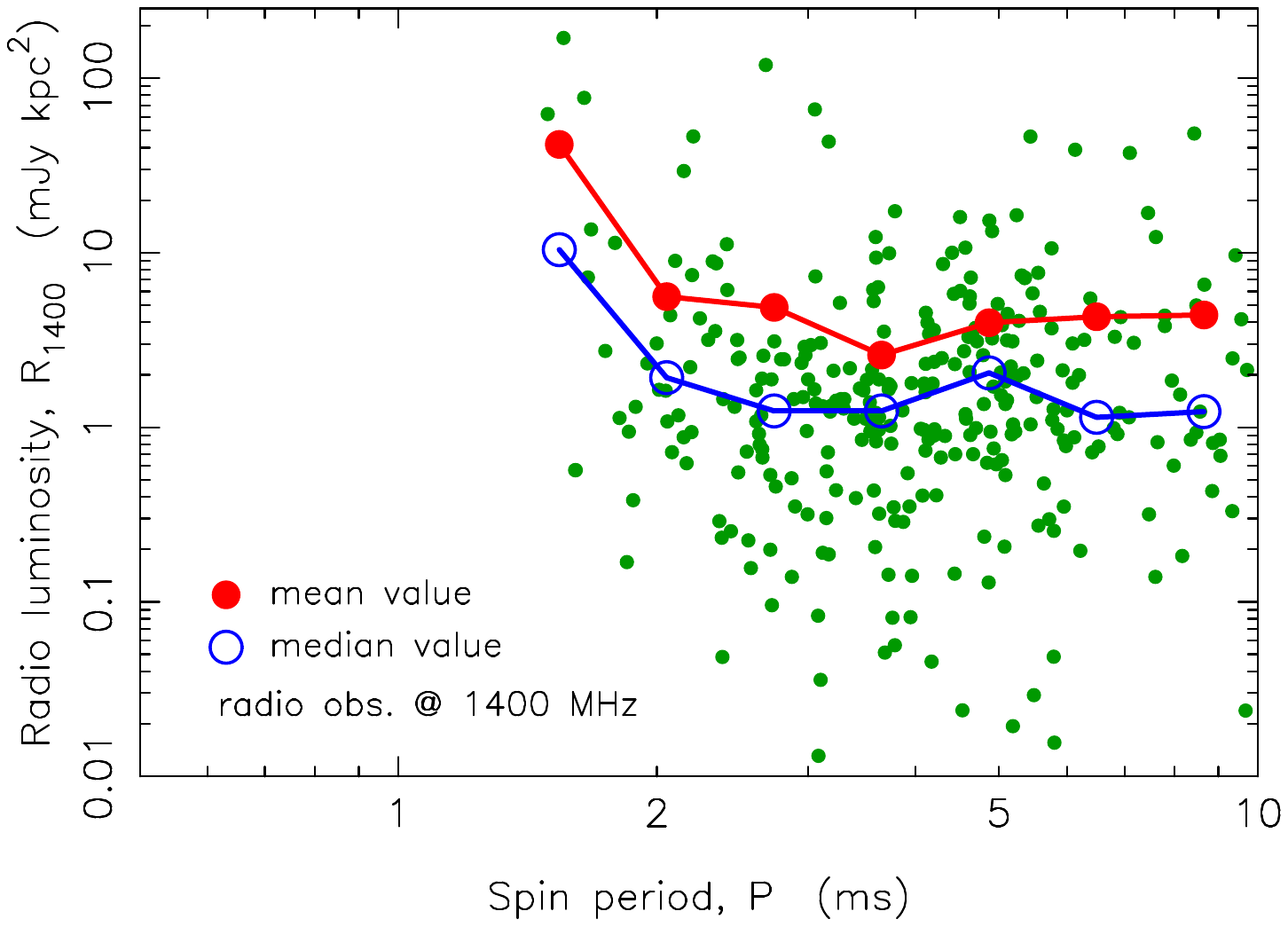} 
\caption{Observed radio luminosity at 1400~MHz ($R_{1400}=S_{1400}\,d^2$) as a function of pulsar spin period ($P$) for 306 MSPs with $P<10\;{\rm ms}$. Mean and median values are shown in bins of width 0.125 in $\log_{10}P$ (solid red points and open blue circles, respectively). There is no evident radio-luminosity selection effect against the fastest-spinning MSPs. Uncertainties in flux-density calibration and distance estimates contribute to the scatter in $R_{1400}$; see text for details. Source: \textit{ATNF Pulsar Catalogue}, July~2025 \citep[][\url{https://www.atnf.csiro.au/research/pulsar/psrcat}]{mhth05}.}
\label{fig:R1400}
\end{figure}

An obvious question is whether there are selection effects against detecting sub-MSPs in terms of their radio flux density. One may ask: why is there not a single sub-MSP candidate with a S/N of 50 in any searches, when there are normal MSPs with S/N of 50? Whether this reflects a decrease in radio emission at very fast spins is important to test.
Figure~\ref{fig:R1400} shows the relevant data, and we notice that the mean radio (pseudo) luminosity at 1400~MHz (i.e. the 1400~MHz flux density times distance squared, $R_{1400}=S_{1400}\,d^2$) does not decrease for the fastest-spinning pulsars approaching $P\sim 1\;{\rm ms}$. We note, however, that this conclusion is subject to uncertainties in the underlying flux-density and distance estimates. In particular, some pulsars rely on nominal flux densities inferred from the radiometer equation rather than absolute flux calibration, and several distances are not based on parallax or VLBI measurements. While these effects increase the scatter in $R_{1400}$, there is currently no evidence for a systematic suppression of radio luminosity at the shortest spin periods. A more stringent test will require restricting the sample to pulsars with well-calibrated flux densities and independently measured distances, including GC sources.

\subsection{Is the fastest-spinning NS in an X-ray binary?}
As discussed in Sect.~\ref{subsubsec:radio_vs_X} (see also Fig.~\ref{fig:spin-histo-all}), there is a systematic difference between the spins of XMPSs and radio MSPs. This difference can be explained, at least in part, by the RLO decoupling phase (Sect.~\ref{subsec:deltaM}), which slows down the newly recycled radio MSP by approximately a factor $\sqrt{2}$. We might therefore expect the fastest-spinning NS to reside in an X-ray binary. 
On the other hand, far more radio MSPs are observable in our Galaxy than accreting NSs, so current discovery statistics favour radio searches. Whereas $\sim 700$ radio MSPs (a rapidly increasing number) have measured $P<10\;{\rm ms}$ \citep{mhth05}, the same is true for only 51 XMSPs \citep{fkg+24}.

\subsubsection{Sco~X-1}\label{subsubsec:scoX1}
Sco~X-1, the first-discovered and brightest persistent X-ray point source in the sky, is a promising source for a fast-spinning NS. It was discovered already in 1962 by three Geiger counters mounted on an Aerobee rocket \citep{ggpr62}. Despite numerous efforts (summarised below), its spin period remains unknown to date. 

Sco~X-1 is a LMXB with an orbital period of 18.9~hr and a slightly evolved low-mass companion star \citep{gwl75}. The X-ray radiation from Sco~X-1 is not only strong but also highly variable. Its variability exhibits two principal states --- corresponding to different mass-accretion regimes onto the NS and defining its motion along the Z-track in colour–colour space --- the so-called normal branch (softer, sub-Eddington) and the flaring branch (super-Eddington). Light-curves and hardness-intensity diagrams are used to track the evolution of the spectral properties of Sco~X-1, while the detection of quasi-periodic oscillations (QPOs) is used to study its temporal evolution \citep{lmm+25}, including two (drifting) kHz~QPO peaks discovered by \cite{vwhc97} which cannot be explained by the beat-frequency model. 

Recent optical searches by \cite{lpi+26} did not reveal any significant coherent pulsation up to a frequency of 1550~Hz (corresponding to a NS spin period of $P=0.65\;{\rm ms}$). This non-detection translates into an upper limit on the background-subtracted pulse amplitude of $9.2\times 10^{-5}$, assuming a 1\% false-alarm probability.

Similarly, searches for GWs from Sco~X-1 using LIGO~O3 data also resulted in non-detections, yielding a GW amplitude constraint of $h_0 < 10^{-25}$ \citep{aaa+22}. 
Subsequent corrected-ephemeris analyses of the same O3 data, using both cross-correlation and hidden-Markov-model methods, confirmed these null results and yielded comparable GW amplitude limits of $h_0<10^{-25}$ \citep{wtw+23,vm25}. 
More recently, an analysis of LIGO~O4a data also yielded a non-detection, but improved the sensitivity substantially, reaching sub-torque-balance upper limits over part of the searched frequency range \citep{abac+26}. 
At the LIGO~O3 sensitivity level, a magnetic-mountain ellipticity of $\varepsilon \gtrsim 10^{-6}$ would be required for detection. At future Einstein Telescope/Cosmic Explorer sensitivities, less deformed NSs will be detectable with ellipticities as small as $\varepsilon \sim 6\times10^{-9}$ \citep{pamm25}.

\subsection{Multiple epochs of accretion for GC MSPs}\label{subsec:multi-acc}
GCs are well-known environments for the dynamical formation of MSPs and other compact binaries \citep[see][for a review]{bd13}. 
An interesting possibility in the context of sub-MSP formation is that MSPs residing in GCs may experience multiple epochs of spin-up through successive dynamical interactions and mass-transfer episodes \citep[see e.g.][and references therein]{vf14}. For a given GC, both the overall stellar encounter rate and the encounter rate per binary must be considered to characterise the resulting pulsar population; that is, the fraction of isolated pulsars, products of exchange interactions, and the number of slow pulsars. Hence, when discussing the possibility of multiple accretion epochs of GC MSPs --- and thereby the prospect of potentially producing a sub-MSP --- one must simultaneously consider the related probability of truncating an ongoing recycling process and leaving behind a relatively slow spinning pulsar. 

Nevertheless, stellar-evolution arguments must also be taken into account. Consider a long-term stable LMXB evolution early in the history of a GC. This may leave behind a fully recycled MSP ($P<10\;{\rm ms}$) after $\sim$~3$-$4~Gyr (i.e. $\sim 9\;{\rm Gyr}$ ago). At that cluster age, all stars with a mass of $\gtrsim 1.4\;M_\odot$ have left the main sequence. This means that for any second epoch of RLO, the donor star has a mass $< 1.4\;M_\odot$ which, as we have argued in detail here, is insufficient to spin-up a NS to become a sub-MSP (let alone the MSP produced in the first epoch likely undergoing propeller-driven spin-down before accreting again). Rare dynamical encounters in GCs may produce ultra-compact or stripped-donor binaries, but these systems are also expected to accrete too little mass, and for too short a time, to form sub-MSPs. Moreover, it is worth remembering the steep dependence of the final recycled spin period on the amount of mass accreted (Eq.~\ref{eq:deltaM}). For example, it takes accretion of another $\sim 0.13\;M_\odot$ to boost the spin of a 1.7~ms MSP to 1.0~ms, comparable to the mass required to spin-up an initially slowly rotating NS to 1.7~ms in the first place.

We conclude that even if multiple, uninterrupted epochs of RLO occur in a GC, producing a sub-MSP appears very unlikely. Perhaps this outcome is not surprising given that the current spin distributions of Galactic field MSPs and GC MSPs are quite similar (Fig.~\ref{fig:spin-histo}, top-right).

\subsection{Differential rotation and instability thresholds}\label{subsec:diff-rot}
Regarding the possibility of differential rotation between the NS crust and a superfluid core \citep[e.g.][]{asc06,hs18}, it is unknown what fraction of the NS’s moment of inertia is effectively coupled to the external spin-down torque. If only the magnetosphere-coupled component spins down, the effective inertia ($I_{\rm eff}$) is smaller. If the coupled fraction grows as the NS spins down (e.g. progressive recoupling of the superfluid), then $I_{\rm eff}$ increases as $\Omega$ decreases and, for dipole-dominated braking, the resulting braking index is $n<3$.

A key parameter for sub-MSPs is the ratio of rotational to gravitational binding energy ($\beta\equiv T/|W|$), which can trigger non-axisymmetric instabilities. As $\beta$ increases from $\sim 0.08$ to 0.24, the expected instabilities range from secular bar-mode (driven by viscosity or GW back-reaction) to dynamical bar-mode (purely hydrodynamical). In differentially rotating stars, however, low-$\beta$ shear/corotation instabilities may set in already for $\beta\sim 0.01-0.10$. As an example, a $1.4\;M_\odot$ NS (with $R=12\;{\rm km}$ and $I=10^{45}\;{\rm g\,cm^2}$) would have $\beta=(0.076,\,0.16,\,0.30)$ for $P=(1.0,\,0.7,\,0.5\;{\rm ms})$ --- subject to the NS EoS and GR corrections. These instabilities can lead to significant spin damping via GW emission (Appendix~\ref{app:GWs}), either from a static quadrupole (``mountain'') or from $r$-modes (a separate CFS-type GW instability with $n=7$), on very short timescales (seconds--minutes), i.e. long-term sub-ms rotation is extremely hard to sustain.

\subsection{NSs born with sub-ms spin periods?}\label{subsec:born-MSP}
While the observed populations of radio MSPs and XMSPs are spinning rapidly due to accretion, one may ask whether some NSs could instead be born with sub-ms spin periods. Although several studies have concluded that NSs are typically born with spin periods $P_0>15\;{\rm ms}$ \citep{hws05}, others point to the possibility that NSs could be born with spins $\sim 1\;{\rm ms}$ \citep{obt+06}. 
For this to occur, the stellar core of a massive star would need to be rapidly rotating before it collapses. There is strong evidence that, under ordinary circumstances, contracting stellar cores lose most of their angular momentum to the expanding and slowly rotating stellar envelopes through magnetic coupling \citep{hws05,slp+08,fpj19,tl21}. However, rotationally induced mixing may lead to chemically homogeneous evolution (CHE) of the most rapidly rotating stars \citep{mae87,lan92}, thereby producing fast-spinning cores at the time of collapse. 
Such models have been proposed to explain long-duration gamma-ray bursts \citep[GRBs][]{yl05,wh06} in the context of the collapsar model \citep{woo93,tcq04}, rapid BH component spins in some in-spiralling double BH merger events detected in GWs \citep[e.g.][]{mlp+16,mdM16,GW231123,pdM25}, and Type~Ic superluminous SNe \citep[SLSNe,][]{alm+18,alab20} in the context of millisecond magnetar birth \citep{woo10,mbm17}. 

The models of \cite{alab20} indeed produce iron cores with angular momenta corresponding to rotation rates near or below 1~ms in heavy NSs. The rotational energies of these NSs can reach $10^{53}\;{\rm erg}$, making them suitable candidates to power SLSNe. \cite{ngb17} derived NS spin periods from a light-curve analysis of 38 SLSNe based on the millisecond magnetar model and found values consistent with those obtained from CHE models, i.e. stretching NS birth spins into the sub-ms regime.
A further possibility for forming a sub-ms NS is via accretion-induced collapse (AIC) of a massive WD \citep{nms+79,mcr26}; see discussion below.

The empirical evidence for sub-ms birth spins is, so far, indirect: the newborn NS spin is inferred by fitting magnetar spin-down central-engine models to observed light curves (e.g. GRB X-ray plateaus; SLSN optical/bolometric light curves; \citealt{rom+13}). Within this framework, some model-dependent fits allow $P_0<1\;{\rm ms}$.
Once born, such sub-ms magnetars would spin down rapidly due to strong dipole radiation on short timescales, preventing detection as long-lived pulsars --- if they emit in radio at all. 
Magnetohydrodynamic (MHD) simulations of rapidly rotating core collapse show that B-field amplification is unavoidable: differential rotation and shear trigger efficient dynamo action and the magnetorotational instability (MRI), leading to exponential field growth on ms timescales. Even modest seed fields are amplified to pulsar-level ($B \sim 10^{12}-10^{13}\;{\rm G}$) or magnetar-level strengths of $B \sim 10^{14}-10^{15}\;{\rm G}$ \citep{dt92,awml03,rgo+16}. Hence, rapid spin-down appears unavoidable. Similarly, powering the high luminosity of SLSNe via extraction of NS rotational energy requires spin-down on a timescale of days to weeks. 

\subsubsection{GW signals from core-collapse SNe, NS mergers and AIC}
If some NSs are indeed born with very short spin periods, this has direct implications for their early GW emission. In this context, the discussion above connects naturally to GW signals anticipated from core-collapse SN explosions and NS mergers. 
In core-collapse SNe, GWs are generated by highly non-spherical mass motions associated with hydrodynamical instabilities \citep[e.g. convection and the standing accretion shock instability, SASI;][]{ott09,mml+23,wft23,vbw+23}. Additional contributions arise from proto-NS oscillations and anisotropic neutrino emission. Recent 3D simulations confirm that these signals are broadband and typically populate the $\sim 1–2000\;{\rm Hz}$ range, thereby encoding information about the explosion dynamics, the proto-NS structure, and the underlying microphysics \citep{llk+26,mul26}. Both SN models investigated by \cite{llk+26} are in reach of current advanced LIGO detectors for Galactic distances. 
\cite{mul26} emphasizes that rapidly rotating or magnetically deformed post-bounce remnants (cf. discussion above) can produce strong emission at several hundred Hz, and that for favorable events next-generation (3G) detectors could extend the reach to Mpc distances for the most extreme cases. 
This makes the question of the birth spin distribution of NSs --- and in particular whether sub-ms spins can be achieved --- directly relevant for GW astronomy.

A related argument applies to NS+NS mergers. Their post-merger remnants are born hot, differentially rotating, and often close to mass-shedding, with sub-ms spin periods. They may also be strongly non-axisymmetric, so that the subsequent redistribution of angular momentum, mode excitation, and possible collapse to a BH are imprinted in the high-frequency GW signal. In particular, the dominant post-merger GW frequencies provide valuable information on the NS EoS and the remnant dynamics \citep{bj12,ber20}.
In this sense, core-collapse SNe and NS mergers probe complementary aspects of the same underlying problem: how compact objects are born spinning rapidly, how efficiently they shed angular momentum, and what role is played by the EoS, magnetic stresses, and GW-driven instabilities.

In general, newborn differentially rotating NSs would tend to approach rigid rotation within seconds after their formation. Likewise, post-merger hypermassive NSs (originating from NS+NS mergers) may initially possess near-mass-shedding sub-ms spins immediately after the merger event \citep[potentially detectable through X-ray follow up,][]{prg+26}, but they are not long-term stable NSs.
Studies including secular losses (GWs, viscous/MHD winds, neutrino-aided outflows) conclude that any NS remnant spins down on seconds--minutes timescales \citep{sh19,rbp20}.

As mentioned above, AIC has also been suggested as a viable channel for producing newborn NSs with ultra-fast spins. In particular, the formation of sub-ms compact stars via AIC has been discussed in the context of self-bound strange-quark stars \citep[e.g.][]{xu05,dxqh09} (see also studies of rapidly rotating strange-quark, self-bound, and hybrid compact stars by \citealt{ghl+99,gsb+01,zpy+06,dpp08}). 
In this scenario, a WD is spun up by long-term accretion in a binary, which is facilitated by the fact that WDs contract as they gain mass.
We note that this represents only one branch of a much broader literature on rapidly rotating exotic compact stars, which is beyond the scope of the present review \citep[e.g.][]{ghl+99,gsb+01,zpy+06,dpp08}.
When such a WD reaches its Chandrasekhar mass --- which may exceed the canonical value of $1.38\;M_{\odot}$ \citep{yl05} --- its collapse likely leads to a rapidly rotating NS \citep{dbo+06}.
In this scenario, the collapse is also expected to produce GW emission, with signals potentially detectable out to $\sim 8\;{\rm Mpc}$ with 3G detectors \citep{lrc23,aor+10}. 

While no electromagnetic transient from AIC has been securely identified, some radio pulsars in GCs exhibit properties (including young characteristic ages) that are difficult to explain without invoking an AIC origin \citep{tsyl13}.
However, whether AIC can produce observable NSs, let alone sub-MSPs, remains unsettled. One issue is that NSs produced via AIC are often predicted to have relatively high B-fields, which would spin-down the newborn NS rapidly (see also Section~\ref{subsec:J0435+3233}).

In light of the scenarios discussed above, 3G GW observatories such as the Einstein Telescope \citep[ET,][]{ET2026}, Cosmic Explorer \citep[CE,][]{CE2019} and DECIGO \citep{DECIGO21} will significantly improve sensitivity across a broad frequency range, enhancing the prospects for detecting both transient and continuous GW signals from (metastable) newborn NSs. In particular, these detectors are expected to extend the reach for core-collapse SN signals beyond the Milky Way and enable detailed studies of post-merger remnants in NS+NS mergers. This will open a new observational window on the birth and early evolution of compact objects, providing valuable constraints on the physical mechanisms discussed in this review.

\subsection{Was PSR~J0435+3233 born as a sub-MSP?}\label{subsec:J0435+3233}
The recently discovered binary MSP J0435+3233 \citep{wwy+26} is a highly unusual system. Its measured spin period derivative ($\dot{P}=4.88\times 10^{-17}$) appears to be at least two orders of magnitude higher than that of typical MSPs with a spin period of $P=3.20\;{\rm ms}$ (Fig.~\ref{fig:spinupline}), resulting in an exceptionally young characteristic age of $\tau\equiv P/(2\dot{P})\simeq 1.04\;{\rm Myr}$. The pulsar is in an 8.0~day orbit with a $\sim 0.33_{-0.05}^{+0.41}\;M_\odot$ companion, most likely a WD. At first glance, this discovery suggests a formation channel distinct from the classical evolutionary path; specifically, \cite{wwy+26} propose that this MSP may be the remnant of an AIC event.

\begin{figure}[ht]
\centering
\vspace*{-0.8cm}\hspace*{-0.4cm}
\includegraphics[width=0.85\textwidth]{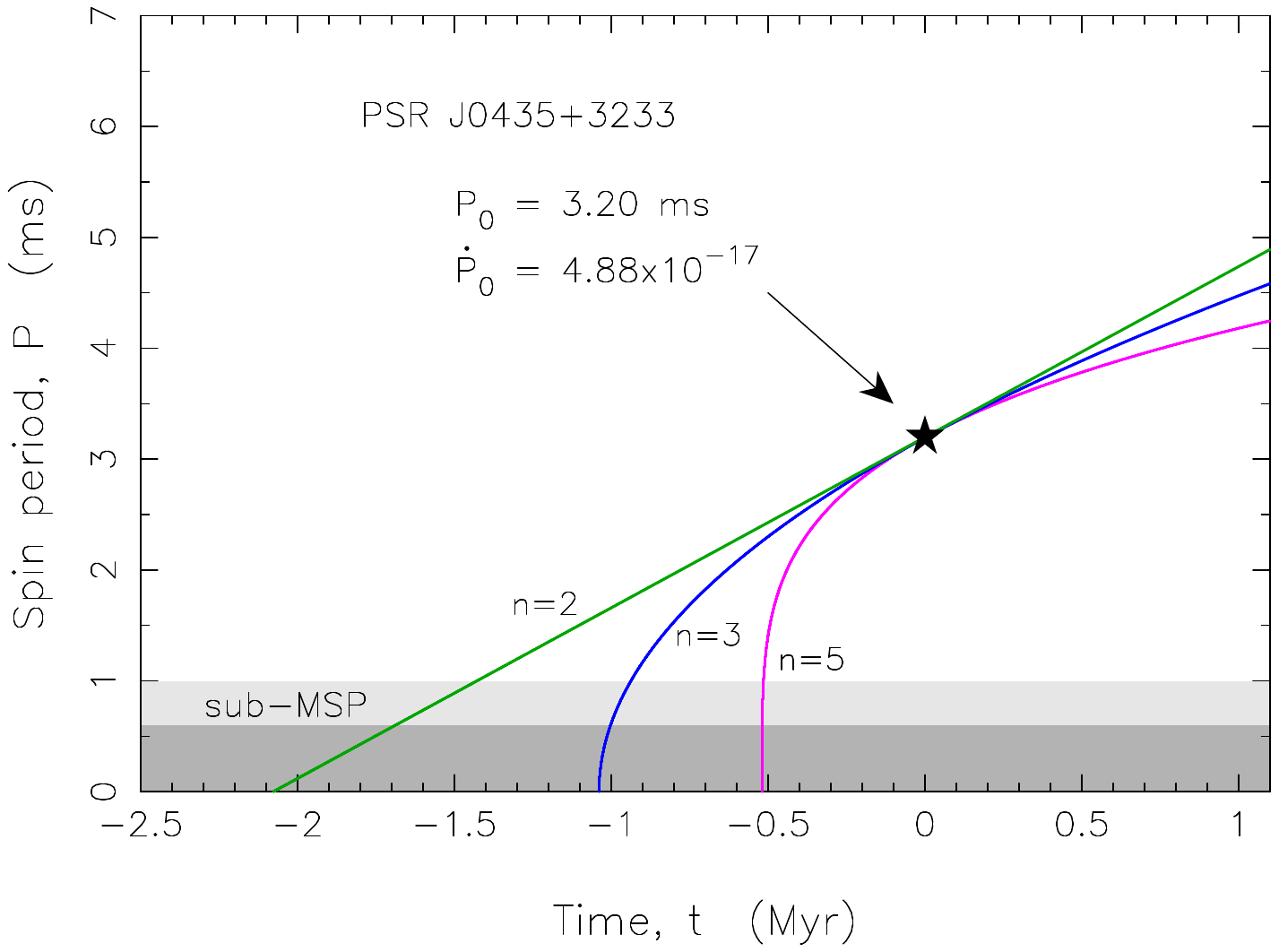} 
\caption{Spin evolution of the anomalous MSP J0435+3233, assuming evolution with a constant braking index of $n=\{2,\,3,\,5\}$ and an intrinsic $\dot{P}=\dot{P}_{\rm obs}$. Its current location in the $P\dot{P}$-diagram ($t=0$) and the inferred spin-down history shown here are consistent with an origin as a sub-MSP produced via AIC (although this interpretation is subject to several caveats; see text for a discussion). The darkest shaded region at the bottom is prohibited by centrifugal break-up ($P\lesssim 0.6\;{\rm ms}$).}
\label{fig:J0435+3233}
\end{figure}

In Fig.~\ref{fig:J0435+3233} we illustrate the past spin evolution of PSR~J0435+3233 along three hypothetical evolutionary tracks assuming constant braking indices of $n=\{2,\,3,\,5\}$ provided the observed $\dot{P}_{\rm obs}$ reflects the intrinsic spin-down. 
Given its remarkably short spin-down timescale, it is conceivable that PSR~J0435+3233 was born with an unusually short spin period, possibly in the sub-ms regime, and has since spun down to its current state.

This proposed scenario, however, raises several concerns. First, AIC is expected to produce a NS with a mass $<1.30\;M_\odot$, which is difficult to reconcile with sub-ms spin periods for most NS EoS (Sect.~\ref{subsec:EoS}). Furthermore, the orbit is highly circular, with an eccentricity of $e \simeq 1.6\times 10^{-4}$, whereas AIC is expected to impart a significant eccentricity $\mathcal{O}(0.1)$ due to instantaneous mass loss associated with the release of gravitational binding energy when the WD implodes to form a NS. It remains unclear how post-AIC accretion or tidal interactions could circularise the orbit (and spin-up the pulsar) on a very short timescale; a detailed formation model is needed.

Finally, if PSR~J0435+3233 is part of a triple system, the measured $\dot{P}_{\rm obs}$ may well be affected by line-of-sight acceleration, such that the intrinsic $\dot{P}$ is significantly smaller (i.e. consistent with a normal MSP). Although a triple configuration was disfavoured by \citet{wwy+26}, it is noteworthy that the ratio $\dot{\nu}/\nu$ is similar to $\dot{\nu}_{\rm orb}/\nu_{\rm orb}$, and likewise for higher-order derivatives, which provide strong evidence for a common acceleration, as expected in a triple system \citep{fcb+26,yhy+26}. Further investigation is required to confirm this interpretation.

\subsection{Comparison to other work}\label{subsec:comparison}
Among the earliest dedicated investigations of sub-MSP formation were the studies by \citet{bpc+99} and \citet{pcg+99}. The former developed a semi-analytical model for the spin evolution of accreting NSs, including magnetic-field decay and relativistic stellar structure, and concluded that sub-MSPs could in principle be produced under favourable conditions. The latter performed population-synthesis calculations and predicted that, in the absence of an additional braking mechanism, the Galactic MSP population should contain a low-period tail extending into the sub-ms regime.

The question of mass-accretion efficiency has also been investigated earlier. \cite{ts99} advocated a NS accretion efficiency, $\epsilon < 0.4$ from a comparison of their calculated evolutionary tracks with e.g. PSR~B1855+09. This result was later confirmed by observations of a number of additional MSPs with relatively light NS mass determinations \citep[e.g.][]{jhb+05,avk+12,aks+16}.
\cite{mly25} recently investigated formation of spider MSP systems from LMXBs and IMXBs (up to $3\;M_\odot$ donor stars) and argued that during the recycling phase, the mass required to spin up a NS to sub-ms is high enough to collapse it into a BH (they applied an upper NS mass limit of $2.5\;M_\odot$). It is evident from their Fig.~8 that even the modelling of spider systems assumed to evolve with $\epsilon \simeq 1$ ($\beta =0$) cannot produce sub-MSPs, not to mention cases for $\epsilon \simeq 0.1-0.3$ ($\beta \simeq 0.7-0.9$). These results are therefore fully consistent with our findings.

\citet{lc11} also conducted a binary-evolution investigation of the possibility of producing sub-MSPs in LMXBs. Using the Eggleton stellar evolution code, they concluded that it is difficult to produce sub-MSPs through this evolutionary channel, owing to either low spin-up efficiency or the onset of unstable mass transfer, in agreement with our conclusions, albeit for somewhat different donor-mass intervals.
We note that in recent years the magnetic-braking prescriptions governing the orbital angular momentum evolution of LMXBs have been heavily debated \citep[e.g.][]{vih19,cthc21,vi21,yc25,neg+26}.

Examples of recent studies focussing on NS spin evolution include those of \citet{kob24,neg+26}, which illustrate the challenges of modelling the coupled evolution of NS spin and accretion-induced magnetic-field decay 
\citep{tv86,smsn89,rom90,sbmt90,gu94,vb95,kb97,zha98,czb01,pm07}. Although there is compelling observational evidence that the surface magnetic fields of NSs are reduced by accretion, the underlying physical mechanism responsible for this field decay, and its efficiency, remain poorly understood.
For a more comprehensive review of NS accretion, binary evolution, and the formation of MSPs through LMXBs and other evolutionary channels, we refer the reader to \citet{tv23}.

\cite{zl25} argued that the absence of sub-MSPs is naturally expected if recycled pulsars are gravity-bound NSs, whose larger radii and magnetic spin-up limits prevent extreme rotation, especially at low masses. By contrast, self-bound compact (strangeon) stars could in principle be spun up to sub-ms periods with the same accreted mass. Accordingly, the non-detection of sub-MSPs is fully consistent with standard NSs, whereas any future discovery --- particularly with a mass below $\sim 1.2\;M_\odot$ --- would provide compelling evidence for a self-bound EoS.

A complementary class of explanations invokes a physical torque ceiling rather than (or in addition to) mass-budget limitations. In particular, \cite{hzf+18} argued that GW losses --- from unstable modes (e.g. r-modes) or non-axisymmetric ``mountains'' --- could balance the accretion torque and thereby account for the absence of sub-MSPs, linking the cutoff to dense-matter microphysics. 
Even within standard recycling, the spin-up may also stall if angular-momentum transfer becomes inefficient during prolonged accretion: \cite{bu15} showed that coupling between accretion-driven magnetic-field decay, thermal evolution, and torque efficiency can impose a limiting minimum spin period that prevents reaching sub-ms spins even for long-lived, high-$\dot{M}$ accretion. 
Finally, dedicated high-time-resolution searches explicitly targeting the (sub)ms regime have not revealed any sub-MSPs \citep[e.g.][]{dpf11}, providing an empirical check that pure selection effects are unlikely to be the sole explanation.

As reviewed by \cite{wac+16}, sub-ms rotation would constitute a qualitatively new regime for NSs, tightly constraining the EoS and the presence of additional spin-down torques such as GW emission. The lack of such objects to date therefore already disfavors scenarios in which extreme spins are easily attained or long-lived.

\subsection{Black holes with sub-ms spins}\label{subsec:BH}
\begin{figure}[ht]
\centering\vspace*{0.8cm}%\hspace*{-0.4cm}
\includegraphics[width=0.86\textwidth]{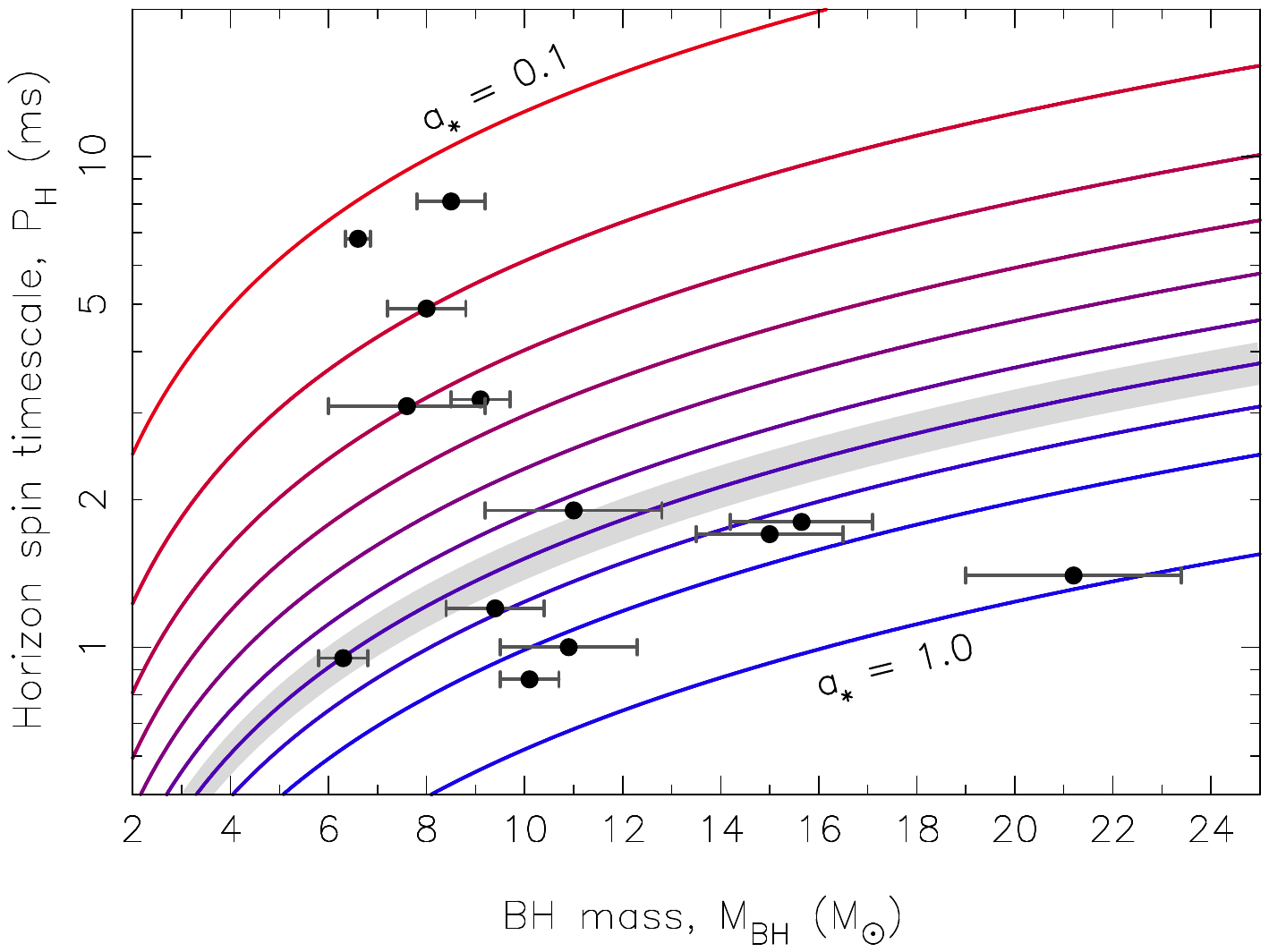} 
\caption{Horizon spin timescale as a function of BH mass. The coloured lines show theoretical curves for fixed spin parameters, $a_\ast=0.1-1.0$ (in steps of 0.1), calculated from Eq.~(\ref{eq:BHspin}). Observational data points correspond to BH X-ray binaries listed in Table~\ref{table:BHspin}, using the central values of $a_\ast$. The grey line indicates typical remnant BH values of $a_\ast^{\rm rem}\sim 0.7$ from NS+NS and BH+BH mergers, see Appendix~\ref{app:BHs}.}
\label{fig:BHspin}
\end{figure}

We conclude this review with a brief (but illuminating) comparison to the spins of another class of compact objects: stellar-mass black holes (BHs).
Unlike NSs, BHs possess no rigidly rotating material surface, and Kerr BH horizon spin, $P_H$, therefore represents a spacetime property rather than a physical rotation of matter. It is derived from the angular velocity of the BH horizon, $\Omega_H$, which characterizes the frame-dragging of spacetime at the event horizon as measured by a distant observer, and it is given by (Appendix~\ref{app:BHs}):
\begin{equation}\label{eq:BHspin}
   P_H \simeq 0.0619~{\rm ms}\;\left(\frac{M}{M_\odot}\right)\frac{1+\sqrt{1-a_\ast^{2}}}{a_\ast} \;,
\end{equation}
where $M$ is the BH mass and $-1 \le a_\ast \le +1$ is the dimensionless spin parameter.
Among the 14 BH X-ray binaries (Table~\ref{table:BHspin}; plotted in Fig.~\ref{fig:BHspin}) with known values of $M$ and $a_\ast$, two qualify as ``sub-ms BHs'', namely GRS~1915+105 ($P_H < 0.86\,{\rm ms}$) and GRO~J1655$-$40 ($P_H = 0.95\,{\rm ms}$). 
Note that the continuum-fitting method infers the BH spin by fitting theoretical accretion-disc models to the observed X-ray continuum spectrum. Since this approach relies on several modelling assumptions, the resulting values of $P_H$ carry systematic uncertainties that are difficult to quantify. For a brief discussion of BH spins resulting from NS+NS and BH+BH mergers, see Appendix~\ref{app:BHs}.

%%%%%%%%%%%%%%%%%%%%%%%%%%%%%%%%%%%%%%%%%%%%%%%%%%%%%%%%%%%%%%%%%%%%%%%%%%%%%%%%%%%%%%%%%%%%%%%%%%%%%%%%%%
\clearpage
\section{Conclusions}\label{sec:conclusions}
In this review, we discussed the existence (or absence) of sub-MSPs and their broad implications for NSs and relativistic astrophysics --- spanning from the dense-matter EoS and the centrifugal break-up limit to implications for strong-field gravity and GW sources. In particular, we set out to assess whether sub-MSPs should exist in the present Galactic population by combining empirical spin and mass measurements with accretion torque and efficiency arguments. Our {\bf main findings} are:

\begin{itemize}
\item \textit{Inefficient recycling dominates.} From systems with well-determined post-recycling masses, we infer typical upper limits to the mass-accretion efficiency of $\epsilon_{\max}\simeq 0.20-0.30$ (Sect.~\ref{subsec:efficiency}). Thus, at least 70--80\% of the mass transferred during the LMXB phase must be lost from the system. 
\vspace{0.2cm}

\item \textit{Sub-MSPs are hindered by a double bottleneck: the relatively massive donors needed to supply sufficient accreted mass evolve too fast for the NSs to reach sub-ms spin.} Reaching $P_{\rm eq}=1.0\;{\rm ms}$ ($0.6\;{\rm ms}$) requires $\Delta M_{\rm eq}\gtrsim 0.25\;M_\odot$ ($\gtrsim 0.50\;M_\odot$). Hence, for realistic $\epsilon\lesssim 0.20$ found in this work, this would imply donor star masses $M_2\simeq 2-4\;M_\odot$ (Fig.~\ref{fig:donor-mass}), i.e. IMXB progenitors. However, such donors drive RLO with X-ray lifetimes $\tau_X\sim 1-10\;{\rm Myr}$, and Eddington-limited accretion caps the accumulated mass at $\Delta M_{\rm NS}\lesssim 0.2\;M_\odot$ --- insufficient to reach sub-ms spin (Sect.~\ref{subsec:deltaM}).\\
We emphasize that this conclusion may change if the very low accretion efficiencies inferred from the sample of low-mass, fully recycled MSPs with He~WD companions (Table~\ref{table:NSmasses}) are not representative of the general population of NSs in binaries.
\vspace{0.2cm}

\item \textit{Magnetospheric conditions alone do not forbid sub-MSPs.} Spin-up bands (Fig.~\ref{fig:spinupline}) allow sub-ms equilibrium for weak $B$ and high $\dot M$. The shortfall arises, however, from the \emph{astrophysical} pathway (low $\epsilon$, high $M_2$, short $\tau_X$), not the equilibrium prescription itself.
\vspace{0.2cm}

\item \textit{Post-formation spin-down is rapid unless $B$ is tiny.} Even if a sub-MSP forms, it spins down to $P>1\;{\rm ms}$ on a timescale of a few 100~Myr for $B\sim 10^8\;{\rm G}$. Only $B\lesssim 10^7\;{\rm G}$ preserves sub-ms spins over several Gyr (Figs.~\ref{fig:Pevol-subMSP} and \ref{fig:isochrones}).
\vspace{0.2cm}

\item \textit{Selection effects are not the primary culprit.}
If sub-MSP luminosities are similar to those of MSPs (the observed pulsar luminosities show no downturn toward $P\sim1$~ms; see Fig.~\ref{fig:R1400}), there are no known selection effects that prohibit the detection of a sub-MSP in the radio domain.
Scattering, while being the predominant cause of sensitivity loss for detecting sub-MSPs, can be mitigated by observing at higher radio frequencies; intra-channel DM smearing can, at least for GCs, be eliminated by coherent dedispersion. Current survey capabilities are at least consistent with the absence of an undiscovered population of hypothetical bright radio sub-MSPs that are isolated or reside in sufficiently wide binaries.\\
In X-rays, reduced pulsed fractions, orbital smearing, and accretion dilution lower sensitivity at very fast spins, without producing a cutoff near $P\sim1\;{\rm ms}$. Existing X-ray searches therefore disfavour a hidden population of bright accreting sub-MSPs.

\smallskip
\item \textit{Sub-MSPs are not expected from current data.} Although statistical extrapolation (Sect.~\ref{subsec:statistics}) cannot strictly exclude the existence of a pulsar with $P<1\;{\rm ms}$, the present data with $\sim 700$ MSPs spinning faster than 10~ms strongly imply that a sub-MSP would be an extremely rare outlier, if it exists at all.
\end{itemize}

\noindent{\bf Implications.} The most straightforward explanation for the null detection of sub-MSPs today is a combination of (i) low accretion efficiency during recycling and short mass-transfer phases in the IMXB channel, and (ii) subsequent magnetic dipole (and possibly GW-assisted) spin-down that quickly moves any nascent sub-MSPs above $1$~ms. GW torques may contribute near the empirical cutoff, but equilibrium-spin physics alone need not invoke them.

\smallskip
\noindent{\bf Observational prospects.} If sub-MSPs exist, they are most likely to be caught \emph{during} accretion, favouring deep, high-cadence X-ray timing of bright atoll sources and transitional systems. 
High-time-resolution observations in the optical band have recently emerged as a promising additional channel for searches for sub-MSPs during accretion states.
In the radio, optimal strategies remain high-frequency, wide-band, coherently de-dispersed searches with sub-$50\,\mu$s sampling. Precision kinematic corrections (parallaxes and proper motions) are essential to establish $B\lesssim 10^7$~G for the fastest MSPs. Future LVK observing runs and next-generation detectors will further improve continuous-wave searches of MSPs, isolated NSs, and XMSPs, providing increasingly stringent tests of GW-regulated spin evolution (Appendix~\ref{app:GWs}).

\smallskip
\noindent{\bf Caveats.} Our quantitative bounds adopt a radiative efficiency $\eta \simeq GM/(Rc^2)\simeq 0.15$ and a conservative minimum NS birth mass, $M_{\rm NS,0}\ge 1.17\,M_\odot$. Moreover, we have not considered scenarios involving strongly super-Eddington mass accretion (e.g. neutrino-cooled hypercritical accretion far exceeding the Eddington limit). 
Modest changes in these assumptions do not alter the qualitative picture: forming and \emph{keeping} a NS at $P<1$~ms is difficult.

\smallskip
In summary, sub-MSPs are not forbidden by basic spin-equilibrium considerations, but they are strongly disfavoured by the combined mass budget, accretion timescale, and subsequent spin-down. If they exist at all, they should be rare, short-lived, and preferentially X-ray selected.

\vspace{1.0cm}
\noindent {\bf Acknowledgements.} 
The authors are especially grateful to Monica Colpi for her thoughtful advice, constructive suggestions, and warm encouragement throughout the review process.
The authors thank Norbert Wex, Gianluca Pagliaro, and Luigi Stella for valuable comments that helped improve the manuscript. V.V.K. acknowledges financial support from the European Research Council (ERC) Starting Grant ``COMPACT'' (Grant Agreement No. 101078094). A.P. acknowledges financial support by INAF (Grants FANS and PULSE-X, PI: Papitto), the Italian Ministry of University and Research (PRIN MUR 2020, Grant 2020BRP57Z, GEMS, PI: Astone), and Fondazione Cariplo/Cassa Depositi e Prestiti (Grant No. 2023-2560, PI: Papitto).

%%%%%%%%%%%%%%%%%%%%%%%%%%%%%%%%%%%%%%%%%%%%%%%%%%%%%%%%%%%%%%%%%%%%%%%%%%%%%%%%%%%%%%%%%%%%5
%\vspace{5.0cm}
%\backmatter
%
%\bmhead{Supplementary information}
%The online version contains supplementary material available at [DOI link].
%
%\bmhead{Acknowledgements}
%The authors thank Norbert Wex and Gianluca Pagliaro for valuable comments that helped improve the manuscript.
%V.V.K. acknowledges continued support from the Max Planck Society.
%
%\bmhead{Funding}
%
%\bmhead{Author contributions}
%T.M.T. initiated and led the project, including conceptualization, formal analysis, methodology, visualization, writing of the original draft, and review and editing.
%V.V.K., R.S. and P.C.C.F. contributed to conceptualization, formal analysis, visualization, writing of the original draft, and review and editing. C.A.N.B. contributed to formal analysis and writing of the original draft.
%S.M.R., A.P. and N.L. contributed to conceptualization, writing of the original draft, and review and editing.
%E.P.J.vdH. and M.K. contributed to design, review and editing.
%
%\bmhead{Competing interests}
%The authors declare no competing interests.
%
%\bmhead{Data availability}
%All data supporting the findings of this study are available within the article and its supplementary information. Additional materials are available from the corresponding author upon reasonable request.
%
%\bmhead{Code availability}
%The code used in this study is available from the corresponding author upon reasonable request.

%%%%%%%%%%%%%%%%%%%%%%%%%%%%%%%%%%%%%%%%%%%%%%%%%%%%%%%%%%%%%%%%%%%%%%%%%%%%%%%%%%%%%%%%%%%%%%%%%%%%%%
\clearpage
\begin{appendices}

\section{Sub-ms black holes}\label{app:BHs}
The angular frequency of a Kerr black hole (BH) horizon is given by \citep{bch73,mtw73}:
\begin{equation}
   \Omega_H = \frac{c^3 \;a_\ast}{2GM \left(1+\sqrt{1-a_\ast^{2}}\right)} \; .
\end{equation}
Thus, the corresponding horizon spin period is:
\begin{equation}
   P_H \equiv \frac{2\pi}{\Omega_H}  = \frac{4\pi\,GM}{c^{3}}\, \frac{1+\sqrt{1-a_\ast^{2}}}{a_\ast} \; .
\end{equation}
Using $GM/c^{3} \simeq 4.9255~\mu{\rm s}\,(M/M_\odot)$, this becomes:
\begin{equation}%\label{eq:BHspin}
   P_H \simeq 0.0619~{\rm ms}\;\left(\frac{M}{M_\odot}\right)\frac{1+\sqrt{1-a_\ast^{2}}}{a_\ast} \; .
\end{equation}
We see that a $10\;M_\odot$ near-maximal Kerr BH ($a_\ast \rightarrow 1$) has a horizon spin period of $P_H \simeq 0.6\;{\rm ms}$. A $2.0\;M_\odot$ Kerr BH (should it exist) would have a horizon spin period of $\simeq 0.12\;{\rm ms}$, which is about an order of magnitude shorter than the current record MSP periods ($1.4\;{\rm ms}$).

Table~\ref{table:BHspin} lists BH horizon spins calculated from measured data using the continuum-fitting method \citep{zcc97,mns14}. The data is plotted in Fig.~\ref{fig:BHspin}. 
(At present, there exists no direct observational analogue to NS kHz QPOs that measures the horizon angular frequency of stellar-mass BHs. While a handful of BH X-ray binaries exhibit high-frequency QPOs at tens to hundreds of Hz \citep{im19}, these signals are thought to originate in the inner accretion flow and do not provide a model-independent constraint on $P_H$.)
Two sub-ms BH sources exist among the known BH X-ray binaries: GRS~1915+105 ($P_H<0.86\;{\rm ms}$) and GRO~J1655$-$40 ($P_H=0.95\;{\rm ms}$).

\begin{table}
\caption{BH horizon spins ($P_H$) measured using the continuum-fitting method \citep[see][and references therein]{tv23}. Among the currently known population of BH X-ray binaries, only two systems (GRS~1915+105 and GRO~J1655$-$40) qualify as sub-ms BHs.}
\label{table:BHspin}
\begin{tabular}{lrll}
\toprule
Source & $P_H$ (ms) & BH spin ($a_\ast$) & BH mass ($M_{\rm BH}/M_\odot$)\\  
\midrule
Persistent X-ray binaries & & & \\
\midrule
Cyg~X-1           & $<1.4^{\ast}$  & $>0.9985^{\ast}$        & $21.2\pm 2.2$   \\
LMC~X-1           & 1.0  & $0.92_{-0.07}^{+0.05}$  & $10.9\pm 1.4$   \\
IC~10~X-1         & $\sim 1.7$  & $0.85_{-0.07}^{+0.04}$  & $15$ (assumed)  \\
M33~X-7           & 1.8  & $0.84\pm 0.05$          & $15.65\pm 1.45$ \\
\midrule
Transient X-ray binaries & & & \\
\midrule
GRS~1915+105      & $<0.86$  & $>0.95$                 & $10.1\pm 0.6$ \\
MAXI~J1803$-$298  & --   & $0.991\pm 0.001$ & --\\
4U~1543$-$47      & 1.2  & $0.80\pm 0.10$          & $9.4\pm 1.0$ \\
GRO~J1655$-$40    & 0.95 & $0.70\pm 0.10$          & $6.3\pm 0.5$ \\
Nova Mus 1991     & 1.9  & $0.63_{-0.19}^{+0.16}$  & $11.0\pm 1.8$ \\
XTE~J1550$-$564   & 3.2  & $0.34_{-0.28}^{+0.20}$  & $9.1\pm 0.6$ \\
LMC~X-3           & $>3.1$  & $<0.3$                  & $7.6\pm 1.6$ \\
H1743$-$322       & $\sim 4.9$  & $0.2\pm 0.3$            & $\sim 8$ \\
MAXI~J1820+070    & 8.1  & $0.13_{-0.10}^{+0.07}$  & $8.5\pm 0.7$ \\
A0620$-$00        & 6.8  & $0.12\pm 0.19$          & $6.6\pm 0.25$ \\
\bottomrule
\end{tabular}
\vspace{0.2cm}
 {\small $^{\ast}$Assuming aligned spin ($\delta= 15^\circ$ yields $a_\ast=0.9696$ \citep{mbo+21} and thus $P_H=1.7\;{\rm ms}$).}\\
\end{table}

The outcome of a NS+NS merger is a fast-spinning BH with $a_\ast^{\rm rem}\simeq 0.70-0.80$ \citep{ksst09,sfh+17}. Assuming a resulting BH mass of $2.5-3.5\;M_\odot$, events like GW170817 and GW190425 would lead to BHs with $P_H\simeq 0.3-0.5\;{\rm ms}$. This may thus also apply to the invisible $2.35\pm 0.19\;M_\odot$ companion star of the eccentric GC MSP J0514$-$4002E \citep{bdf+24}. Hence, NS+NS or BH+NS mergers could leave low-mass BHs with sub-ms $P_H$ values.

Most of the massive BH+BH mergers also leave isolated BHs with spin parameters $a_\ast^{\rm rem} \sim 0.7$ \citep{rbd+08}. From Eq.~(\ref{eq:BHspin}) we see that their $P_H$ values will be larger. For example, typical BH remnants of $50-70\;M_\odot$ will have $P_H=7-11\;{\rm ms}$, and the most massive ones (with high S/N), with $M=100-180\;M_\odot$, result in $P_H=15-27\;{\rm ms}$. 
For a substantial subset of BH+BH mergers detected by LVK, the $a_\ast ^{\rm rem}$ value of the remnant BH is well constrained from full inspiral--merger--ringdown (IMR) analyses, with typical uncertainties $\Delta a_\ast^{\rm rem} \sim 0.05-0.1$. These measurements cluster around $a_\ast^{\rm rem} \sim 0.7$, in good agreement with numerical-relativity predictions.
Only a small subset of LVK events \citep[a handful in GWTC-3,][]{GR25} have sufficient post-merger S/N to permit ringdown-based estimates of the remnant spin. Even for these events, the constraints are significantly weaker with typical uncertainties $\Delta a_\ast^{\rm rem} \simeq 0.2–0.4$, than those obtained from the full IMR signal. The measured remnant spins are nevertheless consistent with the typical theoretically expected range $a_\ast^{\rm rem}\simeq 0.65–0.85$ for most astrophysical BH+BH mergers.

For equal-mass binary mergers ($q \equiv m_2/m_1 = 1$), the limiting remnant spins are $a_\ast^{\rm rem} \simeq \{0.358,\,0.687,\,0.952\}$ for perfectly anti-aligned maximum spins, non-spinning BHs, and perfectly aligned maximum spins, respectively, where $a_\ast^{\rm rem}\simeq 0.687$ reflects the contribution from orbital angular momentum alone.
In the extreme-mass-ratio-inspiral (EMRI) limit ($q\to 0$), for accretion onto a spinning primary BH with spin $\chi_1$, the corresponding limiting values are $a_\ast^{\rm rem} \simeq \{\chi_1, \,\chi_1, \,0.998\}$ \citep{klp10}, i.e. an extreme EMRI with a non-spinning primary, $\chi_1=0$, would leave a non-spinning post-merger BH remnant.

\clearpage
\section{GW emission from accreting MSPs}\label{app:GWs}
Several mechanisms for the emission of GWs from rapidly rotating NSs have been proposed \citep[e.g.][]{las15,and21}. One leading hypothesis is the excitation of unstable r-modes, while other possibilities include non-axisymmetric mass quadrupoles arising during accretion. These may be supported by internal magnetic field stresses, magnetically confined accretion ``mountains'' or variations in temperature, composition, or density within the crust; see e.g. \cite{ucb00,las15,gg18,gaj21,mh25}. 

\subsection{Crustal quadrupoles vs. magnetic mountains}
NSs can emit continuous GWs if they possess a non-axisymmetric deformation, or \emph{quadrupole}, misaligned with their spin axis. Such deformations are often discussed in scenarios where GW losses counterbalance accretion-driven spin-up. The GW torque from quadrupole emission scales steeply with spin frequency ($n=5$): 
\begin{equation}
   |\dot{J}_{\rm GW, quadrupole}| \propto \nu^5 \;. 
\end{equation}

Two broad classes of mechanisms can generate crustal quadrupoles in accreting NSs:
(i) accretion-induced crustal asymmetries of thermal or compositional origin; and (ii) magnetic mountains, in which accreted matter is channelled by the NS magnetic field towards the magnetic poles, creating a ``mountain'' of material.

\begin{enumerate} 
  \item [(i)]\textit{Accretion-induced crustal quadrupoles} --- during accretion, uneven heating or asymmetric nuclear reactions in the crust can create lateral variations in temperature, composition, or density. Such asymmetries may be supported by elastic stresses in the solid crust, leading to a time-independent mass quadrupole that sources continuous GW emission. Typical ellipticities created in accretion-induced quadrupoles due to thermal/compositional effects are $\varepsilon \sim 10^{-8} - 10^{-7}$. No magnetic field is required, as the quadrupole is supported by elastic stresses in the solid crust.\\

  \item[(ii)] \textit{Magnetic mountains} --- require substantially stronger surface magnetic fields ($B \gtrsim 10^{11}$--$10^{12}\;{\rm G}$) such that magnetic pressure supports the deformation against gravity. The resulting ellipticities can reach up to $\varepsilon \sim 10^{-5}$ in extreme cases, though their long-term stability is uncertain due to possible magnetohydrodynamic instabilities or magnetic field burial during prolonged accretion \citep{gaj21}. Although the necessity of such strong pre-accretion fields is debated \citep{mp05}, realistic effects --- including magnetic burial, resistive relaxation, and field screening --- tend to suppress these distortions over long accretion histories \citep[e.g.][]{vm08,rfm25}. Thus, magnetic mountains require fine-tuned conditions to maintain large $\varepsilon$.
\end{enumerate}

Recent work has begun to model both mechanisms within a unified framework, combining magnetic stresses, asymmetric accretion-induced heating, and the elastic response of the crust, rather than treating them separately \citep{bhr+26}.

Magnetic mountains constitute a \emph{subset} of accretion-induced quadrupoles, specifically those supported by magnetic rather than elastic stresses. In contrast, thermally or compositionally induced quadrupoles are purely mechanical and do not rely on the presence of a magnetic field. Both mechanisms can limit spin-up in accreting NSs through GW emission. However, the fastest spinning MSPs all have $B\lesssim10^8\;{\rm G}$ --- values that are much smaller than required for magnetic mountains --- and hence GW emission from magnetic mountains cannot be the reason for the non-detection of sub-MSPs.

The above-mentioned predicted ellipticities are, however, substantially larger than current constraints provided by LVK GW searches. The O4a targeted search for known pulsars \citep[45 objects,][]{aaa+25} finds no detection of continuous GWs and quotes a maximum equatorial ellipticity of $\varepsilon \simeq 8.8\times 10^{-9}$ (95\% C.L.) for the nearby bright MSP~J0437$-$4715 ($P=5.76\;{\rm ms}$), assuming a standard triaxial-ellipsoid model and a canonical moment of inertia.
In a follow-up study, combining data from the LVK observing runs O4a and O4b \citep{aaa+26}, coherent narrowband searches yielded the tightest constraint for PSR~J0534+2200 (the Crab pulsar; $P=33\;{\rm ms}$) with a strain upper limit on the continuous GW amplitude of $\lesssim 2\%$ of its spin-down limit, corresponding to $<0.04$\% of the spin-down power being radiated as continuous GWs.

\subsection{r-modes as a GW source} 
r-modes (Rossby modes) are a class of non-radial (predominantly toroidal) oscillations in rotating stars, restored by the Coriolis force \citep[see][for a review]{ps17}. They are generically unstable to GW emission \citep{has15} through the Chandrasekhar--Friedman--Schutz (CFS) mechanism. The GW torque from r-modes scales even more steeply with spin frequency: 
\begin{equation} 
  |\dot{J}_{\rm GW,\,r-mode}| \propto \nu^7 \;,
\end{equation}
corresponding to a braking index $n=7$ (Sect.~\ref{subsec:spindown}) if r-mode emission dominates the spin evolution.
This steep dependence implies that even modest r-mode amplitudes could halt further spin-up near the observed frequency cutoff. r-mode activity depends sensitively on the NS interior temperature and composition, since viscous damping competes with GW driving. 

Current continuous-wave searches by the LVK Collaboration and others place increasingly stringent upper limits on the amplitude of r-mode oscillations in fast-spinning NSs. 
In particular, the O3 search for the young energetic pulsar PSR~J0537$-$6910 found no signal and excluded a substantial portion of the parameter space in which r-modes could dominate its spin-down under standard assumptions \citep{aaa+21-0537}. More generally, all-sky and directed continuous GW searches have set tight strain upper limits over a broad frequency range, which --- when interpreted in terms of r-mode models with $f_{\rm GW} = \tfrac43\,f_{\rm spin}$ --- significantly constrain persistent large-amplitude r-modes, with typical (95\% C.L.) upper limits of $\alpha \lesssim 10^{-5}\!-\!10^{-3}$ for several young or rapidly rotating NSs \citep{aaa+21-all-sky,aaa+22-all-sky}.

Recent O4-era searches have further improved constraints on continuous GW emission from a variety of source classes. These include targeted searches of known pulsars \citep{aaa+25}, all-sky searches for isolated NSs \citep{lvk26-allsky}, narrowband searches of known pulsars \citep{aaa+26}, and high-frequency Einstein@Home searches extending to GW frequencies of $1686\;{\rm Hz}$ \citep{msm+26}. Collectively, these studies continue to probe parameter space relevant to rapidly rotating NSs, although no continuous GW detections have yet been reported.

Thus, while a direct detection remains elusive, existing non-detections already restrict persistent r-modes to amplitudes well below many theoretical saturation predictions, making long-lived, r-mode--dominated spin-down scenarios increasingly unlikely unless mode amplitudes are extremely small or the emission is transient or intermittent \citep{ril23}.

In summary, while multiple continuous GW emission mechanisms may regulate the spin evolution of accreting NSs, distinguishing among them remains an open challenge. Progress will depend on both improved observational constraints --- such as tighter GW limits from the LVK Collaboration --- and refined theoretical models of NS structure, magnetic fields, and accretion physics.

\clearpage
\section{Large values of $\dot{P}$ for GC MSPs}\label{app:PdotGC}
The difference between measured values of $\dot{P}_{\rm obs}$ for MSPs in the Galactic field population and those belonging to a GC is usually attributed to dynamical and kinematic selection effects. % \citep{phi92,fcl+01,cr05,rgf+21}. 
The measured value of $\dot{P}_{\rm obs}$ of a given MSP is the sum of its intrinsic $\dot{P}$ value ($\dot{P}_{\rm int}$) and various kinematic corrections \citep{phi92}:
\begin{equation} 
\label{eq:Pdotcorr}
  \left( \frac{\dot{P}_{\rm obs}}{P} \right) =  \left( \frac{\dot{P}_{\rm int}}{P} \right)  +
             \frac{\mu^2 d}{c} + \frac{a_{\rm dyn}}{c} \;,
\end{equation}
where the first correction is the Shklovskii term ($d$ is the distance to the MSP, and $\mu$ is its proper motion related to its transverse velocity, $v_\perp =\mu d$) and $a_{\rm dyn}$ is the difference between the accelerations of the pulsar and the Solar System barycentre projected along the line-of-sight \citep[i.e. often expressed by a combination of vertical and Galactic differential rotational accelerations relative to the GC hosting the MSP and motion of the MSP within the GC potential,][]{prf+17}. 
It is for this reason that a trustworthy determination of low surface B-fields in MSPs is challenging since $B\propto \sqrt{\dot{P}_{\rm int}}$ (Eq.~\ref{eq:Bspitkovsky}). 

For most pulsars in the Galactic disc, $a_{\rm dyn}$ is not only relatively small, but can also be reasonably estimated from the pulsar distance estimate using a dynamical model for the Galaxy. 
While for the majority of Galactic-field MSPs the observed period derivative is dominated by the intrinsic spin-down, for nearby low-B-field MSPs, kinematic corrections --- particularly the Shklovskii effect, and in some cases Galactic acceleration --- can be comparable to or even exceed the intrinsic contribution and must therefore be taken into account \citep{lem+15}.

The environment is different in GCs. The large stellar densities in these clusters result in line-of-sight accelerations that contribute to $\dot{P}_{\rm obs}$ at a level far exceeding the values of $\dot{P}_{\rm int}$ one might reasonably expect for MSPs \citep{phi92}. As a consequence, for many dense GCs the values of $\dot{P}_{\rm obs}$ are completely unrelated to the spin-down (characteristic) age of the pulsar: approximately half of the pulsars (those on the far sides of their respective GCs) have negative $\dot{P}_{\rm obs}$; this is a clear indication that the cluster accelerations are dominant. These large GC accelerations cannot be reliably predicted, because the 3D location of the pulsar in the GC is generally unknown: only the positional offset in the plane of the sky is known. Thus, as a general rule, the values of $\dot{P}_{\rm obs}$ for pulsars in GCs should not be interpreted as intrinsic spin-down unless the cluster acceleration can be estimated or constrained.

However, there are cases where reasonable estimates of $\dot{P}_{\rm int}$ can be obtained:
\begin{enumerate}
\item [i)]Using a GC mass model, one can estimate extreme limits on the line-of-sight acceleration at the pulsar’s projected position, $a_{\max}$. If $\dot{P}_{\rm obs}/P \gg a_{\max}/c$, the excess must be intrinsic, since $(\dot{P}_{\rm obs}/P - a_{\max}/c) > 0$. This approach underlies most estimates of ``young'' pulsars in GCs \citep[e.g.,][]{arf+23,wpq+24} and can also constrain $\dot{P}_{\rm int}$ for MSPs in less dense clusters where predicted accelerations are small \citep{lpz+23,lfc+25,lpz+25,fdc+26}.
\item [ii)] For some binary pulsars with long timing baselines, $a_{\rm dyn}$ can be measured directly via the orbital period derivative, $\dot{P}_{\rm orb}$, particularly when GW damping is negligible. This method has yielded $\dot{P}_{\rm int}$ for several binaries in 47~Tucanae, as well as for NGC~1851A \citep{frk+17,dfg+25} and NGC~6752A \citep{cvf+23}, placing them securely within the MSP region of the $P\dot{P}$--diagram (Fig.~\ref{fig:PPdot}).
\end{enumerate}

Although most values of $\dot{P}_{\rm obs}$ for GC MSPs are unreliable, the first method discussed above demonstrates that some GC MSPs have genuinely large intrinsic spin-down rates, notably M28A and NGC~6624A. With spin periods of 3.05~ms and 5.44~ms, their inferred $\dot{P}_{\rm int}$ values are about two orders of magnitude larger than those of typical Galactic-field MSPs, implying characteristic ages of $\tau\sim 25\,\rm Myr$ and unusually large spin-down powers \citep[independently confirmed by their strong $\gamma$-ray emission,][]{faa+11,jkg+13}.
The inferred spin-down ages of these MSPs are thus far shorter than the $\sim 10\;{\rm Gyr}$ ages of their host clusters, posing a formation puzzle. 

In GCs, MSPs are generally produced via dynamically formed LMXBs, which explains the high abundance of MSPs and LMXBs per unit stellar mass compared to the Galactic disc. Once formed, most GC MSPs are otherwise similar to disc MSPs, as shown for 47~Tuc and NGC~1851A.
In the densest GCs, however, LMXBs are likely to be disrupted before recycling is complete \citep[][see also Sect.~\ref{subsec:multi-acc}]{vf14}, leaving behind partially recycled NSs with relatively strong B-fields. Consistent with this picture, such ``young'' pulsars are found only in the densest GCs and occupy a distinct region below the spin-equilibrium line in the $P\dot{P}$--diagram (Sect.~\ref{subsec:bands}).

To illustrate the relative importance of the various contributions, we provide a few numerical examples. Owing to the difference in Galactic acceleration along the line-of-sight, \cite{fcl+01} obtained for 47~Tuc a value of $a_{\rm Gal}/c = (-4.5 \pm 0.2)\times 10^{-19}\;{\rm s}^{-1}$, corresponding to a contribution to $\dot{P}_{\rm obs}$ of $\sim 10^{-21}$ for a 2~ms MSP.
The contribution from the Shklovskii effect is usually negligible compared to that arising from the acceleration within the cluster potential \citep{phi93,prf+17}. For example, for a MSP with a transverse velocity $v_\perp = 20\;\rm km\,s^{-1}$, a distance $d = 5\;{\rm kpc}$, and a spin period $P = 2\;{\rm ms}$, one finds $\dot{P}_{\rm shk} = 1.7\times 10^{-23}$, and thus $\dot{P}_{\rm shk} \ll \dot{P}_{\rm obs}$. For comparison, a typical cluster acceleration experienced by a MSP in 47~Tuc is at least $a_{\rm cluster} \sim {\rm few}\times 10^{-9}\;{\rm m\,s}^{-2}$ \citep{prf+17}, corresponding to a contribution of $\dot{P}_{\rm dyn} \sim \mathcal{O}(10^{-20})$ to $\dot{P}_{\rm obs}$.

To conclude, the high observed values of $\dot{P}_{\rm obs}$ for GC MSPs are generally dominated by dynamical and kinematic selection effects, but in dense GCs they may also reflect truncated recycling.

\clearpage
\section{Numerical examples of mass transfer}\label{app:RLO-numerical}
\subsection{RLO timescales for evolved donors}\label{app:RLO-timescale}
The thermal (Kelvin--Helmholtz) timescale of a subgiant or Hertzsprung-gap donor star with mass $2 \le M_2/M_\odot \le 4$ is shown in Table~\ref{table:KH}. This timescale, which depends on the envelope structure and mass ratio, often represents a lower limit to the duration of the mass-transferring X-ray phase, $\tau_X$, responsible for recycling the NS --- see discussions in Sect.~\ref{subsec:expected-spins}. For such relatively massive and evolved donors, $\tau_X \ll \tau_{\rm torque}$, meaning that recycling is inefficient; consequently, the final post-recycling spin period satisfies $P \gg P_{\rm eq}$.

\begin{figure}[ht]
\centering
\vspace*{-0.8cm}\hspace*{-0.4cm}
\includegraphics[width=0.86\textwidth]{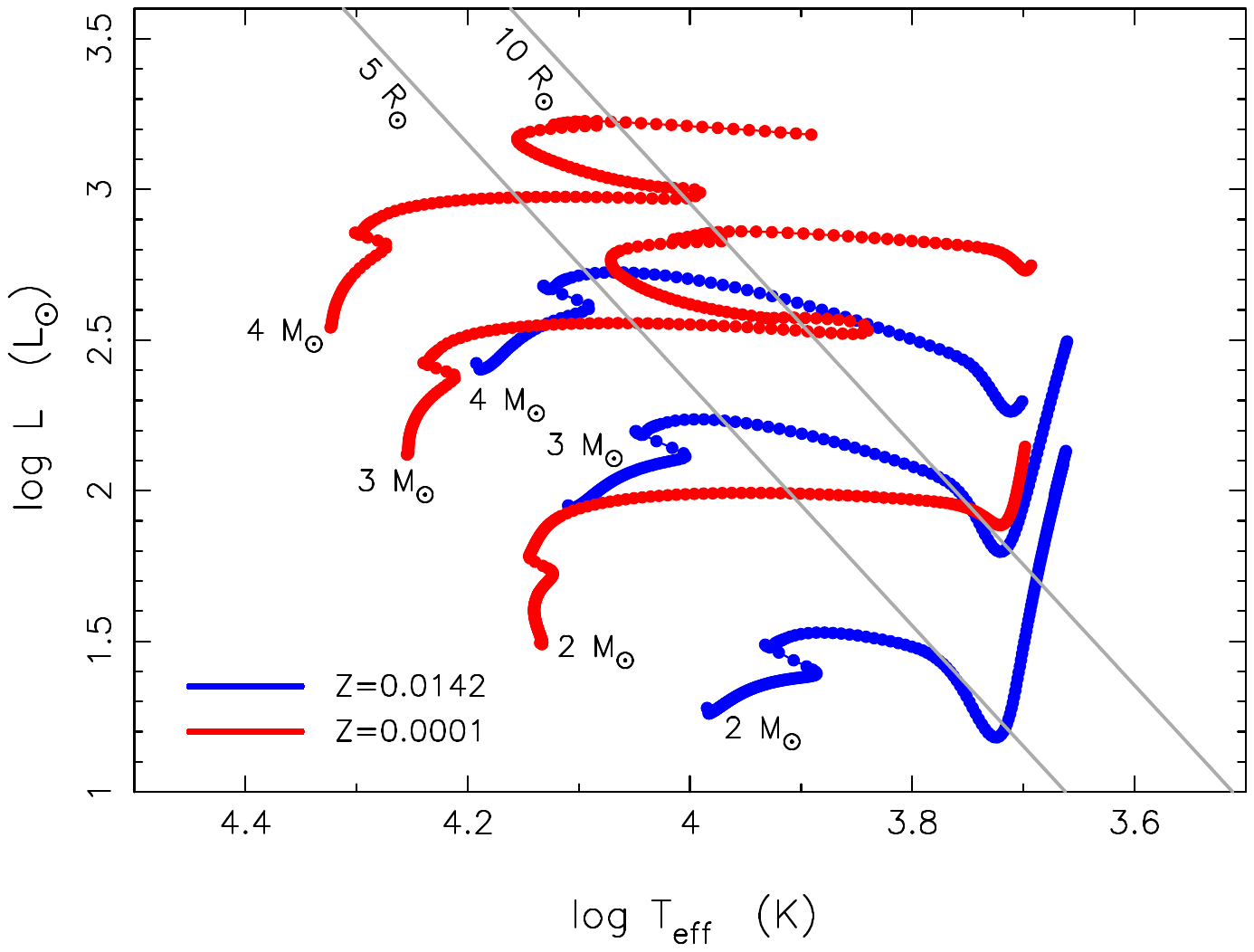} 
\caption{HR diagram of MESA-calculated evolutionary tracks. Six cases are shown for donor stars of masses $2-4\;M_\odot$ and metallicities $Z=0.0001$ and $Z=0.0142$, covering the transition from the subgiant to the Hertzsprung-gap (HG) phase. The combination of the wide metallicity range and the complex morphology of the evolutionary tracks across this transition results in a broad spread of Kelvin-Helmholtz timescales, $\tau_{\rm KH}$ (Table~\ref{table:KH}), in the region where $R\simeq5-10\;R_\odot$.}
\label{fig:HR}
\end{figure}

We calculated these $\tau_{\rm KH}$ values using single-star evolutionary models with MESA version r24.08.1 \citep{pbd+11,pca+13,pms+15,psb+18,pss+19,jbs+23}. We adopt the standard MESA prescriptions for convective mixing and overshooting. Convection is treated using the mixing-length theory with the default parameter $\alpha_{\rm MLT}=2.0$. Overshooting at convective boundaries is handled using the built-in exponential diffusive scheme in MESA, which yields realistic main-sequence core growth consistent with empirical constraints from detached intermediate-mass eclipsing binaries \citep[e.g.][]{cla17}.
Rotation, atomic diffusion, and semiconvection were not included. We also suppress stellar winds entirely, as mass loss is negligible for these masses prior to the red-giant phase.
We assumed, however, a broad range of metallicities between $Z=0.0001$ ($Y=0.249$) and $Z=0.0142$ ($Y=0.2703$), which is the dominant source of the large spread in the resulting values of $\tau_{\rm KH}$.
In most cases, $\tau_{\rm KH}\sim 0.01-1\;{\rm Myr}$ and thus  $\tau_{\rm KH}\lesssim \tau_X \ll \tau_{\rm torque}$. 
An HR~diagram of the calculated tracks is shown in Fig.~\ref{fig:HR}.

\begin{table*}[h!]
\centering
\begin{tabular}{ccccc}
\hline
\hline
\noalign{\vskip 1pt}
\multicolumn{5}{c}{$R=5\;R_\odot$}\\
\hline
\noalign{\vskip 1pt}
$M\;(M_\odot)$ & $T_{\rm eff}\;({\rm K})$ & $L\;(L_\odot)$ &  $\tau_{\rm KH}$ (Myr) & $P_{\rm orb}$ (d)\\
\hline

\noalign{\vskip 1pt}
2 & 5200--8100   & 16--98   & 0.123--0.810 & 2.68\\
3 & 9300--11400  & 170--360 & 0.077--0.117 & 2.07\\
4 & 12200--14600 & 520--940 & 0.051--0.097 & 1.72\\
\hline

\noalign{\vskip 15pt}
%\smallskip
\hline
\hline
\noalign{\vskip 1pt}
\multicolumn{5}{c}{$R=10\;R_\odot$}\\
\hline
\noalign{\vskip 1pt}
$M\;(M_\odot)$ & $T_{\rm eff}\;({\rm K})$ & $L\;(L_\odot)$ &  $\tau_{\rm KH}$ (Myr) & $P_{\rm orb}$ (d)\\
\hline
\noalign{\vskip 1pt}
2 & 4900--5600  & 50--87    & 0.073--0.127 & 7.57\\
3 & 5200--9400  & 64--720   & 0.019--0.255 & 5.86\\
4 & 8400--11800 & 440--1680 & 0.015--0.058 & 4.87\\
\hline

\end{tabular}
\smallskip\smallskip
\caption{Subgiant/HG donors of masses $2-4\;M_\odot$, and a wide spread in metallicities between $Z=0.0001-0.0142$, at fixed radius $R=5\;R_\odot$ (top) and $R=10\;R_\odot$ (bottom). The columns include $T_{\rm eff}$, luminosity $L$, Kelvin--Helmholtz time, 
$\tau_{\rm KH}=GM^2/(2RL)$, and
orbital period $P_{\rm orb}$ assuming that the donor fills its Roche lobe against a $1.4\,M_\odot$ NS. Note, for a given metallicity, $\tau_{\rm KH}$ varies only weakly with donor mass for a fixed $R$, because hotter (more massive) stars are more luminous.}
\label{table:KH}
\end{table*}

\clearpage
\subsection{MESA sample calculations of NS accretion in IMXBs}\label{app:RLO-MESA}
To illustrate the evolution of IMXBs and the associated NS accretion, we use here the standard binary evolution setup in MESA \citep[e.g.][]{pms+15} with a point-mass accretor and the {\em isotropic re-emission} model \citep{vd73,bv91,spv97,tv06,tv23}. The baryonic mass-accretion efficiency is given by $\epsilon =1-\alpha -\beta - \gamma$, where we use $\alpha=\delta=0$ for the fraction of material lost from the donor star in the form of a direct fast wind or via a co-planner circumbinary torus. $\beta$ denotes the fraction of transferred material that is subsequently ejected from the vicinity of the accreting compact object with its specific orbital angular momentum (here a NS).
Hence, the growth rate of the gravitational mass of the accreting NS is given by:
\begin{equation}\label{eq:NS-acc-rate}
  \dot{M}_{\rm NS,grav} = (1-\eta)\;\min\!\left[(1-\alpha-\beta-\delta)\,|\dot{M}_{\rm donor}|,\,\dot{M}_{\rm Edd} \right].
\end{equation}
For example, using $\eta=0.15$, $\alpha=\delta=0$, and $\beta=0.85$ means that the maximum growth rate of the gravitational mass of the NS is restricted to a maximum of $0.85\;\dot{M}_{\rm Edd}\simeq 1.3\times 10^{-8}\;M_\odot\,{\rm yr}^{-1}$.

From an analytical model we can calculate the change in orbital separation following \cite{tau96}:
\begin{equation}
  \frac{a}{a_0}= \left( \frac{q_0(1-\beta)+1}{q(1-\beta)+1} \right) ^{\textstyle\frac{3\beta-5}{1-\beta}}
                 \left( \frac{q_0+1}{q+1} \right) \left( \frac{q_0}{q} \right) ^2 \;.
\label{eq:aa0_ire}
\end{equation}
According to Kepler's third law: $P^2\,\propto\, a^3/M$, and hence we can rewrite the above equations
to yield the change in orbital period \citep[see also][]{kskd01,tv23}:
\begin{equation}
  \frac{P}{P_0}= \left( \frac{q_0(1-\beta)+1}{q(1-\beta)+1} \right) ^{\textstyle\frac{5\beta-8}{1-\beta}}
                 \left( \frac{q_0+1}{q+1} \right) ^2 \left( \frac{q_0}{q} \right) ^3 \;.
\label{eq:PP0_ire}
\end{equation}
The above analytical estimates include only the orbital angular-momentum losses associated with non-conservative mass transfer (isotropic re-emission). The MESA calculations, in contrast, also include tidal spin--orbit coupling and the associated exchange of angular momentum between the donor star and the orbit. As a result, the final orbital periods differ slightly from the pure analytical predictions, even when the same nominal mass-accretion efficiency is adopted.

\begin{figure}[ht]
\centering
\vspace*{-0.4cm}\hspace*{-0.4cm}
\includegraphics[width=0.9\textwidth]{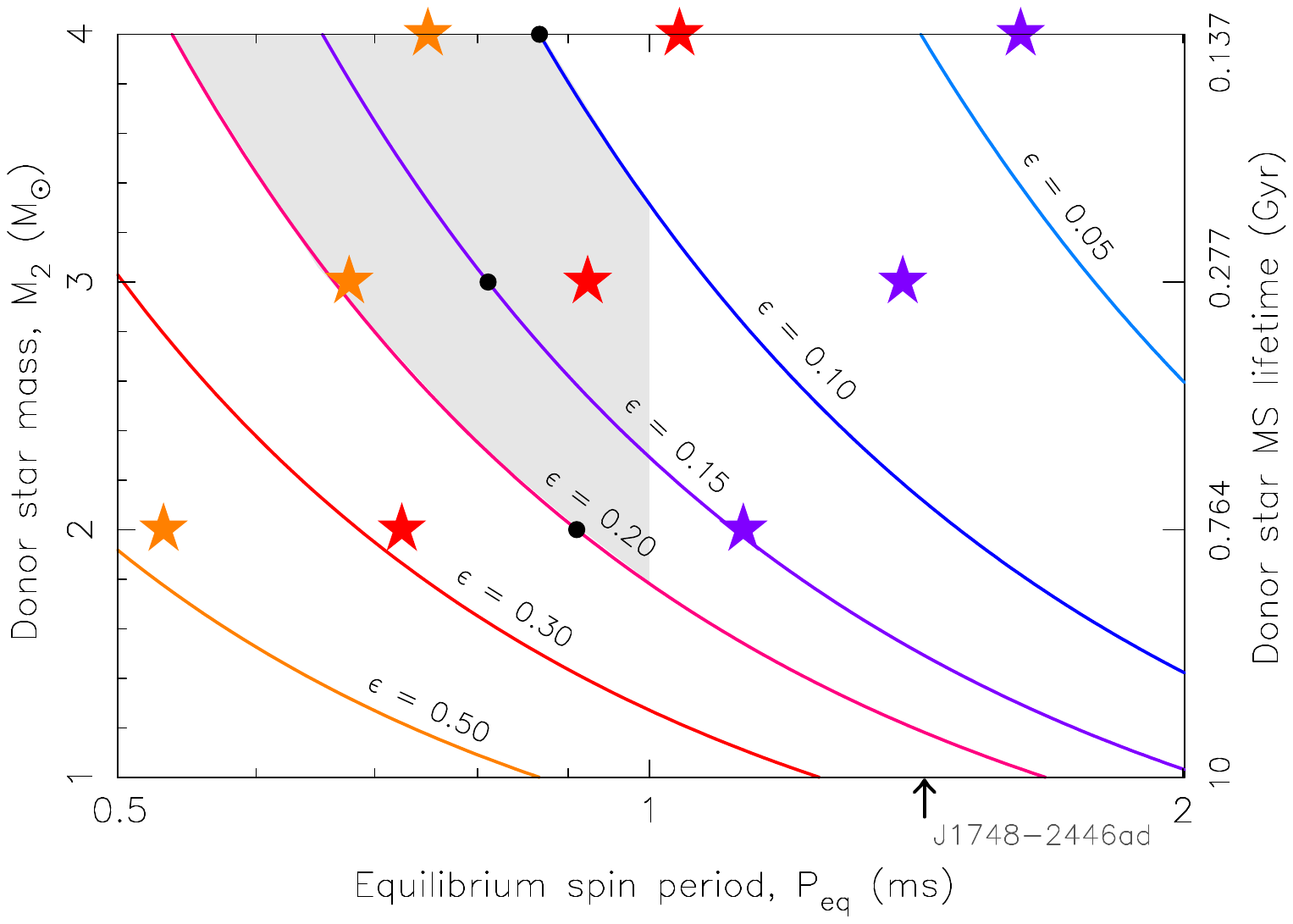} 
\caption{Donor star mass as a function of equilibrium spin period for a number of IMXB models using MESA (shown with star symbols) plotted on top of the analytical models presented in Fig.~\ref{fig:donor-mass}. In addition, three examples of analytical models are marked with black bullet points. Details for all shown models are provided in Table~\ref{table:MESA-IMXB}. 
The IMXB models were calculated with MESA-inlist baryonic accretion efficiencies of $\epsilon_{\rm MESA}=0.50$ (orange stars), $\epsilon_{\rm MESA}=0.30$ (red stars) and $\epsilon_{\rm MESA}=0.15$ (purple stars). However, as a result of super-Eddington mass transfer the {\em effective} mass accretion efficiencies ($\epsilon_{\rm eff}$) are significantly smaller. That explains the often much larger final NS spin periods for the MESA models compared to simple analytical models for $\epsilon=\epsilon_{\rm MESA}$. See text for discussion.}
\label{fig:donor-mass-MESA}
\end{figure}

In Table~\ref{table:MESA-IMXB} we list a number of numerical IMXB calculations with initial donor stars masses of $M_{\rm 2,0}=2-4\;M_\odot$, and MESA-inlist accretion efficiencies of $\epsilon_{\rm MESA}=0.15$, 0.30 and 0.50. 
For $M_{\rm 2,0}=2.0\;M_\odot$ and $3.0\;M_\odot$, the initial (pre-accretion) NS mass was $M_{\rm NS,0}=1.4\;M_\odot$. For $M_{\rm 2,0}=4.0\;M_\odot$, we adopted $M_{\rm NS,0}=1.7\;M_\odot$ to ensure orbital dynamical stability.
The calculated resulting mass-transfer rates are plotted in Fig.~\ref{fig:IMXB-mdot} and the overall outcome of the RLO is shown in Fig.~\ref{fig:donor-mass-MESA}.
Because the mass-transfer rate is highly super-Eddington ($|\dot{M}_2|\gg \dot{M}_{\rm Edd}$), especially for the 3.0 and $4.0\;M_\odot$ donor stars, the resulting {\em effective} mass-accretion efficiencies ($\epsilon_{\rm eff}$) are significantly smaller than $\epsilon_{\rm MESA}$, as a result of the NS mass-accretion rate being capped by the Eddington limit, cf. Eq.~(\ref{eq:NS-acc-rate}).

For example, to produce a sub-MSP, for $M_{\rm 2,0}=3.0\;M_\odot$ and $\epsilon_{\rm MESA}=0.30$, we obtained $\epsilon _{\rm eff}=0.125$ ($P_{\rm eq}\simeq 0.923\;{\rm ms}$). The fact that no sub-MSPs have been discovered so far could therefore suggest that $\epsilon _{\rm eff}$ may in general be less than $\sim 0.10-0.15$ (Fig.~\ref{fig:donor-mass-MESA}). 
We note, however, that the limiting value of $\dot{M}_{\rm Edd}\simeq {\rm a\;few}\; 10^{-8}\;M_{\odot}\,{\rm yr}^{-1}$ is traditionally evaluated under a number of simplified assumptions that may not be fulfilled, so that the true limiting accretion rate could be somewhat higher \citep{tv23}.

\begin{table*}
\centering
\begin{tabular}{r|ccccccc|ccc}
\hline\hline
           & $M_{\rm 2,0}$ & $M_{\rm NS,0}$ & $P_{\rm orb,0}$ & $M_{\rm 2,f}$  & $M_{\rm NS,f}$ & $P_{\rm orb,f}$ & $\epsilon_{\rm MESA}$   & $\epsilon_{\rm eff}$ & $\epsilon$ & $P_{\rm eq}$ \\
\hline
Analytical & 2.0           & 1.4            & 1.0             & 0.25           & 1.70           & 52.1            & --                      & --                   & 0.20       & 0.910 \\
MESA       & 2.0           & 1.4            & 1.0             & 0.277          & 2.05           & 35.7            & 0.50                    & 0.443                & --         & 0.532 \\
MESA       & 2.0           & 1.4            & 1.0             & 0.278          & 1.81           & 37.4            & 0.30                    & 0.282                & --         & 0.725 \\
%MESA       & 2.0           & 1.4            & 1.0             & 0.279          & 1.75           & 37.6            & 0.25                    & 0.240                & --         & 0.811 \\
MESA       & 2.0           & 1.4            & 1.0             & 0.279          & 1.62           & 37.9            & 0.15                    & 0.150                & --         & 1.13 \\
\hline
Analytical & 3.0           & 1.4            & 1.0             & 0.25           & 1.75           & 43.5            & --                      & --                   & 0.15       & 0.811 \\
MESA       & 3.0           & 1.4            & 1.0             & 0.260          & 1.85           & 26.1            & 0.50                    & 0.195                & --         & 0.677 \\
%MESA       & 3.0           & 1.4            & 1.0             & 0.262          & 1.73           & 26.9            & 0.35                    & 0.143                & --         & 0.841 \\
MESA       & 3.0           & 1.4            & 1.0             & 0.261          & 1.69           & 27.6            & 0.30                    & 0.125                & --         & 0.923 \\
MESA       & 3.0           & 1.4            & 1.0             & 0.264          & 1.56           & 28.4            & 0.15                    & 0.071                & --         & 1.39 \\
\hline
Analytical & 4.0           & 1.4            & 1.0             & 0.25           & 1.72           & 22.1            & --                      & --                   & 0.10       & 0.867 \\
MESA       & 4.0           & 1.7            & 1.0             & 0.263          & 2.11           & 33.4            & 0.50                    & 0.130                & --         & 0.750 \\
MESA       & 4.0           & 1.7            & 1.0             & 0.266          & 1.96           & 35.0            & 0.30                    & 0.082                & --         & 1.04 \\
MESA       & 4.0           & 1.7            & 1.0             & 0.267          & 1.84           & 36.9            & 0.15                    & 0.045                & --         & 1.62 \\
\hline
\end{tabular}
\smallskip\smallskip
\caption{Analytical and numerical (MESA) calculations of IMXB systems with initial donor star masses of $M_{\rm 2,0}=2-4\;M_\odot$ and various assumed MESA-inlist baryonic mass-accretion efficiencies ($\epsilon_{\rm MESA}$). See text for further information and the resulting {\em effective} mass-accretion efficiencies, $\epsilon_{\rm eff}$.}
\label{table:MESA-IMXB}
\end{table*}

\clearpage
\section{Estimating radio sub-MSP selection effects}\label{app:radio-selection}
In this section, we explain in detail the methods that went into understanding the selection effects against radio MSPs discussed in Section \ref{sec:selection-effects}. Our approach, as discussed in the main text, is to provide these selection effects purely as a degradation in the S/N of the pulsar, as a function of observable parameters such as spin period and dispersion measure. As part of this publication, we provide a complete posterior distribution of our work, which can be scaled with the sub-MSP luminosity distribution and the telescope of the reader's choice to convert these into physical units. 

Throughout this exercise, we use real astrophysical data obtained as part of the MMGPS UHF and L-band survey with the MeerKAT telescope. 
This survey acquired total-intensity filterbank data with integration times of 9.2~min at L-band and 8.2~min at UHF-band. The data comprise 2048 frequency channels spanning bandwidths of 856~MHz centred at 1284~MHz (L-band) and 544~MHz centred at 816~MHz (UHF-band), with corresponding sampling times of 153~$\mu$s and 120~$\mu$s, respectively.
We inject 40\,000 pulsars into these data with varied distribution of parameters (see main text). To quantify the radio search selection effects, we aim to mimic real search pipelines to capture all the potential biases.
After injection, the filterbanks were processed in two ways, considering the most recent surveys and search techniques available. The subsequent methodology is the same for both L-band and UHF-band data.

\smallskip
The first method used the standard MMGPS search pipeline. This included the following steps:
\begin{itemize}
    \item Radio-frequency interference mitigation using the standard \textsc{PulsarX} \citep{PulsarX} \texttt{filtool}.
    \item Acceleration searching with \textsc{peasoup} at the DM and acceleration ranges in Table~\ref{table:inject} with a DM step size determined by \textsc{DDplan.py} from the \textsc{presto} software.
    % \item Sifting the candidates from injected filterbanks with real candidates from all the other beams of that epoch, obtained from the real time search of the data
    \item The resulting candidate set is cross matched with the injected parameters. All successfully detected injections were then folded using \textsc{PulsarX} \texttt{psrfold\_fil} to measure the recovered S/N. The folding algorithm computes the rms noise directly from the unfolded filterbank data and therefore does not rely on off-pulse windows.
    \item Finally, our AI classifier was applied to the resulting folded diagnostic plots.  
\end{itemize}
\noindent The final detection fractions can be found in the main text for this method. 

\begin{figure}[htbp]
\centering
\includegraphics[width=0.95\textwidth]{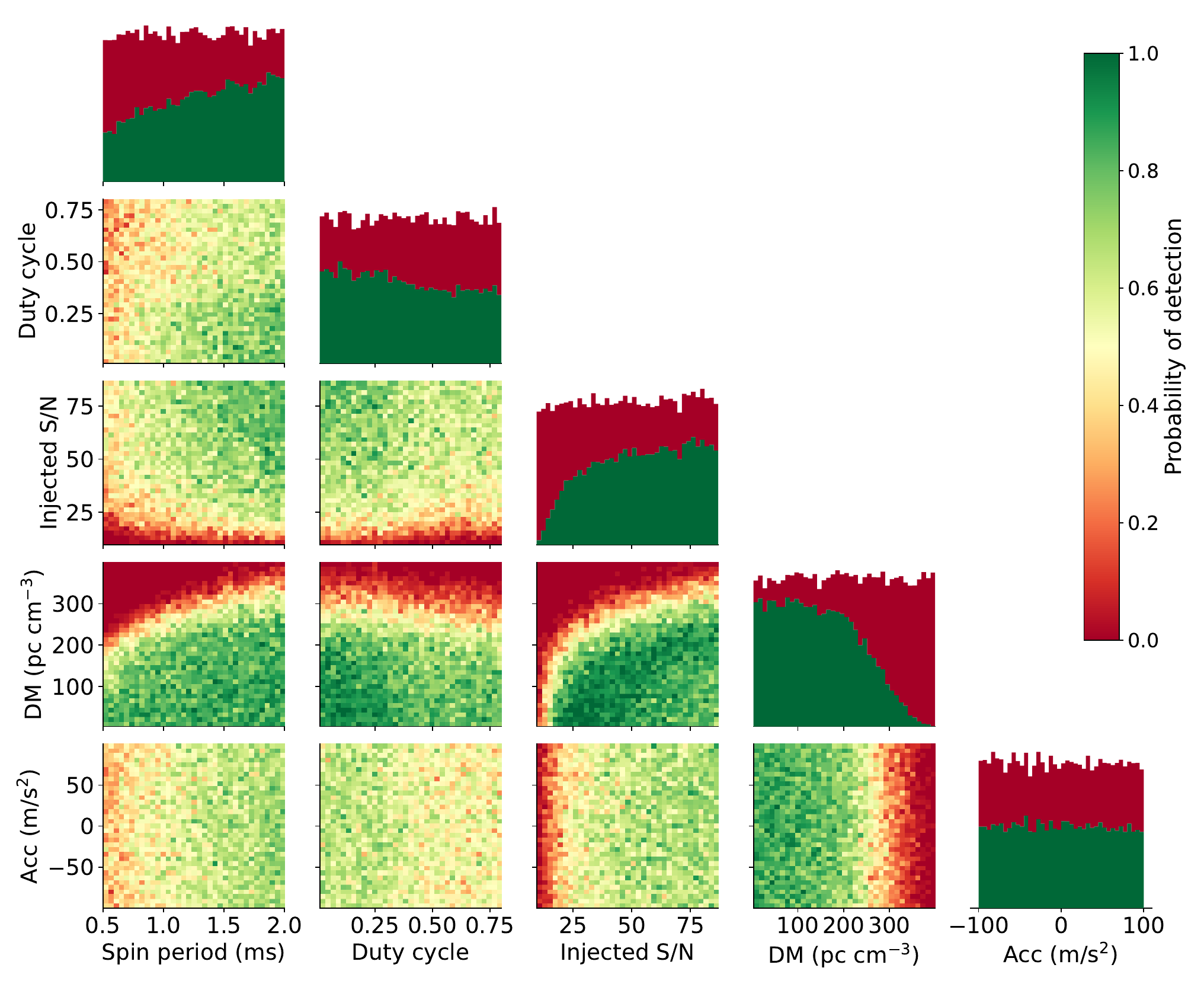}
\caption{Corner plot showing the results of the injection test performed at L-band using the \textsc{presto} software. The diagonal histograms show the distribution of detected (green) and non-detected (red) pulsars. The off-diagonal 2D panels are colour-coded based on the probability of detecting the pulsar by the search pipeline.}
\label{fig:subms-Lband-PRESTO}
\end{figure}

\smallskip
For the second method the injected pulsars were searched again using a \textsc{presto}-based search pipeline. The results are shown in Fig.~\ref{fig:subms-Lband-PRESTO}. This pipeline included the following steps: 

\begin{itemize}
    \item Radio-frequency interference mitigation using the standard \textsc{presto} \texttt{rfifind}.
    \item Acceleration searching with \textsc{presto} \textsc{accelsearch} using a maximum $Z$-value of 200 (see discussion below). This value is chosen because most of the surveys to date have chosen to process the data with this value. The DM ranges are given in Table~\ref{table:inject} with a DM step size determined by \textsc{DDplan.py} from the \textsc{presto} software.
    \item The resulting candidate set is cross matched with the injected parameters. All successfully detected injections were then folded using \textsc{presto} \texttt{prepfold} to measure the recovered S/N.
    \item Finally, our AI classifier was applied again to the resulting folded diagnostic plots.  
\end{itemize}

One characteristic difference in the \textsc{presto}-based search, as opposed to the previous methodology, is that \textsc{presto} uses a frequency-domain resampling technique. 
In this technique, the pulsar acceleration is corrected in the Fourier domain by searching over a fixed number of Fourier frequency bins that account for the drift of signal power across the spectrum (the $Z$-value). 
The range of acceleration searched is not constant, but increases with increasing pulse period. For $Z=200$, this results in the technique being less sensitive to compact binaries in the sub-ms spin period regime. The fraction of pulsars lost as a function of $Z$ can be seen in Fig.~\ref{fig:subms-UHF-presto}.

For longer integrations, higher-order Doppler terms can be partly absorbed with a ``jerk'' search \citep{ar18}, while for observations spanning many orbital cycles ($t_{\rm obs}\gg P_{\rm orb}$), the orbital modulation can instead be recovered with sideband or phase-modulation searches \citep{jcm02,rce03}. Owing to their computational expense and restricted regimes of applicability, these more advanced techniques have so far been used almost exclusively for targeted observations, most notably of globular clusters and unidentified \textit{Fermi} sources \citep{rce03,ar18}. These algorithms have almost exclusively been implemented in \textsc{presto} as Fourier-domain corrections. Hence, the actual range searched for sub-MSPs remains quite modest, since they suffer from the same reduction in parameter space searched as acceleration searches (Sect.~\ref{subsec:scaling-eqn}).

\subsection{Sensitivity to the assumed scattering prescription}\label{subsec:scattering_prescription}
The scattering prescription adopted throughout this work follows the empirical $\tau$-DM relation presented by \cite{TPA-scattering}, which is a frequency-rescaled version of the relation derived by \cite{krishnakumar2015scatter}. While this prescription provides a practical description of pulse broadening, the observed $\tau$-DM relation exhibits substantial intrinsic scatter, and more recent studies have found evidence for variations between different Galactic environments \citep{he2024galactic}. 
This uncertainty is particularly important for sub-MSP searches because pulse broadening rapidly suppresses the harmonic content of the signal. In the Fourier domain, the exponential scattering tail acts as a low-pass filter, with significant S/N losses occurring already for $\tau_{\rm sc}\gtrsim P/2\pi$ --- corresponding to only of order $100\;\mu$s for a sub-MSP.

\begin{figure}[htbp]
\centering
\includegraphics[width=0.95\textwidth]{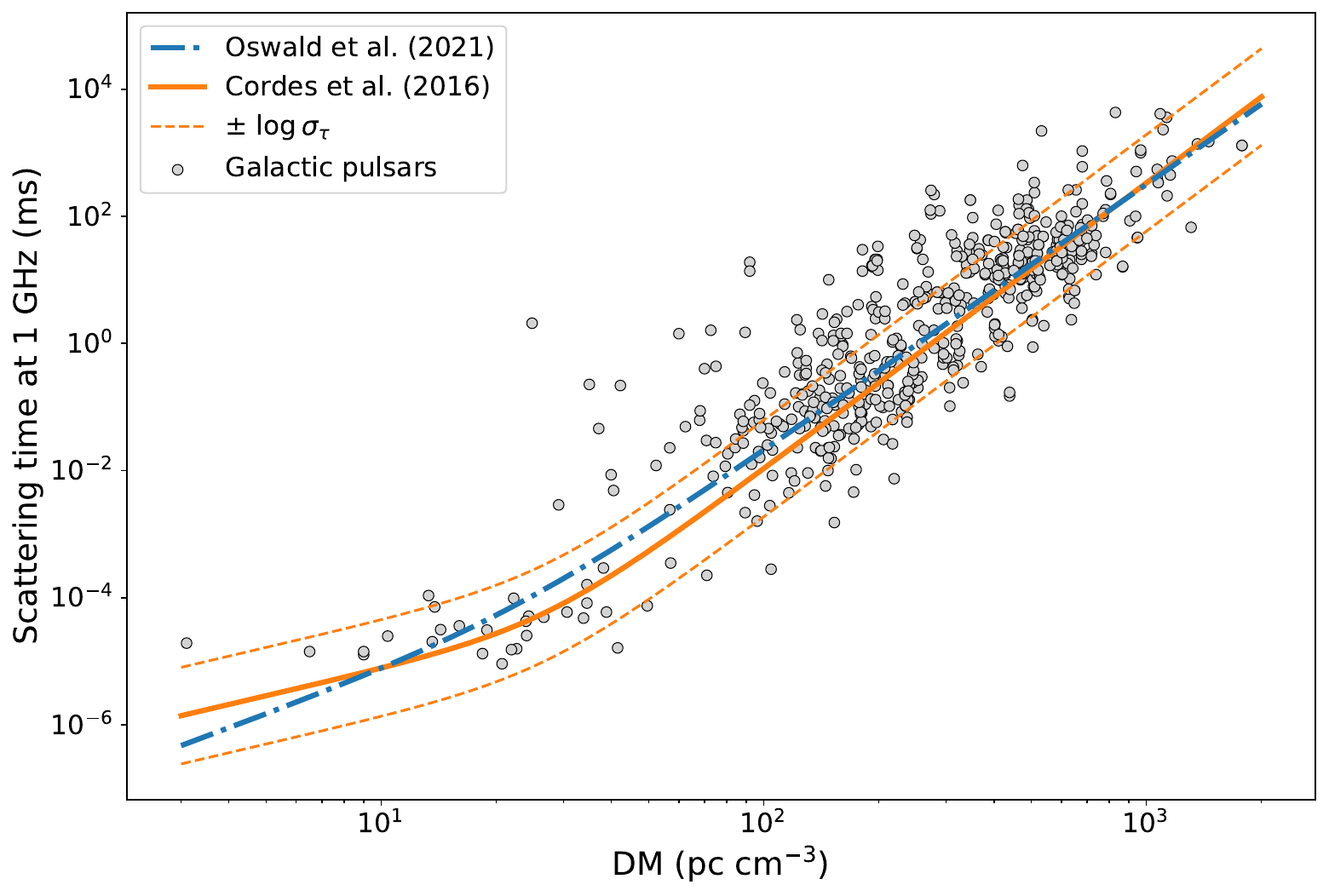}
\caption{Published measurements of pulse broadening time as a function of dispersion measure (DM), together with the empirical $\tau$--DM relations of \citet{cordes2016radio} and \citet{TPA-scattering}. The dashed orange curves indicate the $\pm1\sigma$ in $\log\tau$ scatter about the \citet{cordes2016radio} relation, illustrating the range of scattering times considered in the injection-recovery sensitivity analysis. The substantial scatter in the observational measurements at a given DM highlights the uncertainty associated with empirical scattering prescriptions.}
\label{fig:scattering_psr_population}
\end{figure}

\begin{figure}[htbp]
\centering
\includegraphics[width=0.95\textwidth]{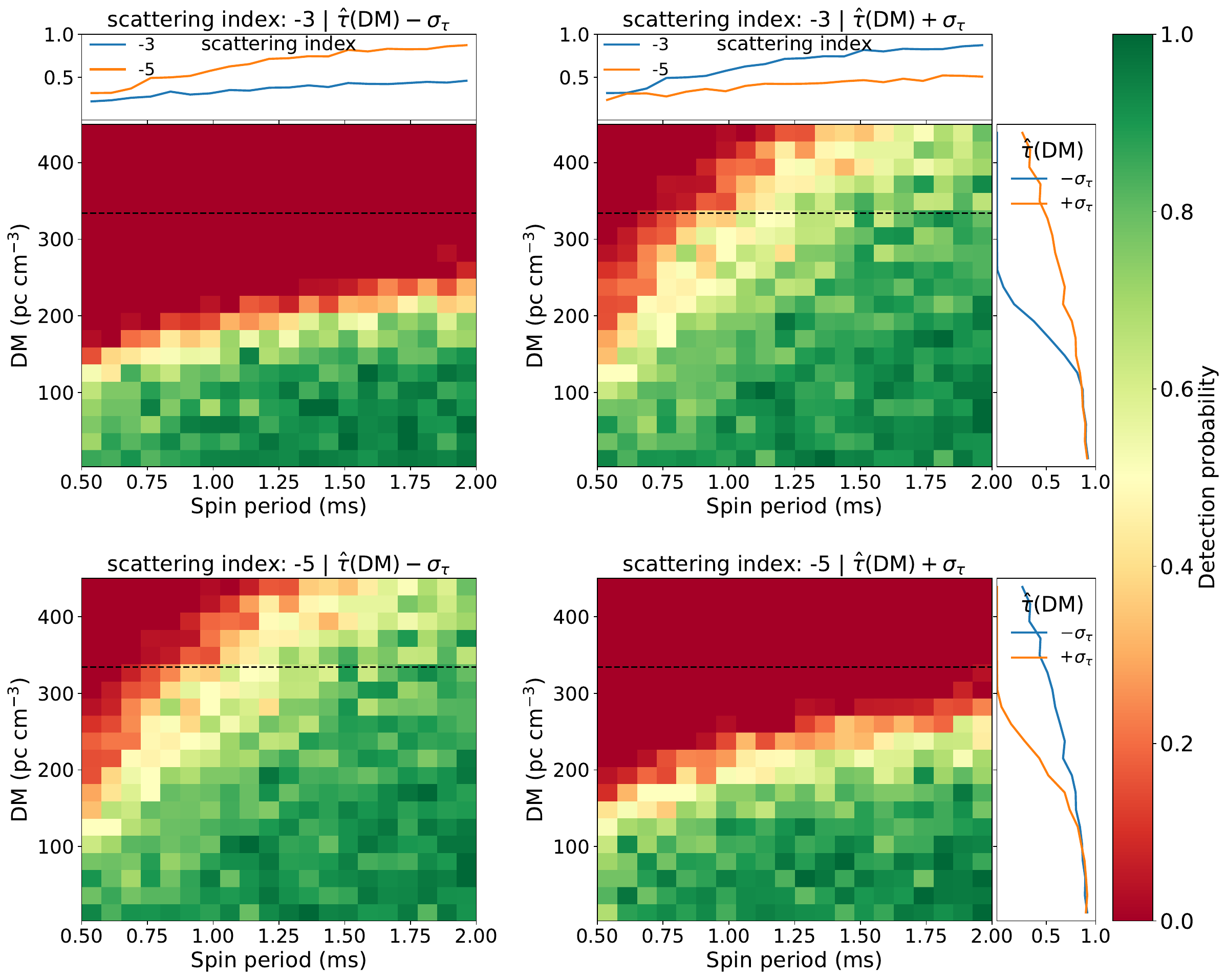}
\caption{Detection probability as a function of injected spin period and DM for four assumed scattering prescriptions at L-band. The left and right columns correspond to scattering times one standard deviation below and above the nominal empirical $\tau$-DM relation, respectively, while the upper and lower rows adopt scattering frequency scalings of $\nu^{-3}$ and $\nu^{-5}$. The marginalized detection probability curves above each column compare the two scattering indices for a fixed $\tau$-DM prescription, while the curves to the right compare the two $\tau$-DM prescriptions for a fixed scattering index. The
dashed black lines show where the data is downsampled in time by a factor of two before the search.}
\label{fig:scattering_comp}
\end{figure}

Fig.~\ref{fig:scattering_psr_population} compares published measurements of pulse-broadening time as a function of DM (\citealt{cordes2016radio}; \citealt{TPA-scattering}, and references therein) with the empirical $\tau$--DM relations of \cite{cordes2016radio} and \cite{TPA-scattering}. The substantial scatter in the observational measurements at a given DM highlights the uncertainty associated with empirical scattering prescriptions. Throughout this work, we adopt the relation of \cite{TPA-scattering} as the baseline scattering model.

To assess the sensitivity of our survey predictions to this assumption, we repeated the injection-recovery analysis using four representative scattering prescriptions. Specifically, we considered scattering frequency scalings of $\nu^{-3}$ and $\nu^{-5}$, and for each scaling adopted scattering times corresponding to the nominal empirical relation of \cite{cordes2016radio}, offset by $\pm1\sigma$ in $\log\tau$. The resulting detection probability maps are shown in Fig.~\ref{fig:scattering_comp}, together with the corresponding marginalized detection efficiencies as functions of spin period and dispersion measure.

These simulations provide an estimate of the systematic uncertainty associated with the adopted scattering prescription and demonstrate how the survey sensitivity varies across a representative range of plausible scattering models.
In addition, estimates of the scattering horizon are further complicated by the limited accuracy of Galactic electron-density models \citep{cl02,ymw17}, whose predictions of $\tau_{\rm sc}$ along individual sightlines carry comparable order-of-magnitude uncertainties.

\subsection{Sensitivity in GC searches}\label{subsec:globular_cluster_search}
\begin{figure}[htbp]
\centering
\includegraphics[width=0.95\textwidth]{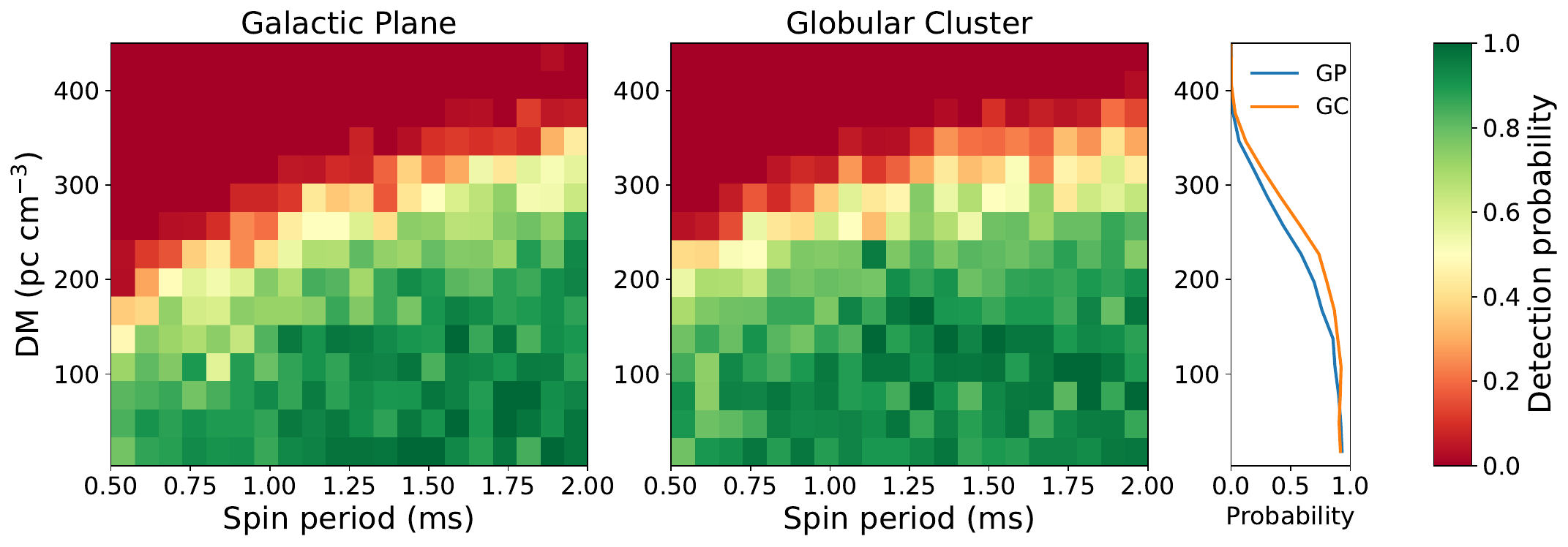}
\caption{Detection probability for coherently dedispersed observations of GCs compared with the Galactic-plane survey configuration. In the GC case, the known cluster DM allows coherent dedispersion to remove intra-channel dispersive smearing, leaving interstellar scattering as the dominant source of pulse broadening.}
\label{fig:gc_search}
\end{figure}

The injection-recovery analysis presented throughout this work assumes the observing configuration of the Galactic-plane survey, in which the search is performed over a wide range of trial dispersion measures using incoherently dedispersed filterbank data. Under these conditions, pulse broadening arises from both interstellar scattering and residual intra-channel dispersive smearing.

For targeted observations of GCs, however, the cluster DM is generally well constrained from previously known pulsars. The data can therefore be coherently dedispersed at the cluster DM, effectively eliminating intra-channel dispersive smearing. In this case, the dominant source of pulse broadening is interstellar scattering.

To assess the impact of coherent dedispersion on the detectability of sub-MSPs, we repeated the injection-recovery analysis assuming that intra-channel dispersive smearing is absent, while retaining the same scattering prescription adopted throughout this work. Fig.~\ref{fig:gc_search} compares the resulting detection probabilities with those obtained for the Galactic-plane search.

The results show only a modest improvement in sensitivity for the coherently dedispersed case. While the absence of intra-channel smearing extends the detectable DM range for the shortest-period pulsars, the overall detection horizon is negligibly increased because pulse broadening is dominated by interstellar scattering over most of the parameter space considered. These results indicate that, for the observing configurations used in this work, scattering rather than residual dispersive smearing is the principal limitation on the detectability of sub-MSPs, even in targeted GC searches.

\subsection{Numerical examples and derivation of scaling equation}\label{subsec:scaling-eqn}
A note on the interpretation of the $Z$-value. 
The theory of Fourier response of binary pulsar signals and the role of higher harmonics in detectability has been discussed in detail by, e.g. \cite{ran01,fsk+04}, with further studies on harmonic sensitivity presented in \cite{blw13}.
As an example, $Z=200$ means allowing for the pulsar signal to drift by up to $\pm 200$ Fourier frequency bins during the observation due to line-of-sight acceleration (i.e. Doppler smearing). 
If detectability requires sensitivity to the $n$-th harmonic of the pulsar spin frequency, the effective Fourier frequency is increased by a factor $n$, and the corresponding maximum line-of-sight acceleration (edge-on orbit) is given by \citep{ran01}: 
\begin{equation}\label{eq:a_max}
    a_{\max}= \frac{Z\,c\,P_{\rm spin}}{n\,T_{\rm obs}^2} \;,
\end{equation}
where $c$ is the speed of light in vacuum, $P_{\rm spin}$ is the pulsar spin period, $T_{\rm obs}$ is the integration time, and $n$ denotes the highest harmonic required for detectability. 
Assuming typical (sub)MSP duty cycles of $\delta = 0.10-0.20$ correspond to summing $n_{\rm eff}\sim 1/\delta \simeq 4-8$ harmonics \citep{lk12}.
If we conservatively sum up to $n=16$ harmonics \citep[following e.g. the MMGPS survey,][]{mmgps} and consider, as an example, a 1~ms pulsar with $Z=200$, this results in $a_{\max}=13\;{\rm m\,s}^{-2}$ for $T_{\rm obs}=9\;{\rm min}$ ($a_{\max}=42\;{\rm m\,s}^{-2}$ for $T_{\rm obs}=5\;{\rm min}$, and $a_{\max}=1040\;{\rm m\,s}^{-2}$ for $T_{\rm obs}=1\;{\rm min}$).

The line-of-sight acceleration of the pulsar in a circular orbit varies with orbital phase as:
\begin{equation}
   a_{\rm los}(t) = a_1\,\Omega^2\,\sin(\Omega t + \phi_0) \;,
\end{equation}
where $\Omega = 2\pi/P_{\rm orb}$ is the orbital angular frequency, $\phi_0$ is an arbitrary phase constant, and:
\begin{equation}
  a_1 = \frac{\mu}{M_1}\,a\,\sin i = \frac{M_2}{M}\,a\,\sin i \; ,
\end{equation}
is the semi-major axis of the pulsar’s barycentric orbit projected along the line-of-sight ($M_1$ is the pulsar mass, $M_2$ is the companion star mass, $\mu =(M_1 M_2)/M$ is the reduced mass (with $M=M_1+M_2$), $a$ is the semi-major axis of the relative orbit, and $i$ is the orbital inclination angle). The maximum line-of-sight acceleration is therefore:
\begin{equation}\label{eq:alosmax}
  |a_{\rm los}|_{\max} = a_1\,\Omega^2 \; ,
\end{equation}
which occurs at conjunction for an edge-on orbit ($\sin i=1$).

Combining this expression with Eq.~(\ref{eq:a_max}) and Kepler's third law ($\Omega=\sqrt{GM/a^3}$), one obtains:
\begin{equation}
    P_{\rm orb}=2\pi\;\left(\frac{n}{Z\,c\,P_{\rm spin}}\right)^{3/4}\,\left( \frac{G\,M_2^3}{M^2} \right) ^{1/4} T_{\rm obs}^{3/2} \; .
\end{equation}

A constant-acceleration search is fundamentally limited by the amount of Doppler smearing that accumulates during the observation. Over long integrations, even modest line-of-sight accelerations can cause the pulsar signal to drift significantly in Fourier space, reducing detectability. Shorter observations accumulate less Doppler smearing and therefore tolerate much higher accelerations. As a result, short integrations are better suited for detecting MSPs in very tight binaries, despite their reduced S/N. Conversely, long integrations are optimal for isolated pulsars or wide binaries, but become increasingly incomplete for compact systems with short orbital periods.

A detection in a constant-acceleration search requires that the instantaneous line-of-sight orbital acceleration approximately satisfies $|a_{\rm los}(t)| \le a_{\max}$. 
Applying $T_{\rm obs}=9\;{\rm min}$, $Z=200$ and $n=16$, searching for a $1.7\;M_\odot$ MSP with $P_{\rm spin}=1\;{\rm ms}$ located in a circular binary with a $0.20\;M_\odot$ He~WD companion thus corresponds to a limiting orbital period of $P_{\rm orb}\simeq 6.0\;{\rm hr}$. Hence, for $P_{\rm orb}<6.0\;{\rm hr}$ it can be difficult to detect this MSP over most orbital phases, unless observed near quadrature or in a more face-on orbit.
For comparison, if $P_{\rm spin}=0.7\;{\rm ms}$ then the limiting orbital period would be about 7.8~hr. 
An identical constraint is obtained for alternative choices
($Z=100$, $n=8$) or ($Z=50$, $n=4$), reflecting the degeneracy between $Z$ and
the number of harmonics summed.
Hence, it is clear that fast-spinning MSPs with small orbital periods occupy a parameter space where most surveys are very insensitive to detecting sub-MSPs.

There are, however, exceptions where radio pulsar surveys are conducted with shorter integration times, including the dedicated FAST Galactic Plane Pulsar Snapshot Survey \citep[GPPS,][]{hww+21} which uses $\sim 5\;{\rm min}$ pointed integrations (and a sensitivity
of a few $\mu {\rm Jy}$ for MSPs). Besides FAST, several radio pulsar surveys have employed short snapshot or drift-scan integrations, including campaigns with Arecibo, GBT, Parkes, and MeerKAT, motivated in part by improved sensitivity to compact binary MSPs. The high sensitivity of FAST in such a short integration time is particularly well suited to detecting MSPs in compact orbits. An extreme case is the CRAFTS transit (drift-scan) survey with FAST \citep{dwq+18} for which $T_{\rm obs}=20\;{\rm s}$. Here, sub-MSPs suffer far less Fourier drift than in long pointings.
The limiting orbital periods for the same binary considered above (applying again $Z=200$, $n=16$, and $P_{\rm spin}=1.0\;{\rm ms}$) are $P_{\rm orb}\simeq 2.5\;{\rm hr}$ ($T_{\rm obs}=5\;{\rm min}$), $P_{\rm orb}\simeq 13.3\;{\rm min}$ ($T_{\rm obs}=1\;{\rm min}$), and $P_{\rm orb}\simeq 2.6\;{\rm min}$ ($T_{\rm obs}=20\;{\rm s}$).

Note, all estimates assume the constant-acceleration approximation. However, for $T_{\rm obs}\gtrsim 0.1\;P_{\rm orb}$ (depending on orbital phase), jerk effects \citep{ar18} will further reduce the detectability.

Finally, we may ask for the minimum spin period for secure detection of a MSP in a tight orbit with $P_{\rm orb}=15\;{\rm min}$ (assuming again the same component masses, $Z=200$, $n=16$, and $T_{\rm obs}=1\;{\rm min}$). Here we find $P_{\rm spin}\simeq 0.85~\;{\rm ms}$. This estimation is relevant for tight-orbit radio MSP binaries before they become ultra-compact X-ray binaries (when the WD fills its Roche lobe) due to GW damping \citep{tau18} --- thereby opening for interesting multi-messenger synergies with combined radio observations and GW detection with LISA --- here corresponding to $f_{\rm GW}\simeq 2.2\;{\rm mHz}$.

\clearpage
\section{Nature of MSP companions}\label{app:companion-nature}
The observational data of radio MSPs used in this review are taken from the ATNF Pulsar Catalogue, version~2.7.0 accessed January~2026 \citep[][{\url{https://www.atnf.csiro.au/research/pulsar/psrcat}}]{mhth05}. To identify the nature of the companion stars, we use the \texttt{BinComp} parameter from the catalogue, which follows the definitions of \citet{tlk12}. For producing Fig.~\ref{fig:spiders}, we supplemented the catalogue with 16 newly discovered MSPs and updated the \texttt{BinComp} classifications for 58~MSPs (Table~\ref{table:BinComp}).
An example of a missing companion type classification in the catalogue is PSR~J1748$-$2446ad -- the fastest spinning known MSP with $P=1.396\;{\rm ms}$ --- which undergoes radio eclipses and has a low-mass main-sequence star (MS) companion, i.e. a redback system\footnote{Redbacks typically have MS companions with masses of $\sim 0.1-0.7\;M_\odot$ and thus lower masses than in the original definition by \citet{tlk12}.}. Black-widows are ultra-light (UL) companions with masses $<0.08\;M_\odot$. 
Most binary MSPs have He WD companions (He), while a smaller number have more massive WD companions (CO), which may be either CO or ONeMg WDs.

Note: In a small number of borderline cases, the \texttt{BinComp} classification necessarily involves some degree of subjective judgement, based on companion-mass constraints, inclination-angle probabilities, the presence or absence of radio eclipses, and other relevant observational information.

\begin{table*}
\centering
\begin{tabular}{cc@{\hspace{1.5cm}}cc}
\hline\hline
Pulsar name & \texttt{BinComp} &
Pulsar name & \texttt{BinComp} \\
\hline
B0021-72E & He & J1737-0314D & MS \\
J0024-7204V & MS & J1737-0314E & MS \\
J0125-2327 & He & J1744-2946 & He \\
J0212+5321 & MS & J1748-2021F & He \\
J0307+7443 & He & J1748-2021H & UL \\
J0406+3039 & He & J1748-2446X & He \\
J0646-5455 & He & J1748-2446Z & He \\
J0653+4706 & He & J1748-2446ad & MS \\
J0838-2827 & MS & J1748-2446ap & He \\
J0843+67 & UL & J1748-2446ar & MS \\
J0955-6150 & He & J1748-2446at & UL \\
J1124-3653 & UL & J1748-2446au & He \\
J1302-3258 & MS & J1748-2446av & He \\
J1312+1810B & He & J1748-2446ax & He \\
J1312+1810D & He & J1750-3703B & He \\
J1312+1810E & He & J1803-4719 & He \\
J1326-4728G & UL & J1823-3021F & MS \\
J1326-4728H & UL & J1824-2452C & He \\
J1326-4728K & UL & J1824-2452H & MS \\
J1326-4728L & UL & J1824-2452I & MS \\
J1342+2822E & He & J1824-2452K & He \\
J1342+2822F & He & J1826-0049 & MS \\
J1402+1306 & He & J1833-3840 & UL \\
J1518+0204F & He & J1946-5403 & UL \\
J1518+0204G & UL & J1947-1120 & He \\
J1546-3747A & MS & J1953+1846E & UL \\
J1551-0658 & He & J2045-6837 & MS \\
J1602-1009 & UL & J2116+1345 & He \\
J1627+3219 & UL & J2133-0049B & He \\
J1647-0156A & He & J2133-0049C & He \\
J1647-0156B & He & J2133-0049D & He \\
J1657-0406A & He & J2133-0049E & He \\
J1657-0406B & He & J2133-0049F & He \\
J1701-3006H & UL & J2133-0049G & UL \\
J1701-3006I & He & J2144-5237 & He \\
J1717+4308B & He & J2150-0326 & He \\
J1720-0533 & UL & J2205+6012 & He \\
\hline
\end{tabular}
\smallskip\smallskip
\caption{Updated \texttt{BinComp} classifications for 74 MSPs that were either undefined in the \textit{ATNF Pulsar Catalogue} or reclassified here for Fig.~\ref{fig:spiders} according to the criteria of \citet{tlk12}. See the main text for further information on the companion types He, MS, and UL.}
\label{table:BinComp}
\end{table*}

\end{appendices}

%%%%%%%%%%%%%%%%%%%%%%%%%%%%%%%%%%%%%%%%%%%%%%%%%%%%%%%%%%%%%%%%%%%%%%%%%%%%%%%%%%%%%%%%%%%%%%%%%%%%%%%%555

\clearpage
\phantomsection
\addcontentsline{toc}{section}{References}
\bibliography{sn-bibliography}

\end{document}